\documentclass[review]{elsarticle} %
\PassOptionsToPackage{hyphens}{url}\usepackage{hyperref}
\usepackage[T1]{fontenc}%

\usepackage[per-mode=symbol]{siunitx}

\usepackage{subfig}

\journal{Computer Physics Communications}

\usepackage{booktabs}
\usepackage{graphicx}
\usepackage{amssymb}
\usepackage[fleqn]{amsmath}
\usepackage{moresize}%

\DeclareMathOperator{\sign}{sign}
\DeclareMathOperator{\erf}{erf}
\newcommand{\phimax}{\ensuremath{{\varphi^\text{max}}}}
\newcommand{\phimin}{\ensuremath{{\varphi^\text{min}}}}
\newcommand{\vecki}{\ensuremath{\vec k_i}}
\newcommand{\veckf}{\ensuremath{\vec k_f}}
\newcommand{\hatki}{\ensuremath{\hat{k}_i}}
\newcommand{\hatkf}{\ensuremath{\hat{k}_f}}
\newcommand{\Ei}{\ensuremath{E_i}}
\newcommand{\Ef}{\ensuremath{E_f}}

\newcommand{\tprime}{\ensuremath{t^\prime}}

\newcommand{\order}[1]{\ensuremath{\mathcal{O}(#1)}}
\newcommand{\thalf}{\tfrac{1}{2}}
\newcommand{\expmhalf}[1]{\ensuremath{\exp\!\left(-\thalf#1\right)}}

\newcommand{\thetabragg}{\ensuremath{\theta_\text{B}}}
\newcommand{\thetak}{\ensuremath{\theta_k}}
\newcommand{\thetan}{\ensuremath{\theta_n}}
\def\LRA{\Leftrightarrow}
\def\RA{\Rightarrow}
\newcommand{\labsec}[1]{\label{sec:#1}}
\newcommand{\refsecnumonly}[1]{\ref{sec:#1}}
\newcommand{\refsec}[1]{Section~\refsecnumonly{#1}}
\newcommand{\reftwosections}[2]{Sections~\refsecnumonly{#1} and \refsecnumonly{#2}}
\newcommand{\refthreesections}[3]{Sections~\refsecnumonly{#1}, \refsecnumonly{#2}, and \refsecnumonly{#3}}
\newcommand{\refsectionrange}[2]{Sections~\refsecnumonly{#1}--\refsecnumonly{#2}}

\newcommand{\labfig}[1]{\label{fig:#1}}
\newcommand{\reffignumonly}[1]{\ref{fig:#1}}
\newcommand{\reffig}[1]{Figure~\reffignumonly{#1}}
\newcommand{\Reffig}[1]{Figure~\reffignumonly{#1}}%
\newcommand{\refthreefigs}[3]{Figures~\reffignumonly{#1}, \reffignumonly{#2}, and \reffignumonly{#3}}
\newcommand{\reftwofigures}[2]{Figures~\reffignumonly{#1} and \reffignumonly{#2}}
\newcommand{\reffigrange}[2]{Figures~\reffignumonly{#1}--\reffignumonly{#2}}
\newcommand{\refsubfigfrommaincaption}[1]{Figure~\protect\subref{fig:#1}}
\newcommand{\Refsubfigfrommaincaption}[1]{Figure~\protect\subref{fig:#1}}%
\newcommand{\reftwosubfigsfrommaincaption}[2]{Figures~\protect\subref{fig:#1} and ~\protect\subref{fig:#2}}
\newcommand{\labeqn}[1]{\label{eqn:#1}}
\newcommand{\refeqnnumonly}[1]{\ref{eqn:#1}}
\newcommand{\refeqn}[1]{Eq.~\refeqnnumonly{#1}}
\newcommand{\reftwoeqns}[2]{Eqs.~\refeqnnumonly{#1} and \refeqnnumonly{#2}}

\begin{document}

\sisetup{range-phrase=-}
\sisetup{range-units=single}

\begin{frontmatter}

\title{Elastic neutron scattering models for NCrystal}
\author[addressess]{T.~Kittelmann\corref{mycorrespondingauthor}}
\author[addressess,addressdtunutech]{X.-X.~Cai\corref{presentrefxx}}
\cortext[mycorrespondingauthor]{Corresponding author. {\em Email address:} \texttt{thomas.kittelmann@ess.eu}}
\cortext[presentrefxx]{Present affiliation: China Spallation Neutron Source, Dongguan, China}

\address[addressess]{European Spallation Source ERIC, Sweden}
\address[addressdtunutech]{Technical University of Denmark, Denmark}

\begin{abstract}
The \texttt{NCrystal} library provides a range of models for simulation of both
elastic and inelastic scattering of thermal neutrons in a range of material
structures. This article presents the available models for elastic scattering,
and includes detailed discussion of their theoretical background, their
implementation, and in particular their
validation. The lineup includes a model for Bragg diffraction in crystal powders
as well as one for incoherent elastic scattering, but the main focus is given to
models of Bragg diffraction in ideally imperfect single crystals: both for the
most widely applicable model of isotropic Gaussian mosaicity, and for a more specific
model of layered single crystals which is relevant for materials such as pyrolytic graphite.  Although
these single crystal models are utilising computationally efficient
approximations where appropriate, attention is given to the provision of precise
and trustworthy results also for the extreme cases of back-scattering, forward-scattering,
and crystals with very large mosaic spreads.  Together
with \texttt{NCrystal}'s other features for crystal structure initialisation and
inelastic physics, the presented models enable realistic modelling of
components at neutron scattering instruments in frameworks like \texttt{Geant4}
and \texttt{McStas}, including monochromators, analysers, filters, support
materials, shielding, and many kinds of samples. As a byproduct of the work,
an improved formula for approximating cross sections in isotropic single crystals with
Gaussian mosaicity is provided.
\end{abstract}

\end{frontmatter}

\begin{small}
\noindent
{\em Manuscript Title:} Elastic neutron scattering models for NCrystal \\
{\em Authors:} T.~Kittelmann and X.-X.~Cai \\
{\em Program Title:} NCrystal                                          \\
{\em Journal Reference:}                                      \\
{\em Catalogue identifier:}                                   \\
{\em Licensing provisions:} Apache License, Version 2.0 (for core \texttt{NCrystal}).\\
{\em Programming language:} \texttt{C++}, \texttt{C} and \texttt{Python}       \\
{\em Operating system:} Linux, OSX, Windows                        
{\em Classification:} 4, 7.6, 8, 17 \\
{\em PACS codes:} \texttt{25.40.Dn}, \texttt{28.20.Cz}, \texttt{02.70.Uu}, \texttt{61.05.F-} \\
\end{small}

\section{Introduction}\labsec{intro}

The software package \texttt{NCrystal}~\cite{ncrystal2019} is an open source
software package, with capabilities for modelling of thermal neutron transport in a variety of
materials. The code is freely available~\cite{ncrystalwww} and released under a
liberal open source license~\cite{apache2}. The initial
release, \texttt{v1.0.0}~\cite{ncrystal2019}, focused on Bragg diffraction in
crystals, while release \texttt{v2.0.0}~\cite{ncrystalv2d0d0zenodo} introduced
realistic modelling of both inelastic and incoherent elastic scattering
processes, as well as the support for certain non-crystalline materials like
liquids. The following releases up until \texttt{v2.2.1}~\cite{ncrystalv2d2d1zenodo}, were
mostly focused on technical changes not affecting the physics modelling (the
notable exception being that atomic data definitions were made more flexible,
allowing for atoms not simply being natural elements).

The publication~\cite{ncrystal2019} provided an in-depth overview
of the \texttt{NCrystal} framework, including issues of configuration and material definition,
and how users might interact with the library: through standalone tools,
interfaces for \texttt{C++}, \texttt{C} and
\texttt{Python} programming languages, or indirectly by using \texttt{NCrystal}
to enhance the physics capabilities of simulation packages such
as \texttt{Geant4}~\cite{geant4a,geant4b,geant4c}
and \texttt{McStas}~\cite{mcstas1,mcstas2}. It also provided a review of the
relevant neutron scattering theory, and focused in particular on the
initialisation of crystal structures and form factors, and included benchmarks
versus existing crystallographic software and powder diffraction
data. It did not, however, go into detailed discussions of the actual
implementation of particular neutron scattering models. The present article improves upon
this situation by providing a thorough review of the models for elastic
scattering presently available in \texttt{NCrystal}, including details of implementation,
performance, and validation.

Although not discussed further in the present article, inelastic scattering does
contribute to a few of the validation plots presented, and it is therefore
worthwhile to briefly
note that the improved support for inelastic scattering introduced
in \texttt{NCrystal} \texttt{v2.0.0} is based on sampling of scattering
kernels~\cite{xxcaisampling2018}, which can be either provided by the users or
calculated on-the-fly by \texttt{NCrystal} from phonon density of state (DOS)
curves. This latter calculation relies on an independent implementation of a method due to
A.~Sj{\"o}lander~\cite{sjolander1958}, which is also used for a similar purpose
in well established
applications like \texttt{LEAPR}~\cite{njoy2012}. The combination of these new
inelastic models and the elastic models described in the present article,
enables highly realistic simulations of typical components at neutron scattering
instruments, including beam filters, monochromators, analysers, detectors,
support materials and many kinds of samples. The realism is increased to
unprecedented levels when \texttt{NCrystal} is used as a backend in an
application \texttt{Geant4}, where the injection of realistic thermal neutron
scattering models nicely complements the pre-existing features for support of
complicated geometries and both nuclear- and electromagnetic physics.

It should be noted that while inelastic models improved greatly
between \texttt{NCrystal} \texttt{v1.0.0} and \texttt{v2.0.0}, the elastic
physics described in the present article is, on the other hand, essentially
identical in the two releases. The one notable exception to this is that the
sampling of incoherent elastic scattering events
in \texttt{NCrystal} \texttt{v1.0.0} was implemented in an inconsistent manner,
resulting in purely isotropic angular distributions and incorrect non-zero
energy transfers. The incoherent elastic \emph{cross sections} were, however,
evaluated correctly also in \texttt{NCrystal} \texttt{v1.0.0}. Where it makes a
difference, all figures in the present article are based
on \texttt{NCrystal} v2.0.0.

After establishing some common formalism and terminology, and reviewing common
features of Bragg diffraction in \refsec{commonbragg}, the next three sections
are devoted to Bragg diffraction under particular models of crystallite
distributions: powders (\refsec{pcbragg}), single crystals with isotropic
Gaussian mosaicity (\refsec{scbragg}), and single crystals with a layered structure
similar to pyrolytic graphite (\refsec{lcbragg}). Along the way, issues such as
theory, details of implementation, and validation are
considered. A noteworthy byproduct of the work on single crystals is the
derivation of a formula in \refsec{scbragg::inteval} (\refeqn{circleintegralapproxformula}) which can be used to
approximate scattering cross sections. This formula is an improvement over the
widely used Gaussian approximation, in that it more accurately incorporates
effects related to large mosaic spreads or configurations approaching
back-scattering. Next, incoherent elastic scattering is discussed in
\refsec{incel}. This section also considers the validity of the commonly assumed
isotropic angular distribution of this process.
Finally, the computational efficiency of all the presented models is discussed
in \refsec{timing}, and possible future developments are considered in
\refsec{outlook}.

\section{Theoretical background and features of Bragg diffraction}\labsec{commonbragg}

The theory of scattering by thermal neutrons in crystals is reviewed
in~\cite[Sec.~2]{ncrystal2019}. There, it is described how neutron scattering in
crystals can be treated under the Born approximation, under certain general
conditions. Notably this includes the assumption that considered crystal systems are ``ideally imperfect'' in the
sense that the crystal structure can essentially be assumed to be perfect inside
tiny crystal grains, or ``crystallites'', but that interference between
interactions in separate grains can be neglected due to their imperfect mutual
alignments. Formulas are also developed
under the assumption that thermal movements of individual atoms around their
nominal positions in the lattice are essentially harmonic and isotropic. Based
on this, thermal fluctuation effects are incorporated via Debye-Waller factors,
that are themselves estimated based on mean-squared-displacements estimated
via the Debye model and a phenomenological parameter, the Debye temperature.
Finally, it is discussed in~\cite[Sec.~4]{ncrystal2019} how \texttt{NCrystal} is
able to derive lists of all relevant reflection planes in a given crystal, along
with their associated parameters like form factors, $d$-spacings, and plane
normals. For more details, readers are referred to~\cite{ncrystal2019} and
references therein, in
particular~\cite{squires_2012,schober2014,marshalllovesery1971}.

The scattering functions or cross sections for coherent elastic scattering in
crystallites, so-called \emph{Bragg diffraction}, can be decomposed
into contributions corresponding to reflections by individual lattice
planes (also known as reflection planes). Such lattice planes are indexed by
Miller indices, $hkl$, and are for a
given crystal associated with a reciprocal lattice vector, $\vec\tau_{hkl}$,
interplanar spacing ($d$-spacing), $d_{hkl}$ and (squared) form factors,
$|F_{hkl}|^2\equiv|F(\vec\tau_{hkl})|^2$.\footnote{Vectors are here and
throughout the text donated with arrows ($\vec{a}$), while the absence of arrows
indicate the corresponding scalar magnitudes ($a\equiv|\vec{a}|$). Additionally,
unit vectors are donated with hats ($\hat{a}\equiv\vec{a}/a$).} The form factors
capture the detailed layout of the crystal unit cell as well as the intrinsic
strength of the relevant neutron-nuclei interactions, and $\vec\tau_{hkl}$ is
the vector of length $\tau_{hkl}=2\pi/d_{hkl}$ which points along the lattice
plane normal, $\hat{n}_{hkl}$. For brevity, the $hkl$ subscripts
might be left out occasionally, e.g.\ writing $\hat{n}$ instead of $\hat{n}_{hkl}$.

The contribution of a given lattice plane $hkl$ to the differential coherent elastic cross
section for scattering the incident neutron with wavevector $\vecki$ and energy
$\Ei$ into the final state with wavevector $\veckf$ and energy $\Ef$, is given
by~\cite[Sec.~2.3]{ncrystal2019}:
\begin{align}
  \frac{d^2\sigma^{hkl}_{\vecki\RA\veckf}}{d\Omega_f d\Ef} = \,& \frac{(2\pi)^3\delta(\Delta{E})}{V_\text{uc}n_a}
  \delta(\vec{Q}-\vec\tau_{hkl})
  |F_{hkl}|^2\labeqn{diffxscohelhkl}
\end{align}
Which is normalised to the number of atoms in the target.
Here, $V_\text{uc}$ is the unit cell volume, $n_a$ the number of
atoms per unit cell, and $\vec{Q}\equiv\veckf-\vecki$ the momentum transfer. The
factor of $\delta(\Delta{E})\equiv\delta(\Ef-\Ei)$ imposes energy conservation, and hence the
scattering is elastic with non-vanishing cross section only when $k_i=k_f$.  The
final factor of $\delta(\vec{Q}-\vec\tau_{hkl})$ is what gives rise to the
rich phenomenology and distinctive patterns of Bragg diffraction. It requires a
momentum transfer equal to the reciprocal lattice vector in question, which
is only possible when the \emph{Bragg condition} is fulfilled:
\begin{align}
  2k_i\geq|\vec\tau_{hkl}|\quad\LRA\quad\lambda\leq2d_{hkl}
  \labeqn{braggcondition}
\end{align}
Where $\lambda=2\pi/k_i$ is the neutron wavelength. When the condition is fulfilled, the scattering angle $\theta$ will satisfy the \emph{Bragg equation}:
\begin{align}
  \lambda=2d_{hkl}\sin(\theta/2)=2d_{hkl}\sin\thetabragg
  \labeqn{braggequation}
\end{align}
Where also the Bragg angle, $\thetabragg\equiv\theta/2$, has been
introduced. For the purposes of the present article, the
complementary angle $\alpha$ is additionally defined as:
\begin{align}
  \alpha\equiv\frac{\pi}{2}-\thetabragg
  \labeqn{alphadef}
\end{align}
With this definition, $\alpha$ is given as the angle between $-\vecki$ and
$\hat{n}$ when the alignment satisfies \refeqn{braggequation}. Thus, Bragg
diffraction can take place only when the plane normal $\hat{n}$ is located on a
cone around $-\vecki$ of opening angle $\alpha$. The intersection of this cone
with the unit sphere defines a circle of radius $\sin\alpha$, which in the
present article will be referred to as the Bragg circle.

As the crystalline systems considered here are at the macroscopic scale composed
of a number of independently oriented microscopic crystallites, one must
average \refeqn{diffxscohelhkl} over the distribution of microscopic crystallite
orientations in order to derive macroscopic cross sections. Denoting such
a \emph{mosaicity distribution} of crystallite orientations with $W$, the total macroscopic cross sections
due to a particular $hkl$ plane is thus given by an integral over neutron final
states and crystallite orientations:
\begin{align}
  \sigma^{hkl}(\vecki)= \,& \int\! d\Omega_{\hat{n}}W(\hat{n}) \!\int\! d\Omega_f\!\int\! d\Ef
  \frac{(2\pi)^3\delta(\Delta{E})}{V_\text{uc}n_a}\delta(\vec{Q}-\vec\tau_{hkl})|F_{hkl}|^2
  \labeqn{sighklki1}
\end{align}
Where $\Omega_{\hat{n}}$ represents the solid angles covered by the
direction of the normal
$\hat{n}\equiv\vec\tau_{hkl}/\tau_{hkl}$ in specific crystallites, and
$W(\hat{n})$ is the density of crystallites expressed as a function of
$\hat{n}$. Denoting the cosine of the angle between
$\hat{n}$ and $-\vecki$ with $\mu$ and a corresponding azimuthal angle
with $t$, $W(\hat{n})$ can be written as $W(\mu,t)$ and \refeqn{sighklki1} becomes:
\begin{align}
  \sigma^{hkl}= \,&
  \int_{-\pi}^{\pi}\!\!\!\!dt\!\int_{-1}^{1}\!\!\!d\mu\,W(\mu,t) \!\int\! d\Omega_f\!\int\! d\Ef
  \frac{(2\pi)^3\delta(\Delta{E})}{V_\text{uc}n_a}\delta(\vec{Q}-\vec\tau_{hkl})|F_{hkl}|^2
\end{align}
If \refeqn{braggcondition} is not satisfied, this trivially evaluates to
zero, as $\vec{Q}$ can then never equal $\vec\tau_{hkl}$. Otherwise it can be evaluated using the identity
$\int{}\!d\Omega\delta(\vec{r}-\vec{a})=2a^{-1}\delta(r^2-a^2)$:
\begin{align}
  \sigma^{hkl}= \,&
  \int_{-\pi}^{\pi}\!\!\!\!dt\!\int_{-1}^{1}\!\!\!d\mu\,W(\mu,t)
\frac{(2\pi)^3|F_{hkl}|^2}{V_\text{uc}n_a}
\frac{2\delta(\tau_{hkl}^2-2k_i\tau_{hkl}\mu)}{\left|\vecki+\vec\tau_{hkl}\right|}\nonumber\\
= \,&
  \frac{(2\pi)^3|F_{hkl}|^2}{V_\text{uc}n_ak_i^2\tau_{hkl}}\int_{-\pi}^{\pi}\!\!\!\!W(\mu=\tau_{hkl}/2k_i,t)\,dt
  \nonumber\\
= \,&
  \frac{\lambda^3\sin\alpha|F_{hkl}|^2}{V_\text{uc}n_a\sin2\alpha}\int_{-\pi}^{\pi}\!\!\!\!W(\mu=\cos\alpha,t)\,dt
\end{align}
Where it was used that $\lambda=2\pi/k_i$, $\cos\alpha=\lambda/2d_{hkl}=\tau_{hkl}/2k_i$ (since
$\tau_{hkl}=2\pi/d_{hkl}$), and $\sin2\alpha=2\sin\alpha\cos\alpha$.
Thus, the resulting cross section can be factorised into two parts:
\begin{align}
  \sigma^{hkl}\equiv q_{hkl}\times g_{hkl}\labeqn{qstrengthandgstrengthfactors}
\end{align}
Here, $q_{hkl}$ represents an intrinsic strength of the interaction:
\begin{align}
  q_{hkl}\equiv\frac{\lambda^3|F_{hkl}|^2}{V_\text{uc}n_a\sin2\alpha}=\frac{\lambda^3|F_{hkl}|^2}{V_\text{uc}n_a\sin2\thetabragg}
  \labeqn{qstrengthfactordef}
\end{align}
And $g_{hkl}$ is a geometrical factor depending on the Bragg angle, mosaicity distribution,
and direction of the incoming neutron. It is given by an integration of the
crystallite densities over the directions where Bragg diffraction is possible
(i.e.\ along the Bragg circle which has a radius of $\sin\alpha$):
\begin{align}
  g_{hkl}\equiv\int_{-\pi}^{\pi}\!\!\!\!W(\mu=\cos\alpha,t)\sin\alpha\,dt=\sin\alpha\int_{-\pi}^{\pi}\!\!\!\!W(\mu=\cos\alpha,t)\,dt
  \labeqn{geomfactordef}
\end{align}

In addition to providing cross section values, implementations of Bragg
diffraction models in \texttt{NCrystal} must also provide Monte Carlo-based
sampling of the direction of $\veckf$ in case of a scattering event. When multiple
reflection planes contribute, one is first trivially selected at random, with a
probability given by its relative contribution to the total cross section. Next,
scattering on the chosen $hkl$ plane can proceed using whatever method is best
suited to the specific mosaicity distribution. One such method proceeds as follows: first an
actual (as opposed to nominal) direction of the plane normal, $\hat{n}$, must be
sampled according to the contribution to the cross section. This corresponds to
selecting a specific orientation of the crystallite in which the neutron scatters. Subsequently, the
neutron must undergo a specular reflection on the plane defined by that normal,
resulting in $\veckf=\vecki+\tau_{hkl}\hat{n}$. The plane normal sampling can
be carried out by sampling a value of $t$ according to the contribution to the
integral in \refeqn{geomfactordef}, or in other words
$P(t)\propto{W}(\mu\!=\!\cos\alpha,t)$.

\section{Crystal powders}\labsec{pcbragg}

The simplest model for Bragg diffraction in \texttt{NCrystal} implements the
so-called powder approximation, in which individual crystallite orientations are
assumed to be completely independent and uniformly distributed over all solid
angles. This approximation is not only suitable for modelling actual crystal
powders, but can additionally be used to approximate interactions in
polycrystalline materials like metals or ceramics --- especially when the level
of correlation in crystallite orientation (``texture'') is low or when the setup
involves sufficiently large geometries or spread in incoming particles that
effects due to local correlations are washed out. Due to its simplicity and
usefulness, implementations of Bragg diffraction under the powder assumption are
widely available (for
instance~\cite{nxslib1,cripo,mcstascompman,vitess1,nxsg4}). The implementation in
\texttt{NCrystal} has a particular focus on efficiency, which is necessary since
the aim is to ensure realistic cross sections also at shorter wavelengths,
potentially requiring very long lists of reflection planes
(cf.~\cite[Sec.~4]{ncrystal2019}).

The mosaicity distribution in the powder approximation is a constant,
$W=1/4\pi$, which when inserted in \refeqn{geomfactordef} yields:
\begin{align}
  g_{hkl}= \frac{\sin\alpha}{2}
  \labeqn{geomfactordefpowder}
\end{align}
and thus:
\begin{align}
  \sigma^{hkl}(\lambda)= \frac{\lambda^3|F_{hkl}|^2}{V_\text{uc}n_a\sin2\alpha} \frac{\sin\alpha}{2} =
\frac{d_{hkl}|F_{hkl}|^2}{2V_\text{uc}n_a }\lambda^2
\end{align}
Unless \refeqn{braggcondition} is not satisfied, in which case $\sigma^{hkl}(\lambda)$
vanishes. The complete cross section for Bragg diffraction in a crystal powder
can thus be written as the following sum over all lattice plane families
satisfying \refeqn{braggcondition}:
\begin{align}
  \sigma^\text{powder}(\lambda) =\,& \frac{\lambda^2}{2V_\text{uc}n_a}
  \sum^{\lambda\leq2d_{hkl}}_{hkl}d_{hkl}|F_{hkl}|^2
  \labeqn{powderxssum}
\end{align}
Although it is straightforward to evaluate \refeqn{powderxssum}, a naive
implementation using it with a list of $N_{hkl}$ lattice planes to provide the
cross section for a particular neutron, would in the worst scenarios result in
evaluation times scaling as \order{N_{hkl}}. As shown in~\cite[Fig.~4]{ncrystal2019},
$N_{hkl}$ might be higher than \num{e6} when including planes with $d$-spacings
down to \SI{0.1}{\angstrom}, and the corresponding computational cost of
an \order{N_{hkl}} algorithm would be unacceptable for many applications.
Despite this, some existing powder cross section implementations, like in
the \texttt{nxslib}-based \texttt{Sample\_nxs}~\cite{nxslib1}
and \texttt{PowderN}~\cite{PowderN:willendrup2006} components of \texttt{McStas},
actually employ \order{N_{hkl}} algorithms. These implementations are,
however, typically focused on physics at longer wavelengths and are mostly used
with much shorter lists of lattice planes. The consequence is of course an
underestimated incoherent elastic cross section at shorter wavelengths, which
would be unacceptable for \texttt{NCrystal}.

The powder cross section code in \texttt{NCrystal} instead employs a
binary search technique, delivering cross section evaluation times scaling
as \order{\log{}N_{hkl}}. Furthermore, as the \texttt{NCrystal} interfaces use
neutron energies rather than wavelengths, the calculations and formula are
carried out directly based on the kinetic energy value, avoiding the square-root
call in the conversion $\lambda=\sqrt{h^2/2mE}$. The binary search involves two
arrays which both have entries ordered by $d$-spacing. The first array contains
values of $h^2/8md^2_{hkl}$, which are the corresponding energy values of a neutron with wavelengths
$2d_{hkl}$. The second array contains corresponding cumulative sums of $d_{hkl}|F_{hkl}|^2$. Thus, the index $i$ found by a binary comparison search for
the neutron energy in the first array can be used in the second array to
immediately evaluate the sum in \refeqn{powderxssum}. Additionally, the arrays
are kept as short as possible by merging entries with identical $d$-spacing
values. \Reffig{beopowderxs}
\begin{figure}
  \centering
  \includegraphics[width=0.85\textwidth]{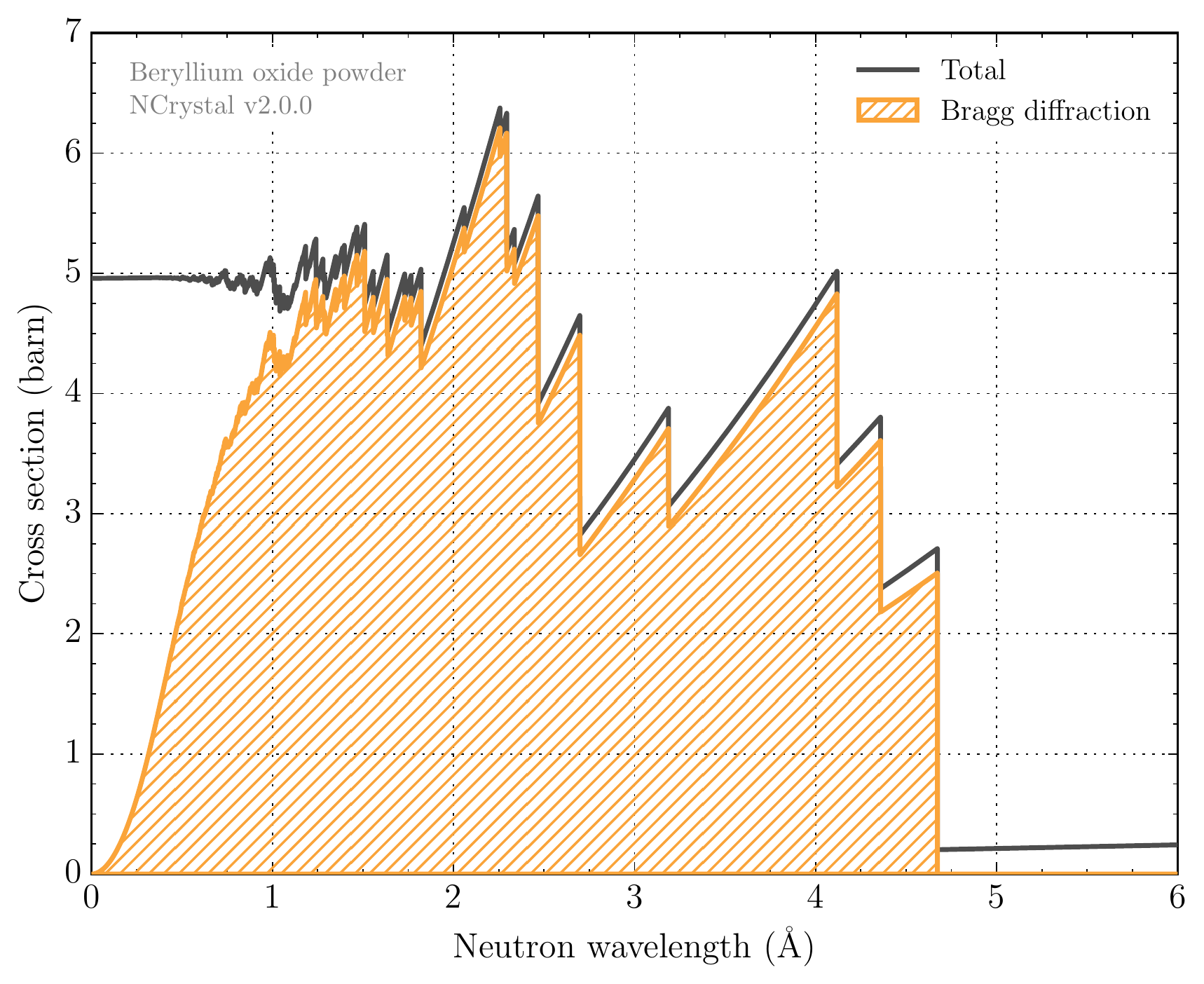}
  \caption{Neutron interaction cross sections in a beryllium oxide powder,
  with characteristic edges at $2d_{hkl}$ in the Bragg diffraction
  component. The total cross section is in this particular case dominated by Bragg
  diffraction and inelastic scattering, while absorption and incoherent elastic
  scattering are negligible.}
  \labfig{beopowderxs}
\end{figure}
shows a typical example of Bragg diffraction
cross section in a crystal powder, as provided by \texttt{NCrystal}.

In case of a scattering event, the relative probability for it to
involve a plane with a particular $d$-spacing value will depend on the total contribution
at that $d$-spacing to the sum in
\refeqn{powderxssum}. The Monte Carlo-based selection of such a $d$-spacing
value is carried out by another binary search, this time in the array with
cumulative sums of $d|F|^2$. The search target is the cumulative
value found at the index $i$ but multiplied with a pseudo random number from the
unit interval, thus ensuring the desired weight when selecting different
$d$-spacings. Specifically, the index $j$ resulting from this second binary
search can be used to access a value in the array of $h^2/8md^2$ values, which can be used to
extract the cosine of the scattering angle:
\begin{align}
   \cos\theta = 1-2\sin^2\thetabragg = 1-2\frac{h^2/8md^2}{E}
\end{align}
It trivially follows from the fact that $W=1/4\pi$ is a constant, that the
azimuthal scattering angle must be uniformly distributed between $-\pi$ and $\pi$.

Now, if the calling code is querying the non-oriented \texttt{NCrystal}
interface method (cf.~\cite[Table~1]{ncrystal2019}), the value of $\theta$ will be provided to the user after an
$\arccos$ evaluation. If instead (as will be most typical), the calling code uses the oriented vector interface,
providing $\hatki$ and expecting a value of $\hatkf$ in return, an
alternative procedure is used. This procedure does not require any expensive trigonometric
function calls, and is as fast as the non-oriented one, despite the fact that it must
sample the azimuthal scattering angle and deal with full directional vectors. First, a vector
is sampled isotropically on the unit sphere using an efficient
algorithm~\cite{marsaglia1972}, and the sampling is redone in the rare case of
providing a vector almost co-linear with $\hatki$. Next, a cross product between the sampled vector
and $\vecki$ yields a vector $\vec{v}$ orthogonal to $\vecki$ but with uniformly
distributed azimuthal scattering angle. Finally, $\hatkf$ is provided as:
\begin{align}
  \hatkf = \cos\theta\hatki  + \sqrt{\frac{1-\cos^2\theta}{|\vec{v}|^2}}\vec{v}
\end{align}
\Reffig{beopowderscatpat}
\begin{figure}[t]
  \centering
  \includegraphics[width=0.9\textwidth]{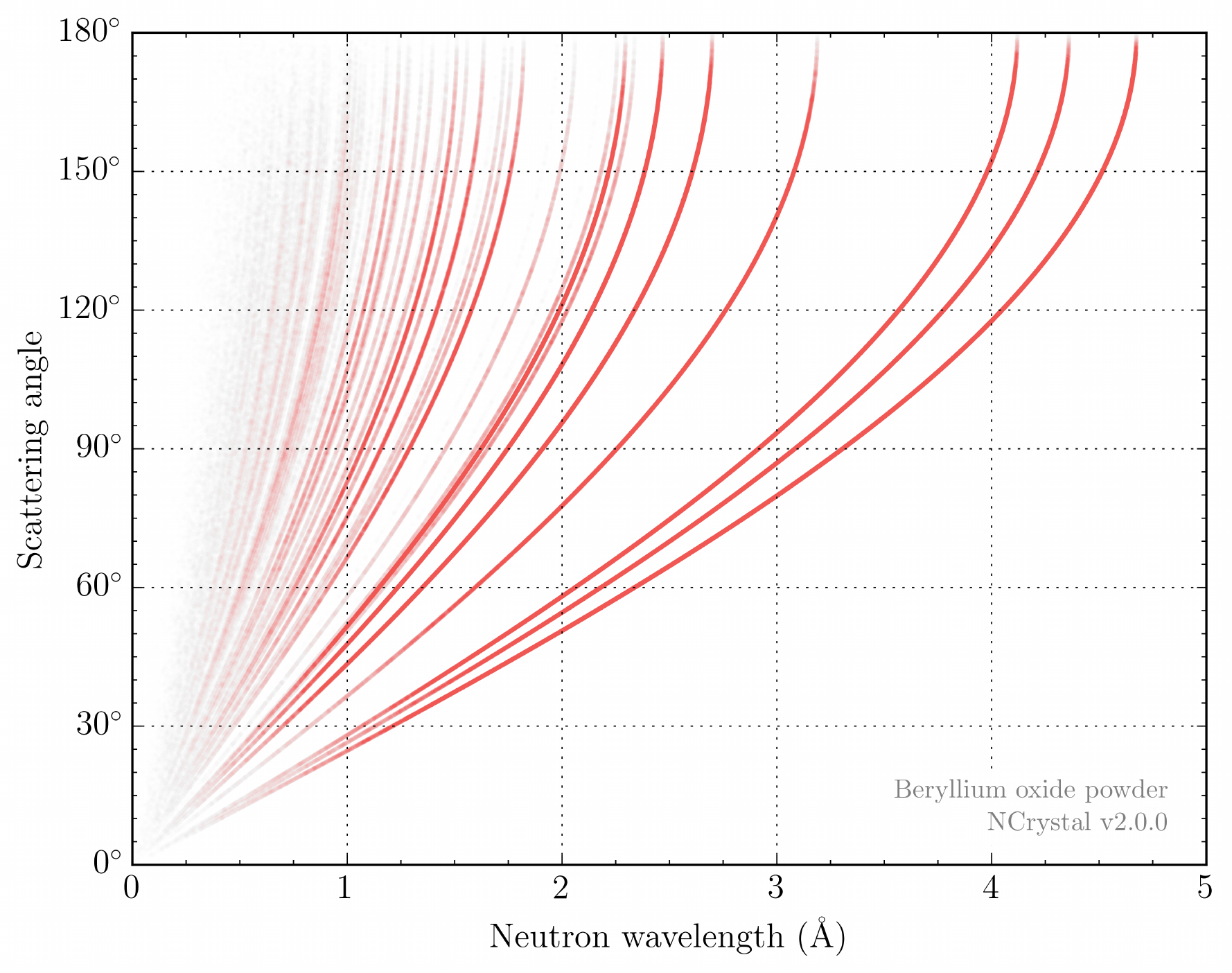}
  \caption{Neutron scattering angles resulting from Bragg diffraction in a
  beryllium oxide powder, for a total of \num{e6} scattering events sampled
  with \texttt{NCrystal}. Statistics at each wavelength are adjusted according
  to the wavelength dependent Bragg diffraction cross section.}
  \labfig{beopowderscatpat}
\end{figure}
shows an example of the sharply defined scattering angles in a crystal powder, as provided
by \texttt{NCrystal}.

Concerning computational efficiency of the powder diffraction code, benchmarks
were run on a typical 2019 portable computer and with a complicated crystal
structure (Corundum) with \order{\num{e6}} reflection planes. They show that
cross sections can be evaluated at a rate of \order{\SI{50}{\mega\hertz}}, while
sampling of scattering events is done at a rate of
\order{\SI{10}{\mega\hertz}}. In both cases, there is only a small dependency on
the wavelength of the incoming neutron. In \refsec{timing} the performance will
be compared in more details with other elastic models in \texttt{NCrystal}.
Finally, as regards validity of results, the code was very thoroughly
benchmarked and validated in connection with the work done to validate the
ability of \texttt{NCrystal} to initialise crystal structures
in~\cite[Sec.~4.4]{ncrystal2019}, to which interested readers are referred.

\section{Single crystals with isotropic Gaussian mosaicity}\labsec{scbragg}

As is often the case in physics, a Gaussian distribution lies at the heart
of the most widely used model for the distributions of crystallites in single
crystal materials utilised for monochromators, analysers, or filters at neutron
scattering facilities. In this model, introduced by
C.~G.~Darwin~\cite{darwin1922}, the deviation of crystallite orientations from
some reference is described by a Gaussian distribution of the angular deviation. The
term \emph{mosaic spread}, or simply \emph{mosaicity}, is commonly used to
indicate either the standard deviation or FWHM value of this angular
distribution, values of which in typical materials range from tens of
arc seconds to several degrees. Such a Gaussian mosaicity distribution is
considered \emph{isotropic} in the sense that the resulting distribution of
crystallite orientations is independent of the choice of reference orientation
--- a feature which is for instance not the case for the distribution discussed
in \refsec{lcbragg}. Obviously it is just the mosaicity distribution
which is considered isotropic in this sense, as derived properties like cross
sections will show a very strong dependency on the direction of the neutron.

The required integrations of angular Gaussian mosaicity distributions
in \refeqn{geomfactordef} are non-trivial and cumbersome at best, and as a
consequence software~\cite{mcstascompman,vitess1,nispmosaicxtalalg} dealing with
single crystals usually employ analytical approximations which are valid only
for small mosaicities and in the absence of back-scattering (scattering
angles close to $\pi$).\footnote{The authors of~\cite{nispmosaicxtalalg}
even goes so far as mistakenly concluding a breakdown of the theory in the
presence of back-scattering due to the factor of $1/\sin2\thetabragg$
in \refeqn{qstrengthfactordef}. However, the divergence from this factor for
$\thetabragg\to\pi/2$ is actually cancelled by the factor of
$\sin\alpha=\sin(\pi/2-\thetabragg)$ in \refeqn{geomfactordef} (the divergence
for $\thetabragg\to0$ is cancelled by the factor of $\lambda^3$
in \refeqn{qstrengthfactordef}).}

For the purposes of \texttt{NCrystal} as a generic simulation backend it is, however, desirable to provide
accurate results for any realistic value of mosaic spread and neutron state ---
preferably without unduly impacting computational efficiency when it can be avoided. For
these reasons, the details of the Gaussian mosaicity distribution and its
integration will be revisited carefully in the following, in order to arrive at
recipes for evaluations which are not only self-consistent and precise, but
which incorporate efficient approximations where possible. In fairness, it
should be noted that the lack of support for large mosaicities and
back-scattering allow some existing software to support features not
available in the model presented here. For instance,
the \texttt{Single\_crystal} component~\cite{mcstascompman} of \texttt{McStas}
supports different values of mosaic spread along different rotation axes, and
allows small Gaussian $d$-spacing fluctuations between crystallites. Given interest and
availability of manpower, such features could at some point be considered for
inclusion in \texttt{NCrystal} as well.

\subsection{Definition of the Gaussian mosaicity distribution}\labsec{scbragg::mosdef}

It is straightforward to simply define a Gaussian mosaicity distribution as:
\begin{align}
   W_\text{simple}(\delta)\propto\expmhalf{\delta^2/\sigma^2}
   \labeqn{gaussianmosdistsimple}
\end{align}
Where the angle $\delta$ signifies the deviation from the nominal reference
orientation, and $\sigma$ the mosaicity as a standard deviation value (the FWHM
mosaicity value is then approximately $2.3548\sigma$).
However, the distribution given in \refeqn{gaussianmosdistsimple} yields
non-zero densities for all $\delta\in[0,\pi]$, and therefore implies a finite
scattering cross section for any direction of the incoming neutron. While
faithful to the concept of a Gaussian having infinite tails, such an idealised
definition implies that the calculation of scattering cross sections for a
particular incident neutron always has to involve the estimation of partial
cross sections for scattering on \emph{all} lattice planes satisfying
\refeqn{braggcondition}. In particular at shorter neutron wavelengths where many
reflection planes must be considered, this can imply very long evaluation times --- with most
effort being spent integrating through truly negligible (and thus completely
uninteresting) parts of the Gaussian tails. Instead, it is desirable to cleanly
truncate the tails at some truncation angle, $\tau$, which when defined at some
reasonable value such as $\tau\equiv5\sigma$, allows for very significant
algorithmic speedups with minimal degradation of realism or precision. Of course, the actual
value of $\tau/\sigma$ could be a configurable parameter, depending on the
precision-to-performance trade-off needed for a particular use case. However,
for consistency and in order to keep configuration as simple as possible, only a single overall
precision parameter, $\epsilon$, is exposed to users, via the configuration
parameter named \texttt{mosprec} (cf.~\cite[Sec.~5]{ncrystal2019}). The idea
behind this parameter is to provide a way for users to easily tell \texttt{NCrystal} what level of accuracy they
(at minimum) require in provided results --- where requests for increased accuracy
obviously might impact computational speed negatively.
Thus, the value of
$\epsilon$, which by default is \num{e-3}, affects not only the value of
$\tau/\sigma$, but also other choices with implications for precision in the
implemented single crystal model: the granularity of lookup tables, the
terminating conditions of numerical integrations, and when certain approximation
formulas can be used instead of numerical integration, as will be discussed
further in \refsec{scbragg::inteval}. For the case of $\tau/\sigma$, it is
adjusted so that a Gaussian in a two-dimensional plane will have $1-\epsilon$ of
its volume inside a radius of $\tau$. Adding also a safety factor of 10\% and
requiring $\tau/\sigma$ to be at least 3, the relationship becomes:
\begin{align}
   \tau = \max\left(3,1.1\times\sqrt{-2\log\epsilon}\right)\times\sigma
   \labeqn{tauoversigmafrommosprec}
\end{align}
Which is shown in \reffig{mosprecntrunc}.
\begin{figure}
  \centering
  \includegraphics[width=0.7\textwidth]{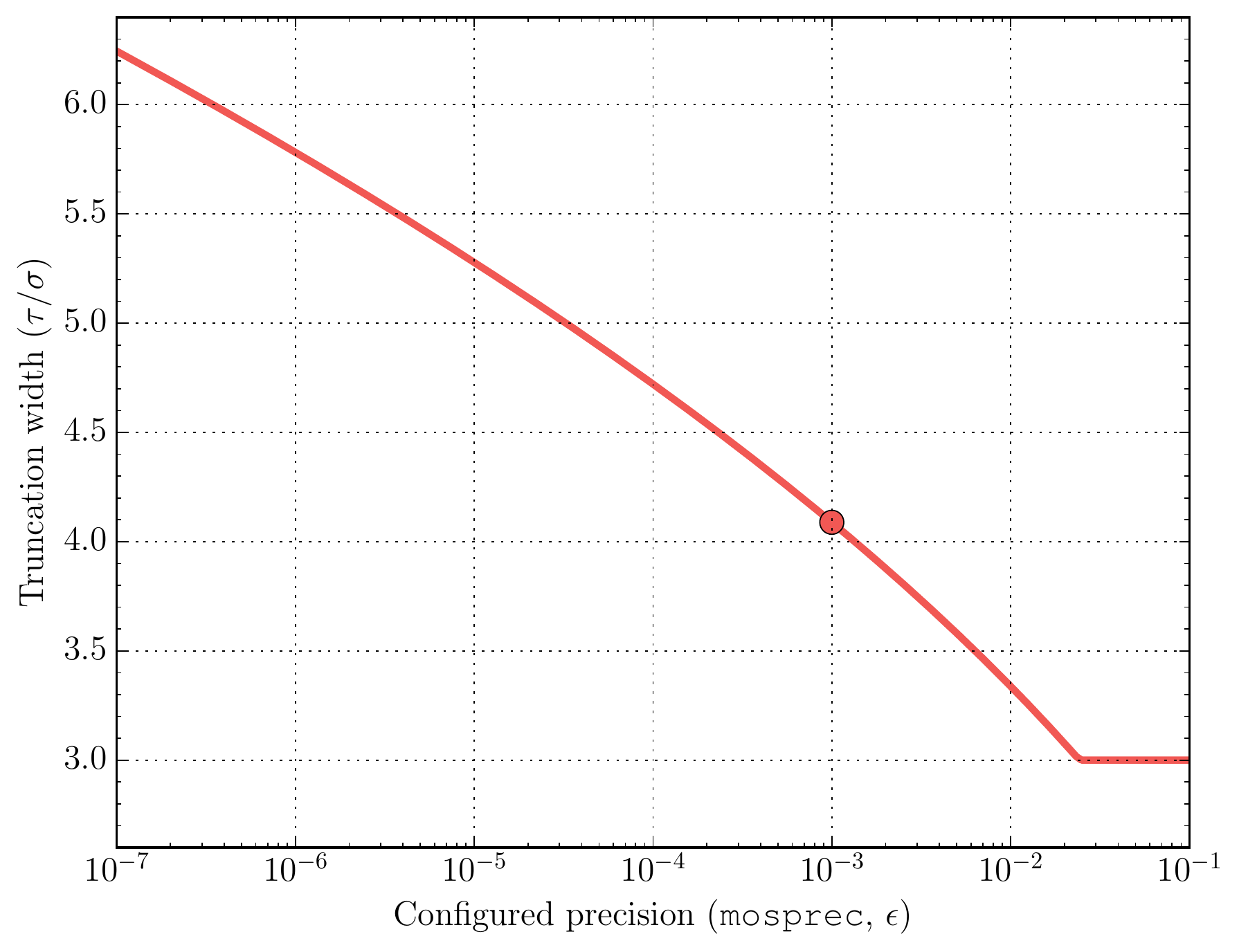}
  \caption{The relationship of \refeqn{tauoversigmafrommosprec} between the $\epsilon$ parameter (accessible to end
    users via the configuration parameter named \texttt{mosprec}) and the
    relative mosaic truncation width $\tau/\sigma$. The red dot indicates the
    default values: $\epsilon=\num{e-3}$ and $\tau/\sigma\approx4.0886$.}
  \labfig{mosprecntrunc}
\end{figure}
For the default value of
$\epsilon=\num{e-3}$, this implies a value of
$\tau\approx4.0886\sigma$. Conversely, a value of $\tau\approx5\sigma$ is
achieved by setting $\epsilon=\num{3.262e-5}$. In order to simplify later
calculations, it is explicitly required
that $\tau<\pi/2$, which guarantees that $\cos\tau$ is positive. At
$\epsilon=\num{e-7}$ this implies that $\sigma$ can at most be $14.4^\circ$,
corresponding to a FWHM mosaicity value of $33.9^\circ$. Fortunately, this value
is far above even the largest mosaic spreads encountered in practice.

Having thus determined both the mosaicity $\sigma$ and the truncation angle
$\tau$, the Gaussian mosaicity distribution will in \texttt{NCrystal} be defined as:
\begin{align}
   W(\delta) = N \expmhalf{\delta^2/\sigma^2}\Theta(\tau-\delta)
   \labeqn{gaussianmosdist}
\end{align}
Where the angle $\delta$ as before signifies the deviation from the nominal reference
orientation, $\Theta$ is the Heaviside step function enforcing the
truncation, and $N$ is a normalisation factor which can be determined by the condition:
\begin{align}
    &\int_0^{2\pi}\int_0^\pi W(\theta)\sin(\theta) d\theta d\phi = 1\hfil&&\nonumber\\
\labeqn{gaussmosnormintegraldef}\RA\qquad &N^{-1} = 2\pi  \int_0^\tau \expmhalf{\theta^2/\sigma^2}\sin(\theta) d\theta\hfil&&
\end{align}
During initialisation of a particular single crystal model,  \texttt{NCrystal}
determines $N$ by numerically
evaluating this integral to a relative error of less than \order{\num{e-12}}.

\subsection{Formulating the geometrical integral}\labsec{scbragg::mosintformulation}

The value of the integral in \refeqn{geomfactordef} with the Gaussian mosaicity distribution
in \refeqn{gaussianmosdist} only depends on $\alpha$, and the relative
angle between the incoming neutron and the nominal normal of the lattice plane
in question, $\hat{n}\equiv\vec{\tau}_{hkl}/{\tau}_{hkl}$. For
convenience this latter angle, $\gamma$, will be defined as the angle between
$\hat{n}$ and $-\vecki$ and a coordinate system specific to each neutron
direction and $hkl$ normal will be adopted, as illustrated
in \reffig{mosintegralon3dsphere}: $-\vecki$ lies
\begin{figure}
  \centering
  \includegraphics[width=0.99\textwidth]{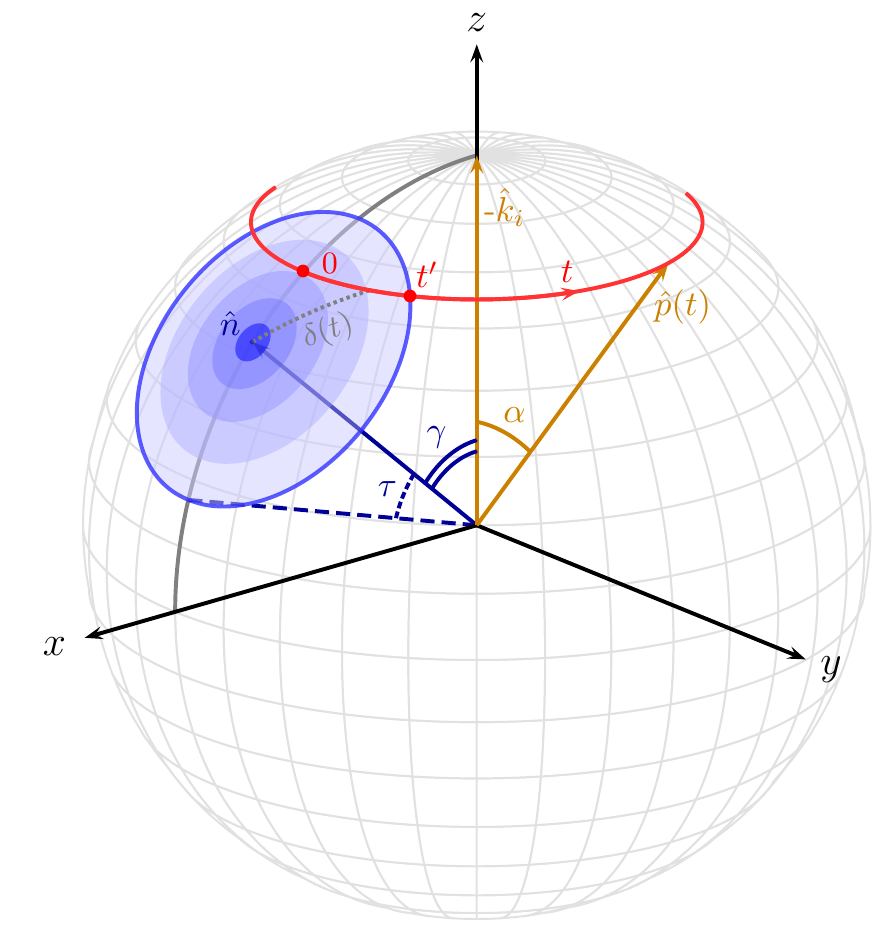}
  \caption{Coordinate system used for calculations related to the interaction of
  neutrons with a single reflection plane in a single crystal with isotropic
  Gaussian mosaicity
  (cf.\ \reftwoeqns{geomfactordef}{gaussianmosdist}). Here, $\hatki$ is the
  incident direction of the neutron, $\hat{n}$ is the nominal position of the
  plane normal, and $\tau$ is the truncation angle of the mosaicity
  distribution --- the strength of which is indicated with blue shaded areas. The
  angle between $-\hatki$ and plane normals satisfying the condition for Bragg
  diffraction is given as $\alpha=\pi/2-\thetabragg$, defining the Bragg circle
  (red). The remaining variables, $\hat{p}(t)$, $t$, $\tprime$, and $\delta(t)$,
  concern a parameterisation of the Bragg circle, which is discussed in the
  main text.}
  \labfig{mosintegralon3dsphere}
\end{figure}
along the positive $z$-axis, and the nominal normal is located in the
$xz$-plane at:
\begin{align}\labeqn{hatndef}
  \hat{n}=(\sin\gamma,0,\cos\gamma)
\end{align}
Bragg diffraction is only possible when plane normals are located at an angle of
$\alpha$ from $-\vecki$, defining a circle as indicated in red
in \reffig{mosintegralon3dsphere}, which can be parameterised as:
\begin{align}\labeqn{poftdef}
  \hat{p}(t)\equiv(\sin\alpha\cos t,\sin\alpha\sin t,\cos\alpha)\quad\text{for}\quad{}t\in[-\pi,\pi]
\end{align}
As mentioned previously, this circle is in the present article referred to as the Bragg circle. With
$\delta(t)$ indicating the angular separation between the nominal normal
position ($\hat{n}$) and $\hat{p}(t)$, \refeqn{geomfactordef} becomes:
\begin{align}
  g_{hkl}\equiv\sin\alpha\int_{-\pi}^{\pi}\!\!\!\!W(\delta(t))\,dt
  \labeqn{geomfactordefscbragg}
\end{align}
Taking the dot product of $\hat{n}$ and $\hat{p}(t)$ yields the relation:
\begin{align}\labeqn{cosd}
\cos{\delta(t)} = \sin\alpha\sin\gamma\cos t+\cos\alpha\cos\gamma
\end{align}
As $\sin\alpha$ and $\sin\gamma$ are both non-negative, it is straightforward
to confirm (with trigonometric addition formulas) what one might also intuitively infer from \reffig{mosintegralon3dsphere}: that the minimal value of
$\delta(t)$ is always attained at $t=0$ and is
$\delta_0\equiv\delta(0)=|\alpha-\gamma|$. Likewise, a maximal value of
$|\alpha+\gamma|$ is attained at $t=\pm\pi$.  Now, with the chosen definitions,
$\delta(-t)=\delta(t)$, so:
\begin{align}\labeqn{xsprinciple}
g_{hkl} = 2N\!\sin\alpha\int_0^{\tprime}\!\exp\!\left[\frac{\big(\arccos\!\left[\sin\alpha\sin\gamma\cos t+\cos\alpha\cos\gamma\right]\big)^2}{-2\sigma^2}\right]dt
\end{align}
where $\tprime$ is the point in $[0,\pi]$ satisfying $\delta(\tprime)=\tau$, unless the Bragg
circle is fully contained within the region of non-zero mosaic density
(i.e.\ $|\alpha+\gamma|<\tau$) in which case $\tprime=\pi$. Of course, if the Bragg circle
does not intersect the region of non-zero mosaic density at all (i.e.\
$|\alpha-\gamma|>\tau$), $g_{hkl}$ is trivially vanishing.

\subsection{Pre-selecting contributing lattice planes}\labsec{scbragg::normsearch}

Before proceeding to evaluate \refeqn{xsprinciple}, it is important to note that
in many typical scenarios, the vast majority of lattice planes
 fulfilling the Bragg condition $2d_{hkl}<\lambda$, will fail the
geometrical requirement $|\alpha-\gamma|<\tau$, and therefore not contribute to the
scattering at all. To avoid wasting resources on evaluating \refeqn{xsprinciple}
for such planes, it is therefore crucial to be able to discard such
non-contributing lattice planes with minimal effort. In practice this
pre-filtering is often where most CPU resources are spent during a simulation of
thermal neutrons in single crystals, with even small optimisations here
benefiting overall simulation time.

In the implementation described here, lattice plane data is first of all grouped
into ``families'' of planes having identical values of $d$-spacing and form
factors, differing only in directions of their plane normals.\footnote{Note that
the strict crystallographic definition of a reflection family only relies on
crystal symmetries. Thus, the families employed here might occasionally consist of more
than one family in the strict crystallographic sense.} For a given
incident neutron, the value of $\cos\alpha$ only needs to be calculated once per
family, and this can be done inexpensively using $\cos\alpha=\lambda\times(1/2d)$,
which costs a mere multiplication when evaluating for a given neutron
state. Additionally, lattice planes $(h,k,l)$ and $(-h,-k,-l)$ form natural
pairs with identical form factors and $d$-spacings, but opposite normals
(loosely speaking they can be thought of as representing the front and back side
of the same plane). Thus, both planes in such a pair will reside in the same
family, and for the sake of evaluating \refeqn{xsprinciple} they will only
differ in having complementary values of $\gamma$: if one plane has a certain
value of $\gamma$, the other will have the value $\pi-\gamma$. It is thus
advantageous to only cache one of two such paired normals in memory and deal
with both paired planes simultaneously.

Now, the requirement for a plane to have a non-zero contribution is:
\begin{align}
  &\delta(0)<\tau\nonumber\\
  \LRA\qquad& \cos(\delta(0))>\cos\tau\nonumber\\
  \LRA\qquad& \sin\alpha\sin\gamma > \cos\tau-\cos\alpha\cos\gamma\labeqn{normalsearchsasggtform}
\end{align}
As $\sin\alpha$ and $\sin\gamma$ are both non-negative, \refeqn{normalsearchsasggtform} is
always satisfied when the right hand side is negative. Thus,
the condition for a neutron to have non-zero scattering cross section with a given
normal can be rewritten as:
\begin{align}
  & (\sin\alpha\sin\gamma)^2>\left[\max(0,\cos\tau-\cos\alpha\cos\gamma)\right]^2\nonumber\\
 \LRA\qquad& (1-\cos^2\!\alpha)(1-\cos^2\!\gamma)>\left[\max(0,\cos\tau-\cos\alpha\cos\gamma)\right]^2\labeqn{normalsearchcheck}
\end{align}
Which is efficient to evaluate, since it involves only cheap non-branching
operations at each side of the inequality. The only trigonometric factor
particular to each pair of normals, $\cos\gamma$, is readily available via a dot
product between the cached normal and $\hatki$, but both $+\cos\gamma$ and
$-\cos\gamma$ must be considered in order to deal with both
of the paired planes being investigated jointly. However, the plane whose normal points into the same hemisphere as
$\hatki$ (i.e.\ $\cos\gamma<0$) can only ever contribute if the plane whose
normal points into the hemisphere opposite to $\hatki$ (i.e.\ $\cos\gamma>0$)
contributes. Formally this follows from the fact that $\cos\alpha$ is
non-negative, so $\cos\alpha\cos\gamma\leq\cos\alpha\left|\cos\gamma\right|$,
and therefore neither plane can contribute unless:
\begin{align}
(1-\cos^2\!\alpha)(1-\cos^2\!\gamma)>\left[\max\left(0,\cos\tau-\cos\alpha\left|\cos\gamma\right|\right)\right]^2\labeqn{combinednormalsearchcheck}
\end{align}
Only when \refeqn{combinednormalsearchcheck} is valid is it possible for either
of the two paired normals to contribute, in which
case \refeqn{normalsearchcheck} can be used to test them individually. In this
respect, note that in case of extreme forward scattering, it is entirely
possible for both of the two paired planes to
contribute to the scattering cross section for a particular neutron.\footnote{It is worth to point out,
that scattering contributions from $-\hat{n}$ does \emph{not} mean that neutrons
can scatter on reflection planes whose normals points into the same half-plane
as the neutron's direction --- which would indeed be impossible. Rather, for
large mosaicities and $\gamma\approx\pi/2$, there is a non-zero probability to
find a crystallite for which the actual (as opposed to nominal) $-\hat{n}$ does
not lie in this forbidden half-plane. And for the case of extreme forward
scattering, $\alpha\approx\pi/2$, there can even be a non-zero probability for finding a crystallite
for which scattering on $-\hat{n}$ fulfils the Bragg condition.}

\subsection{Evaluating the geometrical integral}\labsec{scbragg::inteval}

The evaluation of \refeqn{xsprinciple} is either performed through direct
numerical integration or, when possible, through the application of an appropriate
approximation. In both cases, values of $\sin\alpha$ and $\sin\gamma$ are
computed relatively inexpensively using $\sin=\sqrt{1-\cos^2}$. In case direct
numerical integration is required, the next step is the determination of
the upper limit of integration, $\tprime$.
When $\sin\gamma\sin\alpha\ne0$, one can insert $t=\tprime$ and $\delta(\tprime)=\tau$
in \refeqn{cosd} and get:
\begin{align}\labeqn{cost}
\cos \tprime = \frac{\cos(\tau)-\cos\alpha\cos\gamma}{\sin\gamma\sin\alpha}
\end{align}
The degenerate cases where $\sin\gamma$ or $\sin\alpha$ are vanishing must be
handled separately, but fortunately \refeqn{xsprinciple} is particularly simple to
evaluate in those scenarios: the right hand side vanishes when $\alpha=0$, and
the integrand is independent of $t$ when $\gamma=0$. In the non-degenerate cases, \refeqn{cost} can be
solved for $\tprime$ through a call to $\arccos$ when the right hand side leads
to a value in $[-1,1]$. Given the pre-selection described
in \refsec{scbragg::normsearch}, this only fails when the Bragg circle is fully
contained within the region of non-zero mosaic density, i.e.:
\begin{align}
           &   \delta(\pi) < \tau \nonumber\\
  \LRA\quad&   \cos(\delta(\pi)) > \cos\tau \nonumber\\
  \LRA\quad&  \frac{\cos\tau-\cos\alpha\cos\gamma}{\sin\alpha\sin\gamma} < -1
  \labeqn{scbragg::conditionbraggcircleinside}
\end{align}
In either case, the correct value of $\tprime$ is given by:
\begin{align}
 \tprime = \arccos\left[\max\left(-1,\frac{\cos(\tau)-\cos\alpha\cos\gamma}{\sin\gamma\sin\alpha}\right)\right]
\end{align}

Precise numerical integration of \refeqn{xsprinciple} is potentially very
expensive, as it requires the integrand to be evaluated repeatedly, needing
calls to mathematical functions $\exp$, $\arccos$, and $\cos$ at each sampled value of
$t$. The $\exp$ and $\arccos$ calls are avoided by generating, at initialisation
time, a cubic spline representation of the function
$f(x)=N\expmhalf{[\arccos(x)]^2/\sigma^2}$, covering the interval
$[\cos\tau,1]$, which must then be evaluated with
$x(t)=\sin\alpha\sin\gamma\cos{t}+\cos\alpha\cos\gamma$. The need to evaluate
$\cos{t}$ for multiple values of $t$ is eliminated by choosing a numerical integration
algorithm which always evaluates the integrand at a large number of equidistant
points along $t$. As the points are equidistant, cosine and sine need only be
evaluated at the leftmost point and the interpoint distance, while the cosine
values for all other points can then be generated from those using inexpensive
angular addition formulas. Additionally, it is obviously desirable to choose
an integration algorithm with a fast
convergence, in order to keep the number of evaluations
reasonable. The chosen algorithm is that of Romberg
integration~\cite{romberg1955}, but customised so as to always start by
evaluating the integrated function at 17 equidistant points in $[0,\tprime]$ in one go, and
immediately constructing an accurate estimate from those. Thus, the first 4
layers of the traditional Romberg algorithm are combined in order to avoid
evaluating the integrand at too few points at a time. In the rare cases where convergence is
slow, additional function evaluations will be requested as the Romberg algorithm
dives in to its 5th layer and beyond: first 16 additional equidistant points,
then 32, etc. It is important to note that the very fast convergence
rate of Romberg integration is realised only if the integrand is sufficiently
smooth, which is why the lookup table for $f(x)$ has to be implemented with a
cubic spline, rather than a simpler scheme involving linear interpolation.

While the numerical integration of \refeqn{xsprinciple} is thus performed with
utmost attention to efficiency, it is still preferable to carry out the
evaluation with a cheaper closed-form expression. Although such an
expression is not generally available, it is possible to find one which is a
good approximation under certain very common conditions. Consider
again \refeqn{cosd}: when $\delta(t)$ and $t$ are both reasonably small, their
cosines can be approximated with second order Taylor expansions:
\begin{align}
&1-[\delta(t)]^2/2 ~= \sin\alpha\sin\gamma(1-t^2/2)+\cos\alpha\cos\gamma\nonumber\\
\LRA\qquad &1-[\delta(t)]^2/2 ~= \cos(\alpha-\gamma)- (t^2/2)\sin\alpha\sin\gamma
\end{align}
The magnitude of $\delta_0\equiv\delta(0)=|\alpha-\gamma|$ is strictly smaller
than or equal to $\delta(t)$, so we
can expand the remaining cosine as well, and get:
\begin{align}\labeqn{dtsqapprox}
&1-[\delta(t)]^2/2 ~= 1-\delta_0^2/2-(t^2/2)\sin\alpha\sin\gamma\nonumber\\
\LRA\qquad &[\delta(t)]^2 ~= \delta_0^2+t^2\sin\alpha\sin\gamma
\end{align}
Using this, \refeqn{xsprinciple} can be approximated by:
\begin{align}
g_{hkl} =\,& 2N\!\sin\alpha\int_0^{\tprime} \exp\left(\frac{\delta_0^2+t^2\sin\alpha\sin\gamma}{-2\sigma^2}\right)dt\nonumber\\
=\,& 2N\!\sin\alpha\exp\left(-\thalf\delta_0^2/\sigma^2\right) \int_0^{\tprime} \exp\left(\frac{t^2\sin\alpha\sin\gamma}{-2\sigma^2}\right)dt
\end{align}
Performing a variable change $u(t)\equiv\sqrt{t\sin\alpha\sin\gamma}/\sigma$ gives:
\begin{align}
g_{hkl} =\,& 2N\sqrt{\frac{\sin\alpha}{\sin\gamma}}\sigma\exp(-\delta_0^2/(2\sigma^2)) \int_0^{u(\tprime)}\exp(-u^2/2)du\nonumber\\
      =\,& \sqrt{2\pi}N\sigma\exp(-\thalf\delta_0^2/\sigma^2)\sqrt{\frac{\sin\alpha}{\sin\gamma}}\erf(u(\tprime)/\sqrt{2})
\end{align}
With the error function, $\erf(x)\equiv2\pi^{-1/2}\int_0^x\exp(-s^2)ds$. Using
$\delta(\tprime)=\tau$ and \refeqn{dtsqapprox}, it follows that
$u(\tprime)=\sqrt{(\tau^2-\delta_0^2)/\sigma^2}$, and thus:
\begin{align}
g_{hkl} =\,&  N\sqrt{2\pi}\sigma
  \expmhalf{\delta_0^2/\sigma^2}
  \erf\left(\sqrt{\frac{\tau^2-\delta_0^2}{2\sigma^2}}\right)
  \sqrt{\frac{\sin\alpha}{\sin\gamma}}\nonumber\\
=\,&
  \left(\frac{1}{\sqrt{2\pi}\sigma}\expmhalf{\delta_0^2/\sigma^2}\right)
  \erf\left(\sqrt{\frac{\tau^2-\delta_0^2}{2\sigma^2}}\right)
  \sqrt{\frac{\sin\alpha}{\sin\gamma}}
  \frac{N}{1/(2\pi\sigma^2)}\labeqn{circleintegralapproxformula}
\end{align}
This result can be readily interpreted if one first considers the non-spherical
case of integrating a two-dimensional non-truncated Gaussian field in a planar geometry
along an infinite straight line. The result will depend only on the distance of
closest approach, $\delta_0$, between the line and the centre of the Gaussian,
and is trivially found to be exactly
$(\sqrt{2\pi}\sigma)^{-1}\expmhalf{\delta_0^2/\sigma^2}$. This factor, which we
shall refer to as the simple Gaussian approximation, is present
in \refeqn{circleintegralapproxformula}, along with three correction factors for the spherical geometry
and tail truncation: $\sqrt{\sin\alpha/\sin\gamma}$ is the lowest order
correction for the curvature of the path of integration, the error
function corrects for the truncation, and the factor of $2\pi\sigma^2N$ represents a
correction to the normalisation influenced by both of these effects.

Of the three correction factors, the most interesting is arguably $\sqrt{\sin\alpha/\sin\gamma}$, as it
introduces asymmetries in rocking curves and shifts their peaks slightly away
from the expected positions at $\gamma=\alpha$. It is interesting to note that,
using a different approach, J.~Wuttke~\cite{Wuttke2014} also derived an approximation formula
analogous to \refeqn{circleintegralapproxformula}. This
formula~\cite[Eq.~37]{Wuttke2014} has a different functional form and does not
properly account for normalisation (which is evident from~\cite[Fig.~5]{Wuttke2014}), but it
predicts an asymmetry factor which to lowest order in $\eta=\gamma-\alpha$
expands to the same result as the $\sqrt{\sin\alpha/\sin\gamma}$ factor derived here,
namely $1-\thalf\eta\cot\alpha$.

Although somewhat complicated, \refeqn{circleintegralapproxformula} can be
evaluated rather efficiently since everything except the factor of
$\sqrt{\sin\alpha/\sin\gamma}$ depends only on
$\delta_0^2$, which can be calculated from $\cos\delta_0$. Thus, with $S$ representing a suitable
pre-calculated 1-dimensional cubic spline-based lookup table:
\begin{align}\labeqn{circleintegralapproxformulawithspline}
   g_{hkl} = S(\cos\delta_0)\sqrt{\frac{\sin\alpha}{\sin\gamma}}=S(\cos(\alpha-\gamma))\sqrt{\frac{\sin\alpha}{\sin\gamma}}
\end{align}
As previously mentioned, $\cos\alpha$ and $\cos\gamma$ are cheaply available for
a particular incoming neutron via a multiplication and a dot product respectively,
and their corresponding sinus values are calculated at the expense of a square-root
evaluation. Likewise,
$\cos\delta_0=\cos(\alpha-\gamma)=\cos\alpha\cos\gamma+\sin\alpha\sin\gamma$, so
the entire cross section evaluation
through \refeqn{circleintegralapproxformulawithspline} consists of a few basic
arithmetic manipulations, a cubic spline evaluation, and three square-root
evaluations.

As \refeqn{circleintegralapproxformulawithspline} represents an
approximation, the modelling code must choose when it should be employed, and
when it is necessary to fall back to the full numerical integration of
\refeqn{xsprinciple}. The two approximations invoked above were both related to the
expansion of the cosine function:
\begin{align}\labeqn{cosineexpansions}
   \cos\delta(t) \approx 1-\delta(t)^2/2\qquad\text{and}\qquad\cos t \approx 1-t^2/2
\end{align}
Demanding that these are accurate to a certain precision, amounts to putting a
limit, denoted $\psi$,  on the magnitudes of the cosine arguments. If for instance a precision of
$\epsilon=\num{e-3}$ is required in all Taylor expansions of the cosines, the
arguments must all have magnitudes less than
$\psi\approx\SI{22.6}{\degree}$.\footnote{Typical errors on
final cross sections will actually in general be smaller than those indicated by
$\epsilon$. This is because the shape of the Gaussian mosaic density ensures
that the largest contributions to the cross section will come from areas where
the magnitudes of $t$ and $\delta(t)$ are relatively smaller, and the
Taylor expansions more accurate.}
The first approximation
in \refeqn{cosineexpansions} is valid at the desired precision for all
$0\le{}t<\tprime$ only when $\tau<\psi$, the validity of which is a global
and fixed property of a given crystal. Fortunately, $\tau<\psi$ is satisfied for
most typical \texttt{NCrystal} configurations. On the other hand, it is always possible to find back-scattering
scenarios for which the second approximation in \refeqn{cosineexpansions} breaks
down: when values of $\alpha$ and $\gamma$  are low enough that the Bragg circle
will be entirely located within an angle of $\tau$ from $\hat{n}$,
$\tprime$ becomes equal to $\pi$. To be more precise, it follows from  \refeqn{cost} that the condition
$\tprime<\psi$ is equivalent to:
\begin{align}\labeqn{tprime_approx_condition}
  \cos\psi\sin\gamma\sin\alpha+\cos\alpha\cos\gamma < \cos\tau
\end{align}
Which can indeed never be satisfied for the extreme back-scattering case of vanishing
$\alpha$ and $\gamma$. So in conclusion, when $\tau<\psi$ and
\refeqn{tprime_approx_condition} is satisfied, the code will evaluate the
cross section using \refeqn{circleintegralapproxformulawithspline}, and
otherwise it will fall back to the Romberg-based numerical integration outlined
previously. In all cases providing results that are accurate at the desired level,
defaulting to $\epsilon=\num{e-3}$. As mentioned in \refsec{scbragg::mosdef},
users of \texttt{NCrystal} can modify this default value through the
configuration parameter named \texttt{mosprec}.

\subsection{Sampling scattering events}\labsec{scbragg::genscat}

In order to sample an outgoing neutron direction ($\hatkf$) in case of a
scattering event, a particular $hkl$ plane is first sampled randomly among those
contributing, with the sampling based upon their contributions to the total cross
section. These contributions are usually already available, as they were
determined and cached during a previous cross section calculation.

Having thus selected the particular $hkl$ plane involved in the scattering, a
crystallite orientation must then be sampled in order to provide an actual (as
opposed to nominal) normal of the plane on which to scatter. This is first done
in the coordinate system of \reffig{mosintegralon3dsphere} by sampling a value
of $t$ in $[-\tprime,\tprime]$ according to the value of the integrand
in \refeqn{xsprinciple} and using the resulting normal given by \refeqn{poftdef}
to perform a specular reflection of $\hatki$ into $\hatkf$. Finally, a
suitable rotation matrix is applied to the resulting $\hatkf$ vector in order
to rotate it into the coordinate system used by the calling code. The actual
sampling of a $t$ value proceeds via acceptance-rejection
sampling~\cite{vonneumann1951}, relying on the fact that the density function,
the integrand of \refeqn{xsprinciple}, attains its largest value at $t=0$. This
maximal value can thus be used as a constant overlay function for the rejection
sampling. As splined lookup tables of the integrand were already prepared for
the purpose of numerical integration of \refeqn{xsprinciple}, these are reused
during the sampling, which in practice is found to have acceptance ratios from
15\%--35\%, depending on the scenario.
The resulting sampling speeds are acceptable, and usually insignificant, as will
be discussed further in \refsec{timing}.

\subsection{Effective model at very short \texorpdfstring{$d$}{d}-spacing}\labsec{scbragg::sccutoff}
As described in~\cite{ncrystal2019}, great care is taken to construct
lists of reflection planes which include very large number of entries at shorter values of
$d$-spacing, down to a threshold of \SI{0.1}{\angstrom} for most materials.\footnote{For the case of some
very complicated crystal structures such as yttrium oxide, this threshold is
instead \SI{0.25}{\angstrom}, but can in any case be controlled by
the \texttt{dcutoff} configuration parameter as explained
in~\cite[Sec.~5]{ncrystal2019}.} Although experimental features of Bragg diffraction are typically
more prominent at wavelengths longer than \order{\SI{1}{\angstrom}}, the
combined effect from the very high number of very weakly scattering planes at
shorter wavelengths, can still represent a non-negligible correction to the total cross
section. A correct total cross section at shorter wavelengths is necessary
in order to prevent regions of artificially low cross section, with adverse
effects on simulations of e.g.\ beam filters or shielding. While the treatment
of very large numbers of weak planes poses no problems for the efficient
\order{\log{}N_{hkl}} modelling of crystal powders or polycrystals discussed
in \refsec{pcbragg}, the single crystal modelling is a different matter
entirely. Despite the efforts at optimisation described in
\reftwosections{scbragg::normsearch}{scbragg::inteval}, the resource
requirements involved in cross section calculations still scales with the number
of planes fulfilling the Bragg condition, and as such will be very significant
for short wavelength neutrons. As an example, a user interested in simulating
long wavelength neutrons being reflected by a single crystal monochromator, might
end up spending almost all processing time dealing with the few neutrons at
shorter wavelength which invariably end up in the monochromator (for instance due to
inelastic scatterings in the monochromator itself). Although it would be
possible to circumvent the problem by simply ignoring planes with shorter
$d$-spacings, it is preferable to instead approximate their contribution with a
cheaper model which retains the most important contributions to neutron
scattering.

The chosen mitigation strategy is therefore to approximate the mosaic distribution of
lattice planes with very short $d$-spacings as being completely isotropic,
allowing them to be treated with the efficient powder model discussed
in \refsec{pcbragg}. This approximation, equivalent to averaging the cross
sections of the affected lattice planes over an isotropic distribution of
incoming neutrons, is deemed appropriate as the very high number of weak planes
at lower $d$-spacings anyway tend to have normals in many directions, thereby
washing out non-isotropic effects. By default
the approximation is applied to planes with $d$-spacings less
than \SI{0.4}{\angstrom}, but users with particular needs can of course modify
the threshold (or disable the approximation entirely) using an appropriate
configuration parameter, named \texttt{sccutoff}. To illustrate the effectiveness of
\begin{figure}
  \centering
  \subfloat[Germanium]{\labfig{sccutoffage}\includegraphics[width=0.79\textwidth]{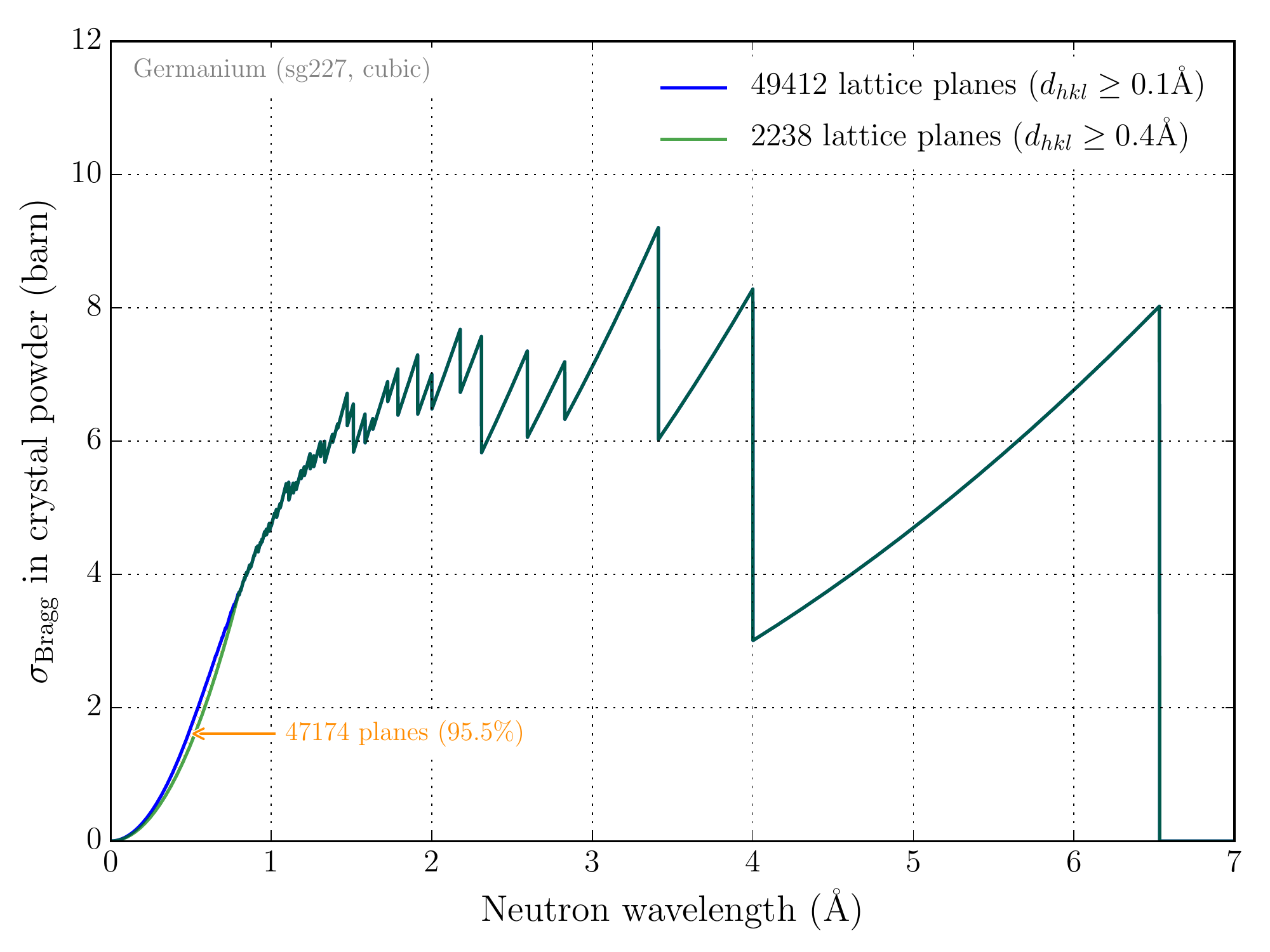}}\\
  \subfloat[Corundum]{\labfig{sccutoffbsapphire}\includegraphics[width=0.79\textwidth]{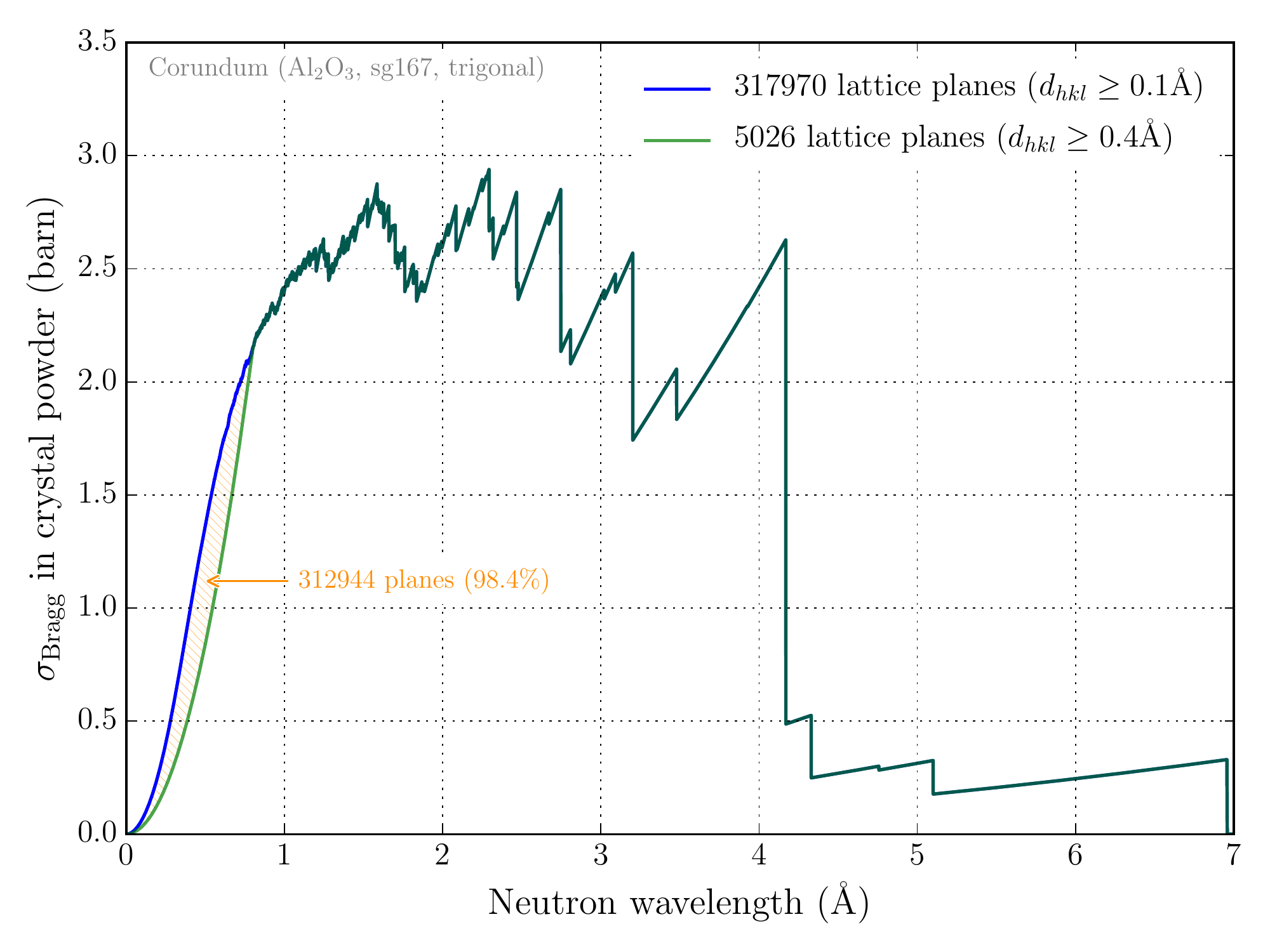}}
  \caption{Bragg cross sections in a crystal powder, based on reflection planes
    with $d$-spacing above respectively \SI{0.1}{\angstrom} and
    \SI{0.4}{\angstrom}. The number of reflection planes above and between the two $d$-spacing
    threshold values are indicated.}  \labfig{sccutoff}
\end{figure}
the approximation, consider \reffig{sccutoffage}, in which is shown the Bragg
diffraction cross section of a Germanium powder. Here, the curve resulting from
the inclusion of all lattice planes with $d$-spacings down
to \SI{0.1}{\angstrom} (the default) is compared with the one due to planes with
$d$-spacings above \SI{0.4}{\angstrom}. The conclusion, which is typical for
most crystal structures, is that while the isotropic approximation ends up being
applied to
the vast majority of planes, the affected planes are all so weakly scattering
that even their combined contribution only amounts to a tiny fraction
of the cross section. As will be shown in \refsec{timing}, this in
practice means that single crystal modelling is made 1--2
order of magnitudes faster for short wavelength neutrons, at very little cost of
realism. Even in the most pessimistic cases, like the corresponding plot for a
Corundum powder shown in \reffig{sccutoffbsapphire}, the trade-off between speed
and accuracy provided by the approximation is likely to be appropriate for all
practical use cases.

\subsection{Validation}\labsec{scbragg::validation}

The work done to validate the single crystal model, as implemented
in \texttt{NCrystal}, can essentially be divided into two primary
categories. Firstly, it is verified
in \reftwosections{scbragg::val::mpmathcmp}{scbragg::val::consistencychecks}
that the implementation actually provides results consistent with the chosen
model: Bragg diffraction as described in \refsec{commonbragg} in a material with
the mosaicity distribution defined in \refsec{scbragg::mosdef}. Secondly, it is
verified in \refsec{scbragg::val::cmpothers} that it is able to reproduce
results from accepted existing models, in the domains of their validity. In
addition to this, benchmark numbers for computational efficiency are presented
and discussed in \refsec{timing}.

\subsubsection{Comparison with reference implementation}\labsec{scbragg::val::mpmathcmp}

As should be evident from \refsectionrange{scbragg::mosdef}{scbragg::sccutoff},
the actual implementation of the model in \texttt{NCrystal} is rather
complicated. This is of course a result of the desire to optimise the
implementation for speed of evaluation, in order to provide end-users with a
better trade-off between number of neutrons simulated, computational resources
spent, and precision. A reasonable concern is of course that some applied approximations might
be too crude, or some code too complicated, jeopardising the validity of the
results. Fortunately, validation work does not suffer from the same requirement for
computational efficiency, allowing for the implementation of inefficient reference
models, that are both less complicated and more precise.
The reference results in the present section are obtained
with \texttt{mpmath}~\cite{mpmath}, a \texttt{Python} module which allows
floating-point arithmetic with arbitrary precision, and which includes both
mathematical functions, and utilities for numerical
integration. With \texttt{mpmath}, \refeqn{xsprinciple} can be implemented in
just a few lines of high-level code, and results evaluated to 100 significant
digits (at the cost of very long evaluation times). When selecting validation scenarios, special attention was given to using not only ``easy''
configurations, but also those with back-scattering ($\alpha$ comparable to
$\sigma$) and forward-scattering ($\pi-\alpha$ comparable to $\sigma$).

To verify first the evaluation of \refeqn{xsprinciple} a series of rocking
curves were constructed for various scenarios, showing the effect of varying
the neutron incidence angle $\gamma$, on the estimated cross sections through
$g_{hkl}$. Note that in order to remove
trivial differences by design due to the
$\epsilon$-dependency of the relative mosaic truncation angle
(cf.\ \reffig{mosprecntrunc}), a fixed truncation angle of $\tau=5\sigma$ was
used for all curves.
Three configurations will be discussed in the following, but figures
with the results of other configurations are available in the appendix
in \reffigrange{gosint_appendix_first}{gosint_appendix_last}. First,
\reffig{gosint_bigmos}
\begin{figure}
  \centering
  \includegraphics[width=0.78\textwidth]{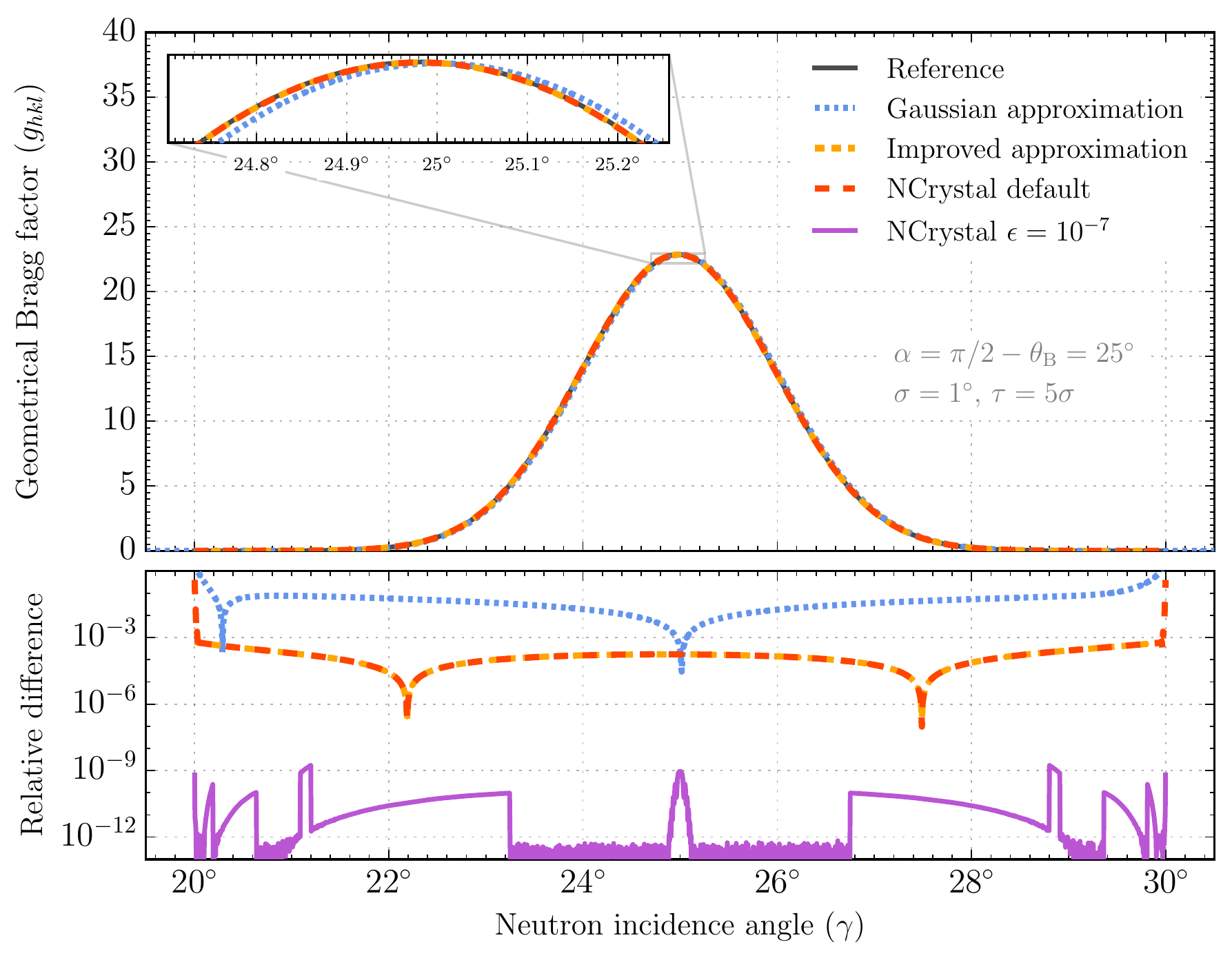}
  \caption{Various approaches to evaluation of \refeqn{xsprinciple} as a
  function of $\gamma$ when $\sigma=1^\circ$ ($2.35^\circ$ FWHM), $\alpha=25^\circ$, and
  $\tau=5\sigma$, compared with a high precision reference curve. For
  comparison, a simple Gaussian approximation as well as the improved
  approximation of \refeqn{circleintegralapproxformula} are both
  evaluated to high precision and included in addition to the results from \texttt{NCrystal}. The numbers provided for
  the implementations in \texttt{NCrystal} are provided both for the default
  settings and for a setting with improved precision ($\epsilon=\num{e-7}$).}
  \labfig{gosint_bigmos}
\end{figure}
shows the results for a crystal with a relatively
large mosaicity of $\sigma=1^\circ$ and with the Bragg angle
far from back-scattering.  At these large mosaicities, the simple Gaussian
approximation of evaluating $g_{hkl}$ as $(\sqrt{2\pi}\sigma)^{-1}\expmhalf{\delta_0^2/\sigma^2}$ is
not able to provide very accurate results, while the
improved approximation in \refeqn{circleintegralapproxformula} is able to provide
results accurate to 3--4 digits. It is clear from the overlapping curves in \reffig{gosint_bigmos} that
\texttt{NCrystal} with default settings is utilising this approximation. There
is, however, a minor breakdown in accuracy around the edges at
$\gamma\approx20^\circ$ and $\gamma\approx30^\circ$, which is due to artefacts
caused by the limitations of the cubic spline used in
\refeqn{circleintegralapproxformulawithspline} to implement the evaluation
of \refeqn{circleintegralapproxformula}. As the decrease in
accuracy happens only at the edges, where $g_{hkl}\approx0$, this is an
acceptable price to pay for the increased computational
efficiency. When \texttt{NCrystal} is configured for increased accuracy,
$\epsilon=\num{e-7}$, the code automatically falls back to numerical
integration, with 9 significant digits correctness at all angles.

Next, \reffig{gosint_stdcond}
\begin{figure}
  \centering
\includegraphics[width=0.78\textwidth]{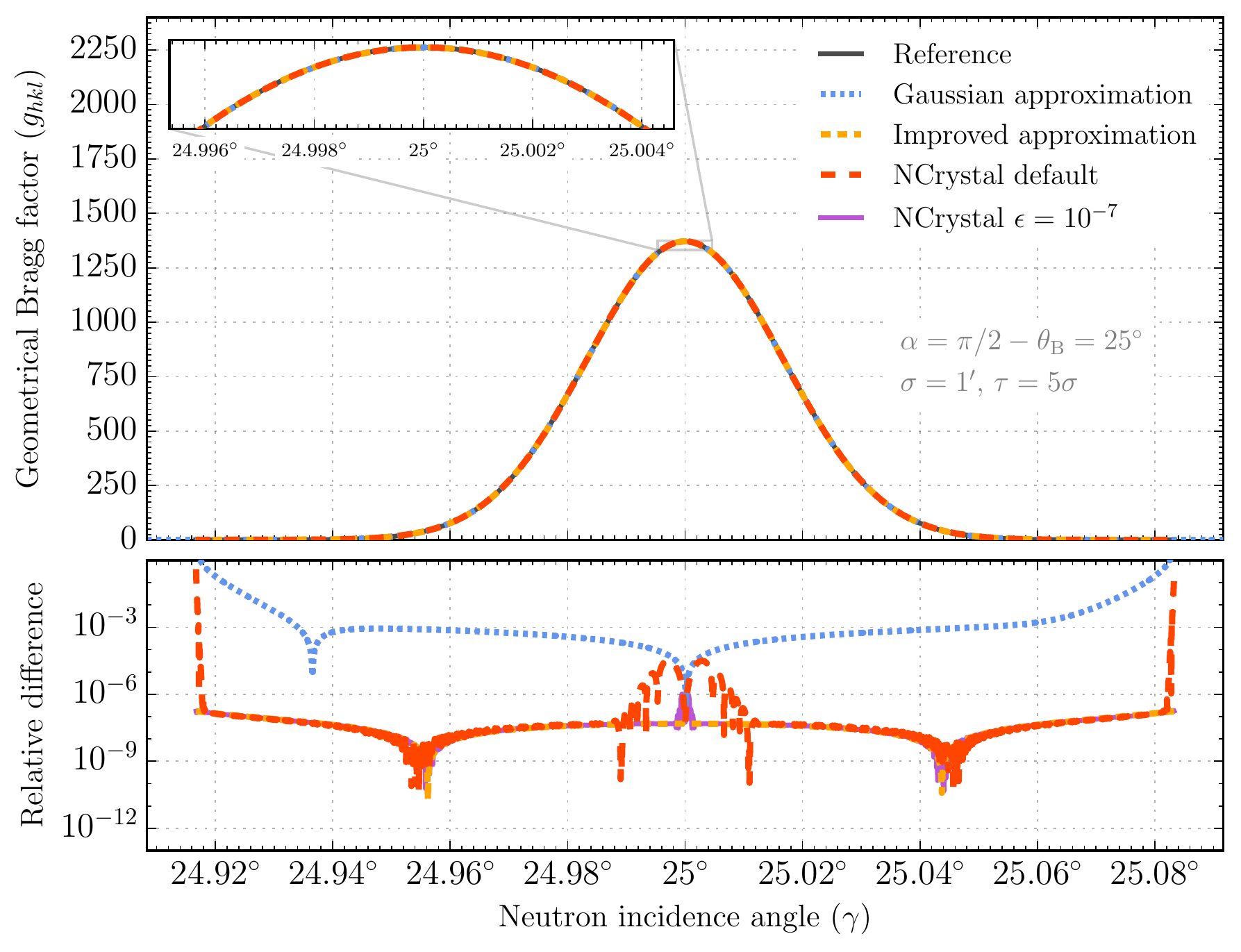}
  \caption{Same curves as in \reffig{gosint_bigmos}, but for a lower mosaicity
  of  $\sigma=1^\prime$ ($0.039^\circ$ FWHM)}
 \labfig{gosint_stdcond}
\end{figure}
shows the same scenario but with a much smaller
mosaicity of $\sigma=1^\prime$. Here, both approximations have more ideal
conditions, and the improved approximation of \refeqn{circleintegralapproxformula} is able to provide 7 significant
digits over the entire range of incidence angles. Consequently, even when \texttt{NCrystal} is configured for increased
accuracy, $\epsilon=\num{e-7}$, the approximation is utilised, completely
foregoing any fall back to numerical integration. Again some artefacts appear as
a consequence of the usage of cubic splines in \texttt{NCrystal}'s implementation of the
approximation, and this time also in the central region. Fortunately, the
required accuracies are still achieved, except for a few points where
\texttt{NCrystal} with the $\epsilon=\num{e-7}$ configuration only provides 6
significant digits. All in all, this is still considered to be an acceptable
price for the efficiency gains provided by the cubic splines.

Finally, \reffig{gosint_backscat}
\begin{figure}
  \centering
\includegraphics[width=0.78\textwidth]{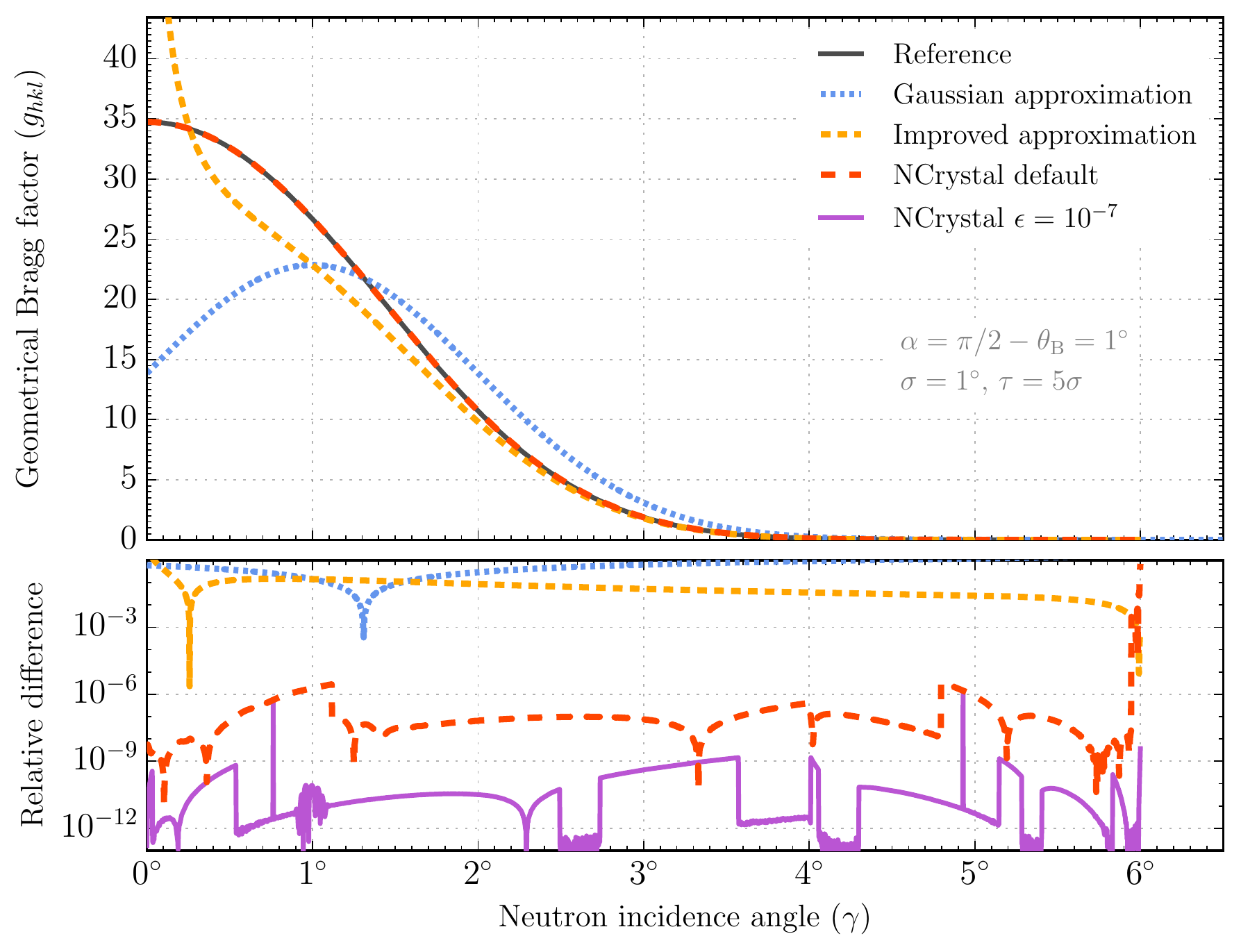}
  \caption{Same curves as in \reffig{gosint_bigmos}, but in back-scattering
  configuration with $\alpha=\sigma$.}
  \labfig{gosint_backscat}
\end{figure}
shows a back-scattering scenario. Although the
improved approximation still outperforms the simple Gaussian approximation in
this case, neither approximation is able to adequately reproduce the reference
results. Consequently, \texttt{NCrystal} falls back to numerical integration
with highly accurate results. Again the usage of cubic splines, this time for
the integrand of \refeqn{xsprinciple}, leads to acceptable inaccuracies at the
edge where $g_{hkl}\approx0$.

In order to evaluate the scattering event sampling code described
in \refsec{scbragg::genscat}, \texttt{mpmath} is again used to provide reference
distributions for comparison with the outcome of \texttt{NCrystal} simulations
involving a single reflection plane only. The reference curve is constructed by
using \texttt{mpmath} to evaluate the mosaicity distribution defined
by \reftwoeqns{gaussianmosdist}{gaussmosnormintegraldef} on a very large number
of points along the Bragg circle. Specifically, with the Bragg circle
parameterised by \refeqn{poftdef}, as $\hat{p}(t)$ for $t\in[-\pi,\pi[$, the
angular distance $\delta(t)$ is calculated from $\hat{p}(t)$ to $\hat{n}$
defined by \refeqn{hatndef}, and $W(\delta(t))$ evaluated. Additionally,
since \texttt{NCrystal} code always treats the pair of planes $(h,k,l)$ and
$(-h,-k,-l)$ simultaneously, the reference curve is evaluated by calculating the
angular distance to $-\hat{n}$ as well, and adding the contribution from
scattering on both $\hat{n}$ and $-\hat{n}$.

For simplicity, the \texttt{NCrystal} simulation is set up in a scenario in
which the neutron and plane normals are oriented similar
to \reffig{mosintegralon3dsphere}, with the neutrons impinging from the positive
$z$-axis on a pair of reflection planes with normals $\pm(\sin\gamma,0,\cos\gamma)$,
and with wavelength adjusted according to the desired Bragg angle. It is
trivially confirmed that the simulation provides scatterings with the expected
scattering angle of exactly $2\thetabragg$, so the validation should only verify the
distribution of azimuthal scattering angles. In the chosen reference frame,
these are simply extracted from the simulation results as
$\varphi=\sign(\hat{k}_f^y)\arccos(\hat{k}_f^x)$, which directly corresponds to
the parameter $t$ in the Bragg circle parameterisation.

\Reffig{scscatvalstd}
\begin{figure}
  \centering
  \includegraphics[width=0.8\textwidth]{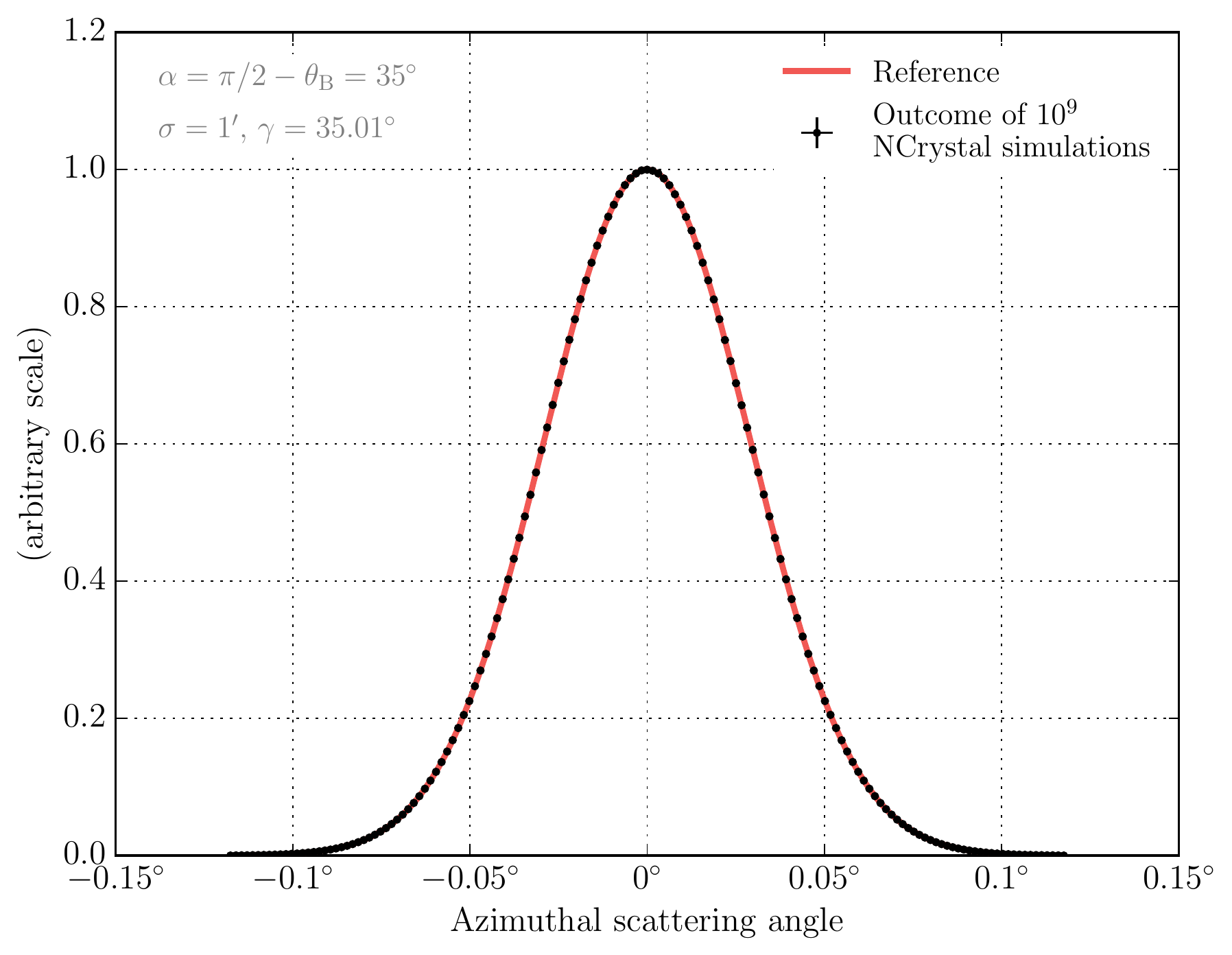}
  \caption{Azimuthal scattering angle distribution with \texttt{NCrystal}
    default settings, compared with a high precision reference curve, for a
    scenario where $\sigma=1^\prime$ ($0.039^\circ$ FWHM), $\alpha=35^\circ$, and
    $\gamma=35.01^\circ$. Statistical uncertainties are smaller than the shown points.}
  \labfig{scscatvalstd}
\end{figure}
shows the outcome from a scenario with neither
back- or forward-scattering and a low mosaicity of $\sigma=1^\prime$. In this
typical scenario, the distribution of azimuthal angles is essentially Gaussian,
and the outcome of \texttt{NCrystal} simulations is in perfect
agreement with the calculated reference.

Next, \reffig{scscatvalbackscat}
\begin{figure}
  \centering
  \includegraphics[width=0.8\textwidth]{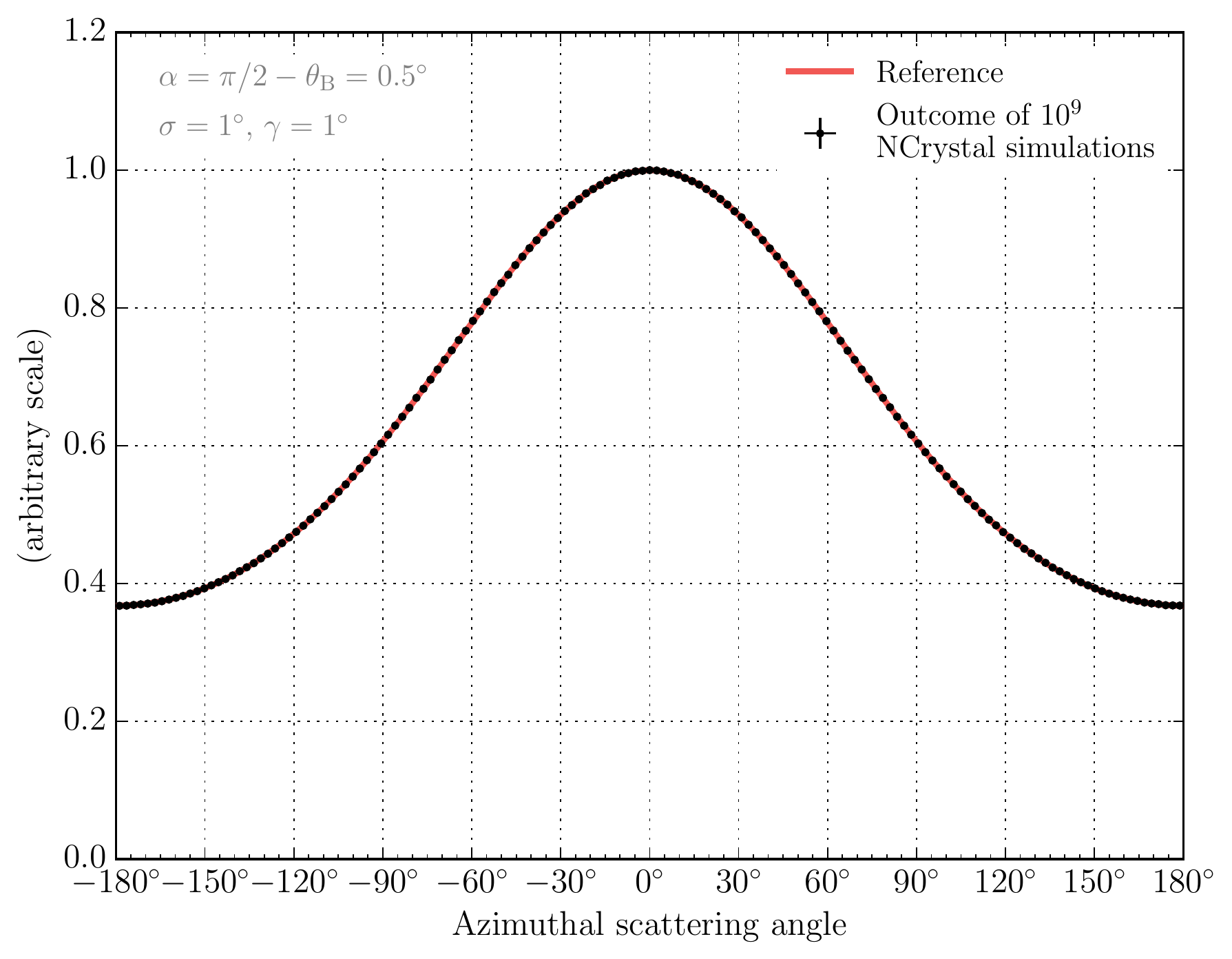}
  \caption{Azimuthal scattering angle distribution with \texttt{NCrystal}
    default settings, compared with a high precision reference curve, for a
    back-scattering scenario where $\sigma=1^\circ$ ($2.35^\circ$ FWHM), $\alpha=0.5^\circ$, and
    $\gamma=1^\circ$. Statistical uncertainties are smaller than the shown points.}
  \labfig{scscatvalbackscat}
\end{figure}
investigates a scenario with back-scattering and a
higher mosaicity of $\sigma=1^\circ$. Here, all parts of the Bragg circle are
close to $\hat{n}$, with $\delta(t)<\tau$ everywhere. Thus, all azimuthal
angles are possible, with those near 0 favoured. Again, \texttt{NCrystal}
simulations perfectly reproduce the calculated reference.

Finally, \reffig{scscatvalantinormal}
\begin{figure}
  \centering
  \includegraphics[width=0.8\textwidth]{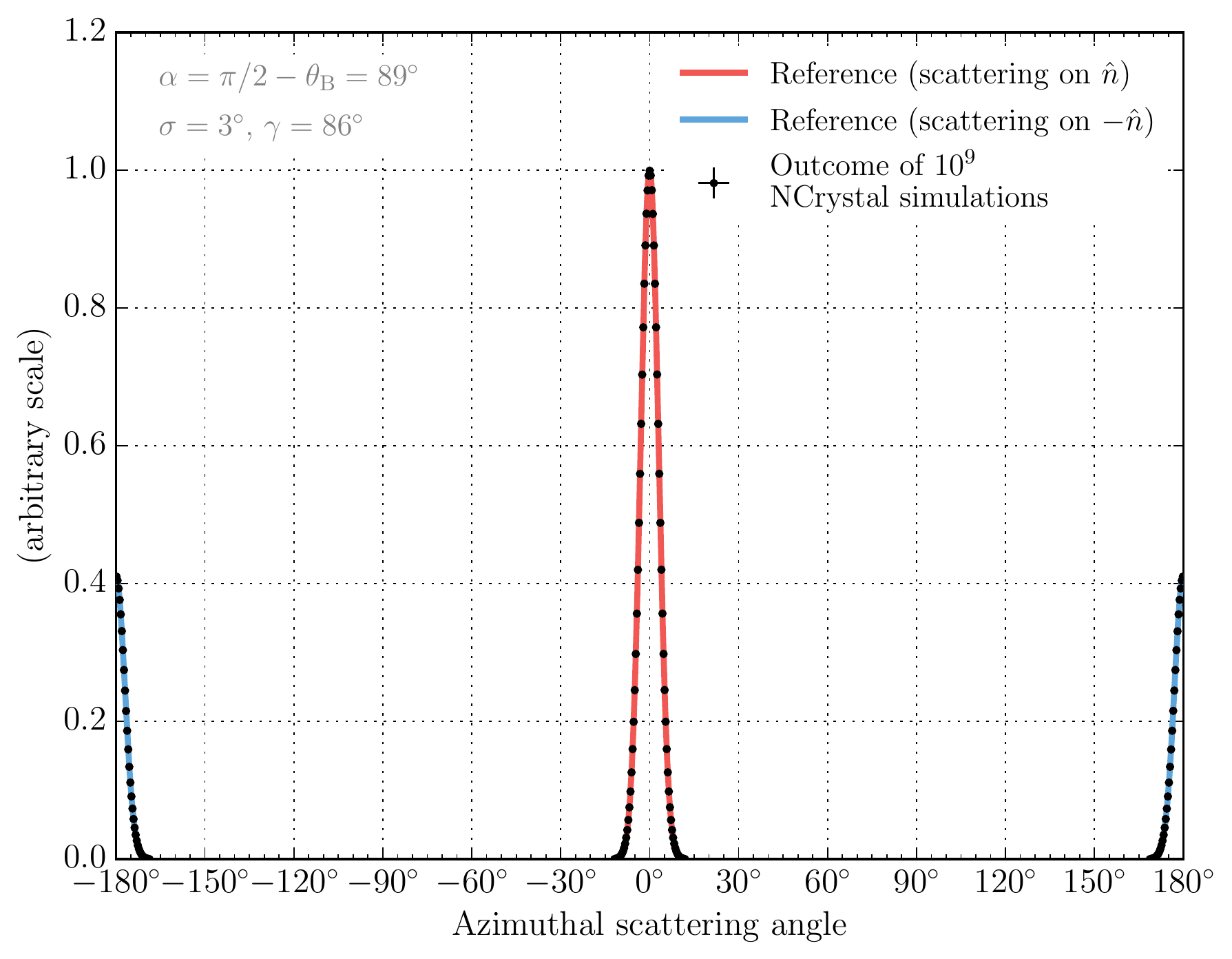}
  \caption{Azimuthal scattering angle distribution with \texttt{NCrystal}
    default settings, compared with high precision reference curves, for a
    forward-scattering scenario where $\sigma=3^\circ$ ($7.06^\circ$ FWHM), $\alpha=89^\circ$, and
    $\gamma=86^\circ$. Statistical uncertainties are smaller than the shown points.}
  \labfig{scscatvalantinormal}
\end{figure}
investigates a scenario with
forward-scattering, using an even higher mosaicity of $\sigma=3^\circ$ for
visualisation purposes. Scattering on $\hat{n}$ again produces a bell-curve
around 0, but this time there is a non-zero contribution for scattering on
$-\hat{n}$ as well, giving rise to events scattered to $\varphi\approx\pm\pi$.
Also in this case, the \texttt{NCrystal} simulations perfectly reproduce the
calculated reference.

\subsubsection{General consistency checks}\labsec{scbragg::val::consistencychecks}

Irrespective of mosaic distribution, any model of Bragg diffraction which is
implemented in a self-consistent manner and supports neutrons at any incidence
angle, must be able to pass a few generic consistency checks.

First of all, the fundamental symmetry between $(h,k,l)$ and $(-h,-k,-l)$ planes
means that a neutron which just scattered in a Bragg diffraction process on the
$(h,k,l)$ plane will have non-zero cross section for a subsequent scattering on the
$(-h,-k,-l)$ plane, bringing it back to its original direction. Thus, in the
absence of other reflection planes or physics processes, a neutron in an
infinite material will keep reflecting back and forth between the two planes in
a ``zig-zag'' or ``ping-pong'' walk between the two sides of the
planes.\footnote{As was noted in \refsec{scbragg::normsearch}, it is in
scenarios involving extreme forward scattering possible
for both $(h,k,l)$ and $(-h,-k,-l)$ to simultaneously contribute to the scattering
cross section. In such cases, the ``zig-zag'' walk will strictly
speaking also include the occasional ``zig-zig'' or ``zag-zag''.} It was
verified for a range of scenarios that this ``zig-zag'' walk is reproduced
in \texttt{NCrystal} simulations by letting neutrons scatter \num{e10} times in
a wide range of scenarios, covering both large and small mosaicities as well as
both forward- and back-scattering.

The second consistency check carried out consists of submitting a single crystal
material to an isotropic illumination with neutrons. Regardless of mosaic
distribution, this should produce the same average wavelength-dependent
scattering cross section as found in a crystal powder. While the concept is
simple, this benchmark is actually a rather powerful test of a mosaic model,
as it probes all aspects of the cross section calculations. For instance,
improper handling of back-scattering or failure to account for scattering on
$-\hat{n}$ in forward-scattering will adversely impact the
results. Likewise, the test is sensitive to a range of problems in the
pre-selection procedure (cf.\ \refsec{scbragg::normsearch}), inconsistencies
between overall normalisation and truncation scheme, or failure to fall-back to
a full numerical integration when required.

\Reffig{powderscbragg}
\begin{figure}
  \centering
  \includegraphics[width=0.9\textwidth]{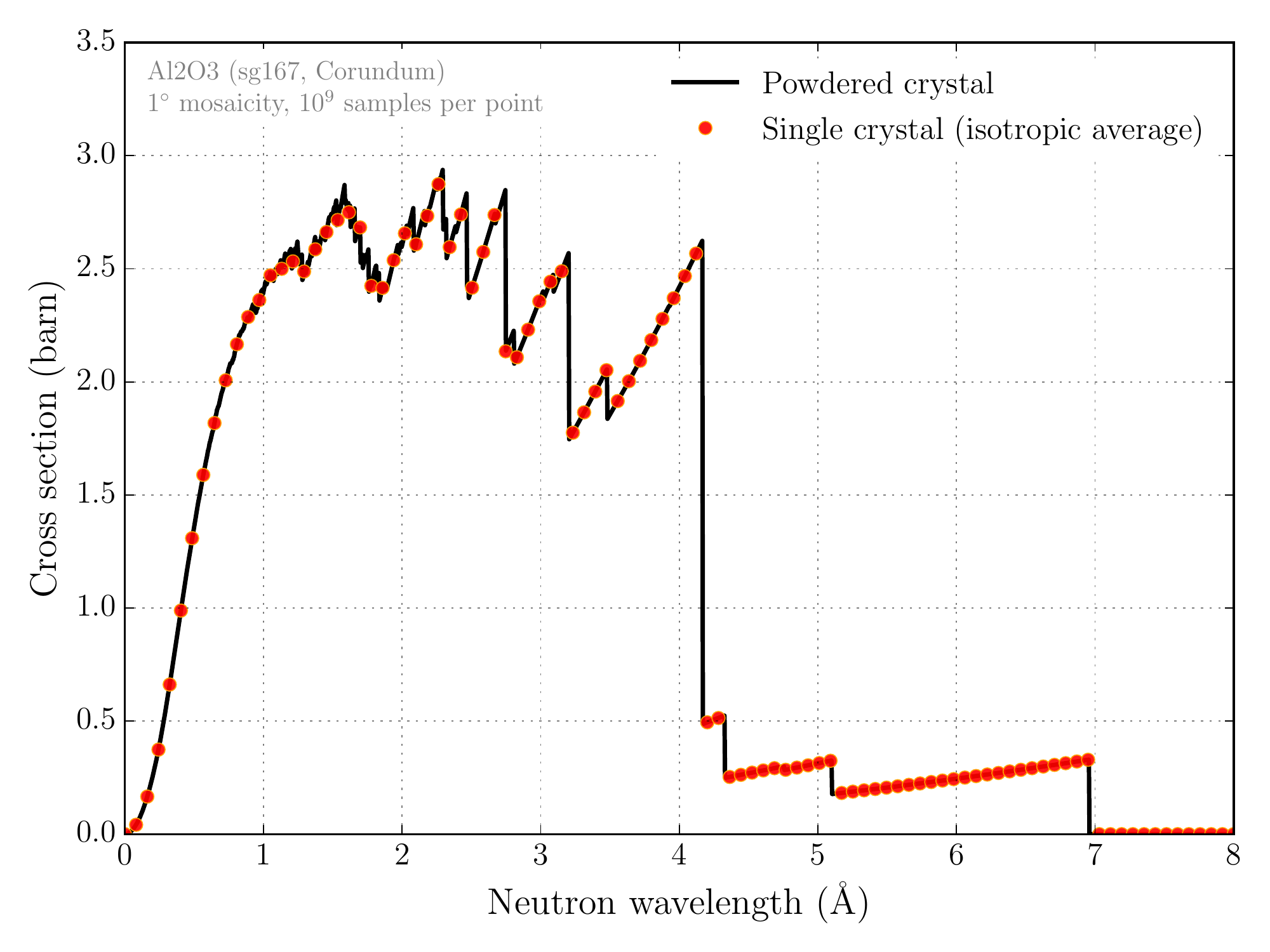}
  \caption{Single crystal corundum Bragg diffraction cross section provided by \texttt{NCrystal},
           averaged over an isotropic sample of \num{e9} neutrons at each
           wavelength point. The FWHM mosaicity is $1^\circ$. For reference the cross section curve of the same
           crystal as a powder is shown.}
  \labfig{powderscbragg}
\end{figure}
shows an example of such an isotropically averaged single
crystal cross section as a function of wavelength, showing excellent agreement
with the cross sections predicted for the same crystal as a pure powder
(cf.\ \refeqn{powderxssum}). More systematically, \reffig{powderscbraggreldiff}
\begin{figure}
  \centering \includegraphics[width=0.9\textwidth]{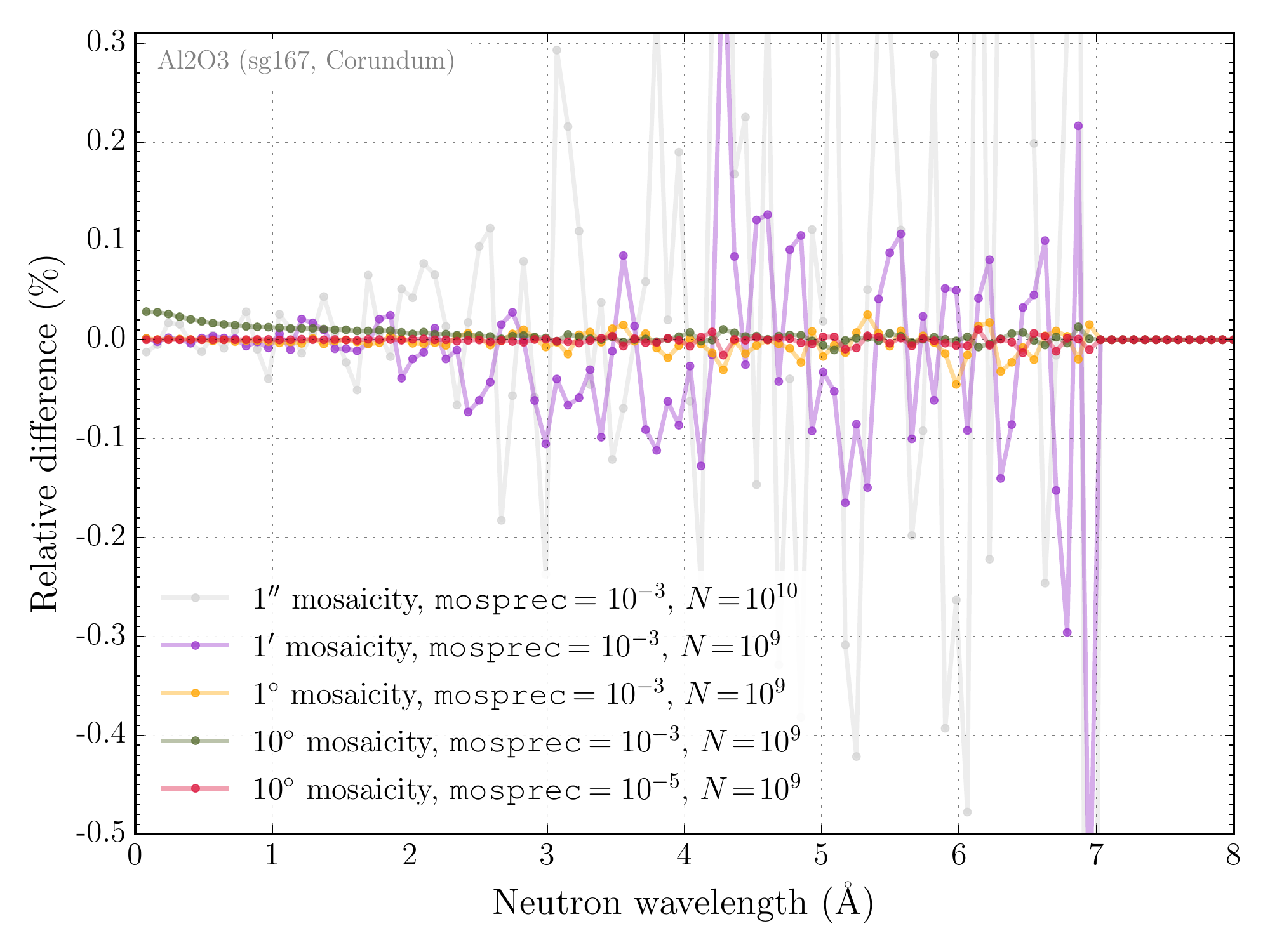} \caption{
  Relative difference between isotropic averages of single crystal cross
  sections provided by \texttt{NCrystal} and the equivalent powder cross
  sections. Results are shown for various FWHM mosaicities, sample sizes and for
  both default ($\epsilon=\num{e-3}$) and increased
  ($\epsilon=\num{e-5}$) accuracy settings, as indicated.}
  \labfig{powderscbraggreldiff}
\end{figure}
shows the relative differences between the isotropic averaged single crystal
cross sections and the equivalent powder cross sections, for various
settings. At lower mosaicities, neutrons in the isotropic sample are more likely
to have vanishing cross section for scattering on a given reflection plane,
leading to significant statistical fluctuations in the curves for the lower
mosaic spreads. Likewise, at longer wavelengths only a few planes satisfy the
Bragg condition (cf.\ \refeqn{braggcondition}), with a resulting increase in
statistical fluctuations. All in all, \reffig{powderscbraggreldiff} shows no
sign of significant discrepancies, except for showing an \order{\num{e-4}}
discrepancy for the $10^\circ$ mosaicity curve at the shortest wavelengths. This
is acceptable as it is still an order of magnitude more precise than the
target \order{\num{e-3}} precision with that setting. Increasing the requested
level of precision to $\epsilon=\num{e-5}$ makes this discrepancy disappear,
thus verifying that the source of the small disagreement is indeed well-controlled
inaccuracies arising from the numerical approximations used, and not a sign of
more fundamental problems.

\subsubsection{Reflectivity benchmarks}\labsec{scbragg::val::cmpothers}

An important benchmark for \texttt{NCrystal}'s single crystal model is the
ability to reproduce established results. Based on the availability of high
precision reference results, the chosen figure of merit is the fraction of
neutrons in a monochromatic pencil beam which are reflected by a slab of single
crystal. The slab has infinite transversal dimensions, and a thickness which is
finite but not necessarily small. In such a crystal the details of the
``zig-zag'' walk discussed in \refsec{scbragg::val::consistencychecks} will in
general have a significant impact on the probability for both reflection and
transmission, as will the detailed evaluation of cross sections and sampling of
scattered directions. For compatibility with the reference results, it is
additionally assumed that only one particular reflection plane contributes to
the cross sections, that this plane is parallel to the surface of the slab, and
that all inelastic and incoherent scattering events are essentially counted as
absorption events, ending further simulation of the neutron in question.

Based on the Darwin-Hamilton equations~\cite{Hamilton1957},
Sears~\cite{Sears1997a} derived analytical closed-form solutions to the
reflectivity under the restriction that wavevectors never scatter out of the
plane spanned by the plane normal and initial neutron direction --- essentially
amounting to a requirement that the azimuthal scatter angle always be strictly 0
in the interactions. This is certainly an approximation at best, and one which completely breaks
down for back-scattering and scattering at grazing incidence
(cf.\ \reftwofigures{scscatvalbackscat}{scscatvalantinormal}). J.~Wuttke~\cite{Wuttke2014}
generalised the Darwin-Hamilton equations in order to lift this azimuthal
restriction of the scattering direction, and under certain approximations
developed analytical results for reflections on a slab. He also provided a
Monte Carlo code able to perform numerical integrations needed to support also
large mosaicities and provide a reference for his analytical
expressions. That code does, however, also explicitly exclude back-scattering
and scattering at grazing incidence. In order to compare the outcome
of \texttt{NCrystal}-based simulations with these existing solutions, a Monte
Carlo stepping simulation was carefully set up, emulating the restrictions of
the references, in order to predict idealised reflection probabilities:
inelastic and incoherent elastic scattering events were treated in the same
manner as absorption events, all but a single reflection plane was removed from
the simulation, the normal of which was made to coincide with the surface normal
of the slab. Finally, scenarios with back-scattering or grazing-incidence
scattering were avoided, and in particular values of neutron incidence and
Bragg-angles which were also used in~\cite{Wuttke2014} were favoured --- on the
assumption that they might have been particularly well validated. In order to
extract reference results, Sears' formulas were directly evaluated, while
Wuttke's Monte Carlo programme was downloaded and executed. Where relevant,
parameters were set to emulate scattering on a Germanium-511 plane.

First, the rocking curves in \reffig{wuttkesearsrocking_mos0d5_tb45}
\begin{figure}
  \centering
  \includegraphics[height=0.7\textwidth]{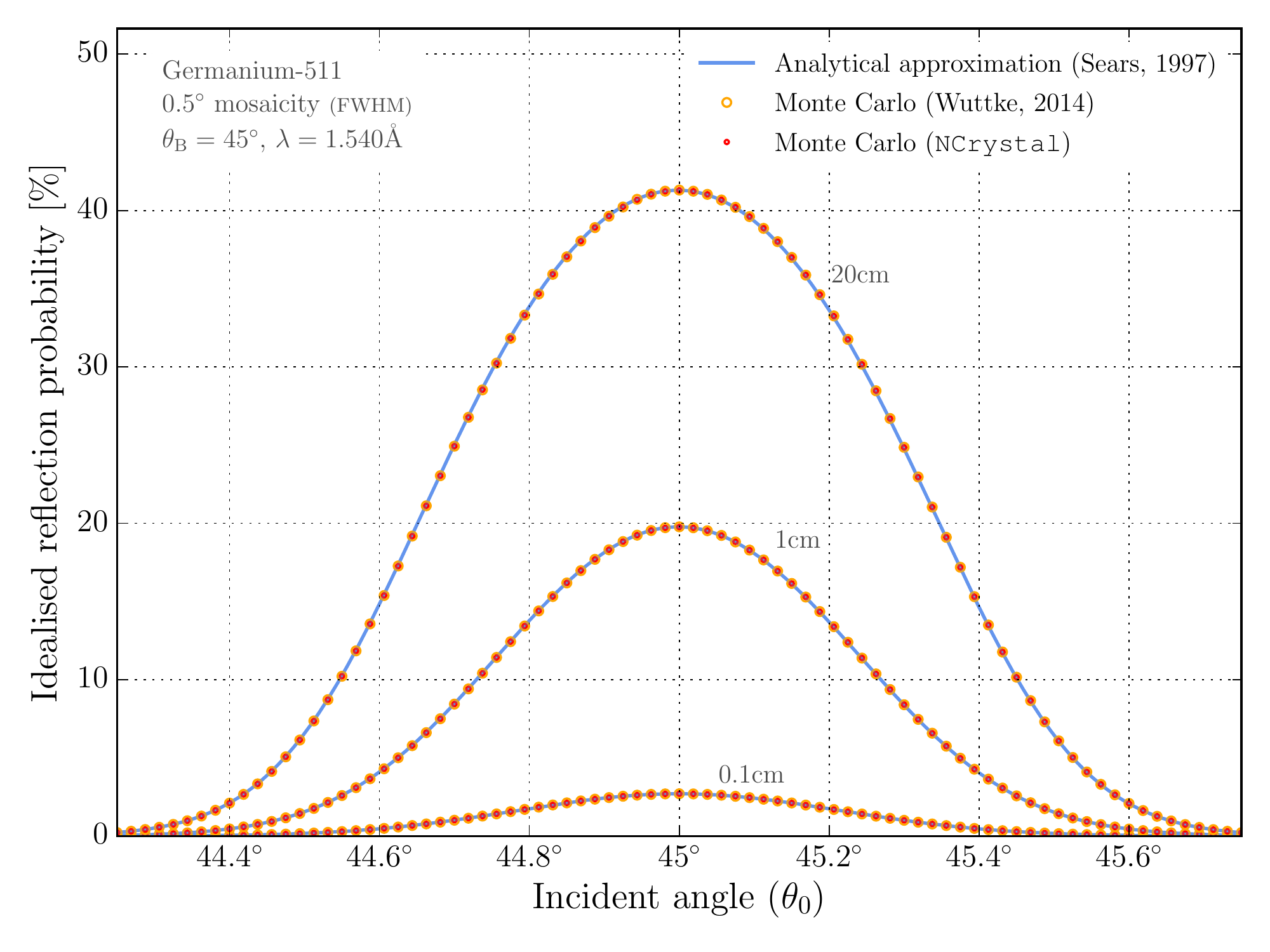}
  \caption{Idealised reflection probabilities as a function of neutron incidence
  angle predicted by \texttt{NCrystal}, Sears~\cite{Sears1997a}, and
  Wuttke~\cite{Wuttke2014}.  Curves are shown for three different slab
  thicknesses, for a collimated beam of monochromated neutrons impinging on a
  slab of Germanium, with the 511 reflection plane parallel to the slab surface,
  a FWHM mosaicity of $0.5^\circ$ and a Bragg angle of $45^\circ$.}
  \labfig{wuttkesearsrocking_mos0d5_tb45}
\end{figure}
show the resulting reflection probabilities as a function of neutron incidence angle for
a FWHM mosaicity of $0.5^\circ$, a Bragg angle of $45^\circ$ and three different
slab thicknesses. For these moderate parameters, the results
from \texttt{NCrystal} and the two references are all in excellent agreement.

Lowering the mosaicity to $0.01^\circ$
in \reffig{wuttkesearsrocking_mos0d01_tb45},
\begin{figure}
  \centering
  \includegraphics[height=0.7\textwidth]{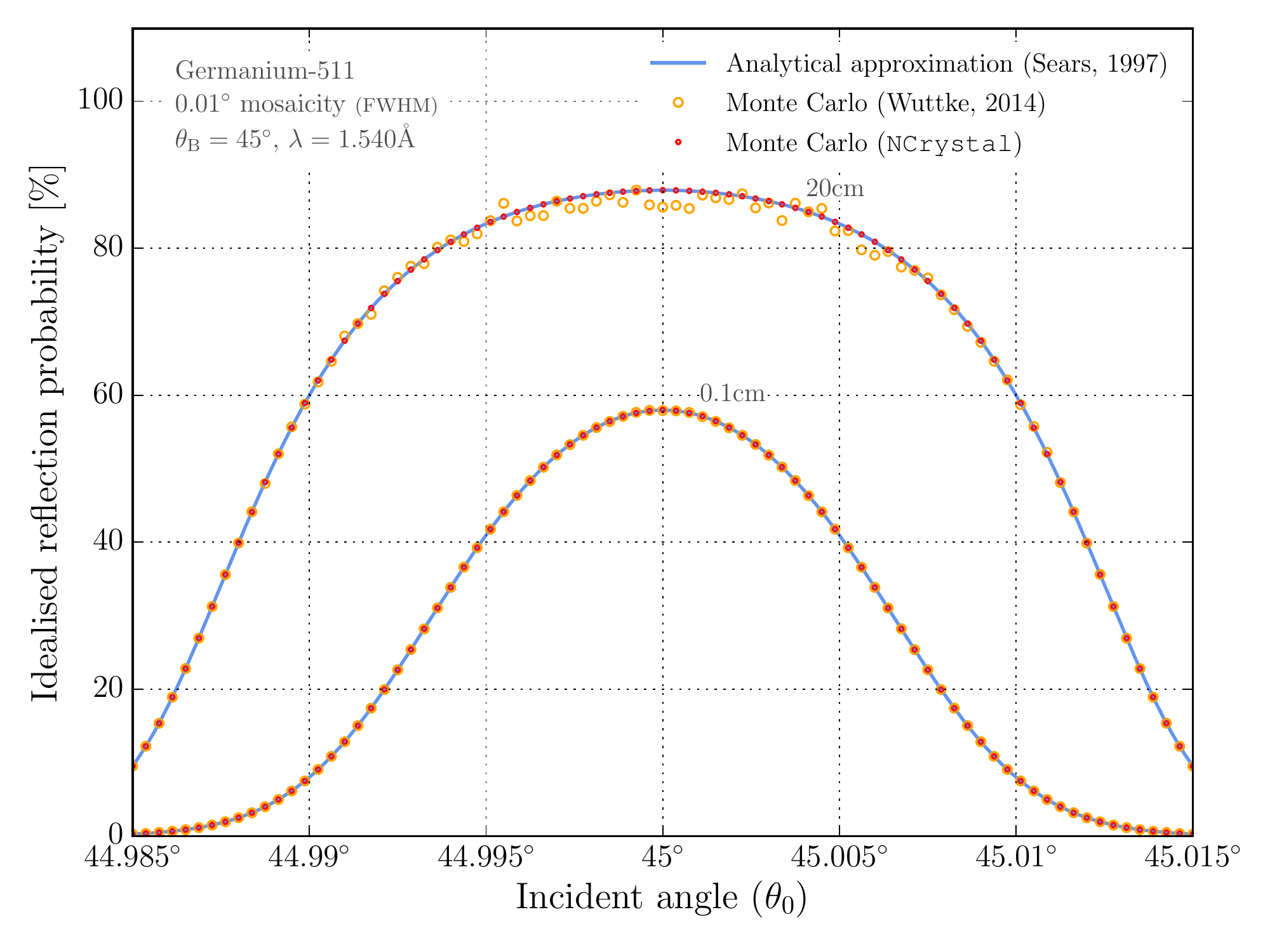}
  \caption{Same curves as in \reffig{wuttkesearsrocking_mos0d5_tb45}, but for a
  smaller FWHM mosaicity of $0.01^\circ$. For clarity the curve for a slab
  thickness of $\SI{1}{\centi\meter}$ is not shown, due to overlaps with the
  $\SI{20}{\centi\meter}$ curve.}
  \labfig{wuttkesearsrocking_mos0d01_tb45}
\end{figure}
once again results in good
agreement. The exception is that the results from Wuttke's simulations seem to
suffer from numerical instabilities on the central parts of the curve for a
thicker slab, where multiple reflections are significant. This is perhaps not
too surprising given that Wuttke's code was primarily developed in order to
study crystals with mosaic spreads much larger than $0.01^\circ$. On the other
hand, such low mosaicities ideally satisfy the assumptions behind Sears'
model, which is in perfect agreement with the \texttt{NCrystal} results.

This is in stark contrast to the situation
in \reffig{wuttkesearsrocking_mos4_tb80},
\begin{figure}
  \centering
  \includegraphics[height=0.7\textwidth]{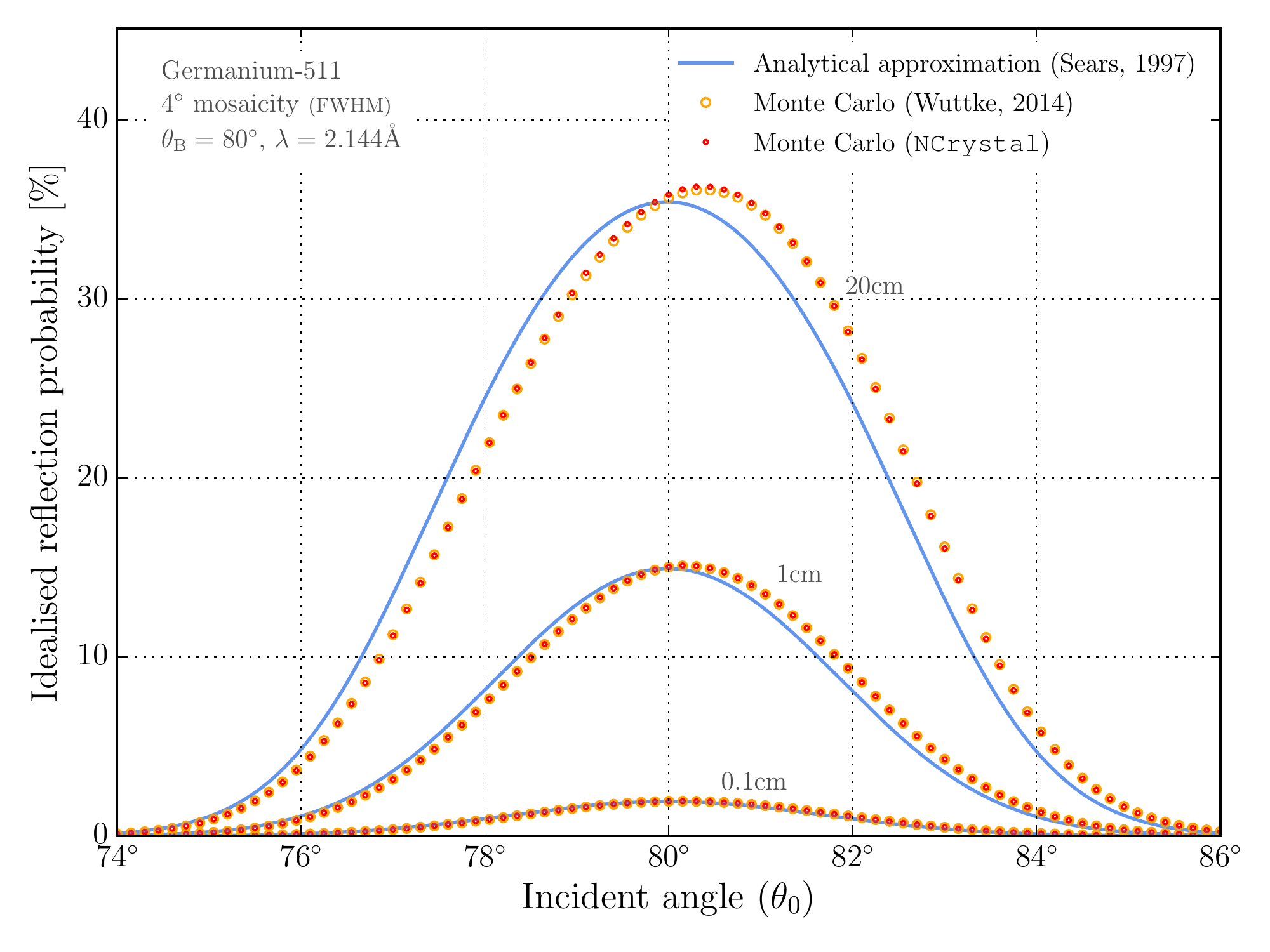}
  \caption{Same curves as in \reffig{wuttkesearsrocking_mos0d5_tb45}, but for a
  larger FWHM mosaicity of $4^\circ$ and an increased Bragg angle of $80^\circ$.}
  \labfig{wuttkesearsrocking_mos4_tb80}
\end{figure}
where the mosaicity is increased to
$4^\circ$ and the Bragg angle to $80^\circ$. In this scenario, the
integration of the mosaicity function along the Bragg circle is highly
influenced by the curvature of the path of integration, inducing strong
asymmetric effects in the rocking curves and the detrimental effect of the
in-plane scattering assumption in Sears' model increases --- in particular for
thicker slabs where multiple reflections are significant. Once again, the
\texttt{NCrystal} results are in good agreement with the most trustworthy
reference curve, this time the one from Wuttke's code. For reference, rocking
curve comparison for more parameters are available in the appendix in
\reffigrange{appendix_wuttkesearsrocking_the_first}{appendix_wuttkesearsrocking_the_last}.

As a complement to the rocking
curves, \reffig{wuttkesears_reflvthickness_thetabragg80}
\begin{figure}
  \centering
  \includegraphics[height=0.7\textwidth]{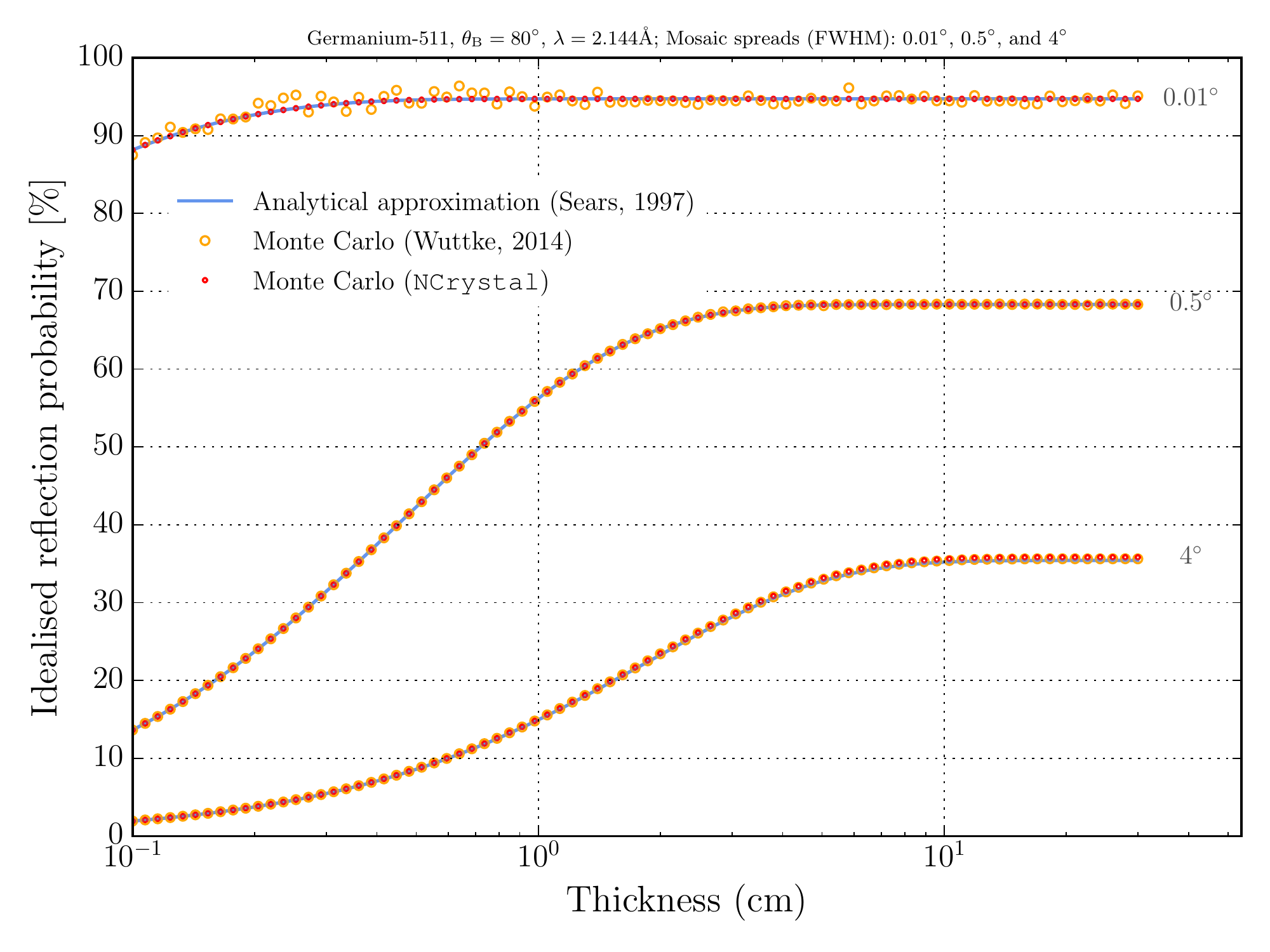}
  \caption{Idealised reflection probability as a function of slab thickness
  predicted by \texttt{NCrystal}, Sears~\cite{Sears1997a}, and
  Wuttke~\cite{Wuttke2014}. Curves are shown for three different mosaicities, for
  a collimated beam of monochromated neutrons impinging on a slab of Germanium,
  with the 511 reflection plane parallel to the slab surface, and neutron
  incidence angle and Bragg angle both $80^\circ$.}
  \labfig{wuttkesears_reflvthickness_thetabragg80}
\end{figure}
shows the reflection
probabilities as a function of slab thickness, for neutron incidence and
$\thetabragg$ both $80^\circ$ and for 3 different mosaicities. Although
the discrepancies between the different models are less pronounced than in the
rocking curves, the qualitative conclusions are essentially similar. For a
mosaicity of $0.5^\circ$, \texttt{NCrystal} and the two reference curves are all
in excellent agreement. For the smaller mosaicity of $0.01^\circ$, there is
again excellent agreement between \texttt{NCrystal} and Sears' model, while
Wuttke's code is plagued by numerical instabilities. For the larger mosaicity of
$4^\circ$, there is good agreement between \texttt{NCrystal} and Wuttke's code,
while Sears' model suffers from its approximations --- especially at larger slab
thicknesses, which are more affected by out-of-plane scattering. For
reference, similar plots for Bragg angles of $10^\circ$ and $45^\circ$ are
available in the appendix
in \reftwofigures{appendix_wuttkesears_reflvthickness_tb10}{appendix_wuttkesears_reflvthickness_tb45}
respectively.

\section{Layered single crystals for pyrolytic graphite}\labsec{lcbragg}

The isotropic Gaussian mosaicity distribution discussed in \refsec{scbragg}
provides an appropriate description for modelling of a wide range of single crystal
materials encountered at neutron scattering facilities and elsewhere. However,
exceptions do exist, with some materials exhibiting radically different
crystallite distributions. One such material which is of particular importance
for instrumentation at neutron scattering facilities is pyrolytic graphite, with
important and frequent deployment as both monochromators and analysers targeted
at selecting longer wavelengths~\cite{riste1969,shapiro1972}, as well as tunable
beam filters intended to remove short wavelength contamination in monochromated
spectra~\cite{bergsma1967,frikkee1975}.

The structure of graphite can be described as stacks of two-dimensional graphene
sheets, in which carbon atoms are bound strongly into hexagonal grids,  with
a weaker binding between the sheets. This layout at the microscopic level
ultimately gives rise to anisotropic features in the macroscopic mosaicity
distributions. With suitable manufacturing
methods~\cite{moore1964,ubbelohde1968}, the axis orthogonal to the graphene
sheets, the lattice $c$-axis, will be distributed along a preferred direction,
suitable for description with a Gaussian mosaicity distribution like the one
discussed in \refsec{scbragg::mosdef}. The rotation \emph{around} this axis,
defined for instance by the direction of the lattice $a$-axis, will however be
uncorrelated between different crystallites. Thus, the associated rotation angle
will be uniformly distributed in $[-\pi,\pi]$, giving rise to features in
scattering interactions akin to those observed with crystal powders.

In order for \texttt{NCrystal} to model pyrolytic graphite, a specialised model
of layered single crystals is implemented as described
in \refsectionrange{lcbragg::defs}{lcbragg::sampling}. An alternative reference
model for validation work is described in \refsec{lcbragg:altrefimpl},
and actual validation work is then carried out in \refsec{lcbragg::validation}.

It is worth noting that while the presented models were developed in order to
support studies involving pyrolytic graphite~\cite{milan2020arxiv}, it could
in principle also be used to model other materials like hexagonal boron
nitride~\cite{Pease:a00642} which exhibits similar layered structure. Under
certain conditions relating to the neutron velocity and rotational speed, it
could even be used to describe an isotropic Gaussian single crystal which is
placed on a spinning sample holder.

\subsection{Geometrical integral and definitions}\labsec{lcbragg::defs}

\texttt{NCrystal} effectively models a layered single crystal as consisting of a
large number of small crystals with isotropic Gaussian mosaicity distributions,
but rotated randomly and uniformly around a symmetry axis $\hat{L}$, which is
normal to the crystal layers in their nominal position. In pyrolytic graphite,
$\hat{L}$ is thus aligned along the nominal direction of the $c$-axis in unit
cell coordinates, and the user must complete the configuration of a particular
setup by specifying the coordinates of $\hat{L}$ in the laboratory frame
(cf.~\cite[Sec.~5]{ncrystal2019} for how this is done). Specifically, if
$g^\text{SC}_{hkl}(\alpha,\gamma)$ represents the geometrical weight for
scattering on a given crystal plane in a crystal with an isotropic Gaussian
mosaicity distribution given by \refeqn{xsprinciple}, then the equivalent factor
in a layered single crystal is defined to be:
\begin{align}
  g^\text{LC}_{hkl}(\alpha,\hatki) &= \frac{1}{2\pi}\int_{-\pi}^\pi g^\text{SC}_{hkl}(\alpha,\gamma(\hatki,\varphi)) d\varphi
  \labeqn{lcbraggxsdef}
\end{align}
Where $\varphi$ represents the rotation around $\hat{L}$, the offset of which is
in principle arbitrary and can be defined in whatever manner is most
convenient. For instance, one could define it so that the lattice $a$-axis in pyrolytic
graphite would coincide with a certain direction in the laboratory frame when
$\varphi=0$. However, the present discussion will instead adopt a convention
relative to the direction of the neutron, $\vecki$, and specific to each
$hkl$-plane, which minimises the distance between the plane normal and $-\vecki$
at $\varphi=0$. More specifically, the particular coordinate system shown in
\reffig{lcbraggon3dsphere}
\begin{figure}
  \centering
  \includegraphics[width=0.99\textwidth]{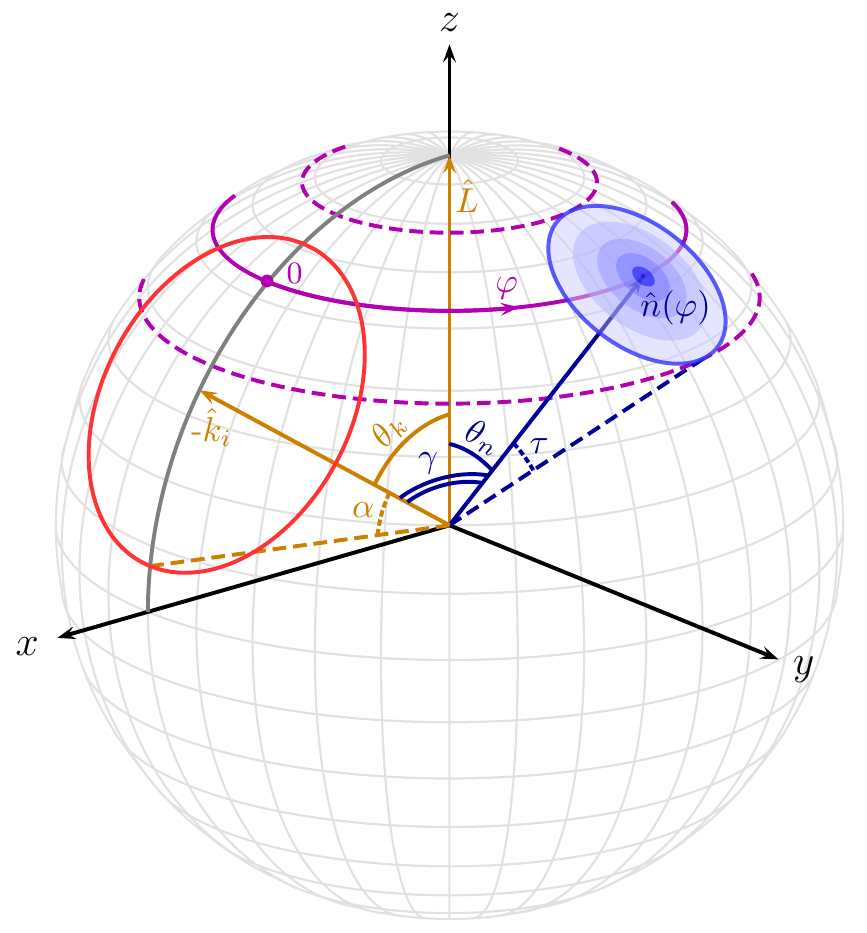}
  \caption{Coordinate system used for calculations related to the interaction of
  neutrons with (a group of) reflection planes located at an angle $\thetan$ to
  the symmetry axis $\hat{L}$ in a layered single crystal (cf.\ \refeqn{lcbraggxsdef}). Here, $\hatki$ is the
  incident direction of the neutron, $\hat{n}(\varphi)$ is the nominal position of the
  plane normals as a function of rotation around $\hat{L}$, and $\tau$ is the truncation angle of the mosaicity
  distribution, the strength of which is indicated with blue shaded areas. The
  angle between $-\hatki$ and plane normals satisfying the condition for Bragg
  diffraction is given as $\alpha=\pi/2-\thetabragg$, defining the Bragg circle
  (red). Finally, $\thetak$ is the angle between $-\hatki$ and $\hat{L}$, and
  $\gamma$ is the angle between $\hat{n}(\varphi)$ and $-\vecki$.}
  \labfig{lcbraggon3dsphere}
\end{figure}
is adopted: $\hat{L}$ is placed on
the positive $z$-axis, and $\hatki$ is placed in the $xz$-plane with coordinates:
\begin{align}
  -\hatki = ( \sin\thetak,\, 0,\, \cos\thetak )
\end{align}
Where $\thetak$ is the angle between $-\hatki$ and $\hat{L}$. If $\thetan$ is
the angle between $\hat{n}$ and $\hat{L}$, the nominal coordinates of
the normals are then given by:
\begin{align}
  \hat{n}(\varphi) = ( \sin\thetan\cos\varphi,\, \sin\thetan\sin\varphi,\, \cos\thetan)
\end{align}
With these definitions, $\gamma$ is straightforward to determine from the dot
product of $-\hatki$ and $\hat{n}(\varphi)$:
\begin{align}\labeqn{lcbragg::cosgamma}
  \cos\gamma = \sin\thetak\sin\thetan\cos\varphi+\cos\thetan\cos\thetak
\end{align}

Note that, due to the symmetries between the $(h,k,l)$ and $(-h,-k,-l)$
reflection planes, the physics should be unchanged under the transformation
$\hat{L}\to-\hat{L}$, and as long as the code is written so as to deal with the
two sets of equivalent planes at $\thetan$ and $\pi-\thetan$ jointly, it can
therefore always be assumed for simplicity that both $\thetak$ and $\thetan$ lie
in $[0,\pi/2]$. Additionally, all $hkl$-planes with similar $\thetan$ and
$d$-spacing can be treated concurrently, as a single group of planes whose
combined scattering strength is simply the sum of the contributing planes
$q_{hkl}$ values, thus effectively reducing both memory usage and time spent
processing planes during simulations. Additionally, the joint processing of the
groups of planes at $\thetan$ and $\pi-\thetan$ benefits from the fact that the
group of planes at $\pi-\thetan$ can only ever contribute if the group $\thetan$
contributes. For simplicity, the discussions will from this point onward mostly
ignore direct mention of the planes at $\pi-\thetan$, as the required
modifications to the equations are mostly trivial. Similarly, the discussions
will ignore the degenerate cases $\thetan\approx0$ and $\thetak\approx0$, as
they are trivially solvable: the $\thetan\approx0$ case (relevant for $00l$
planes in pyrolytic graphite) can be treated with the model developed
in \refsec{scbragg}, and when $\thetak\approx0$ the integrand
in \refeqn{lcbraggxsdef} becomes independent of $\varphi$.

The particular coordinate system defined in \reffig{lcbraggon3dsphere}
ensures the symmetry $\gamma(\varphi)=\gamma(-\varphi)$, implying that for purposes of
evaluating \refeqn{lcbraggxsdef}, it is possible to restrict $\varphi$ to the
interval $[0,\pi]$ and use instead:
\begin{align}
  g^\text{LC}_{hkl}(\alpha,\hatki) &= \frac{1}{\pi}\int_0^\pi g^\text{SC}_{hkl}(\alpha,\gamma(\thetak,\thetan,\varphi)) d\varphi
  \labeqn{lcbraggxsdefpositivevarphi}
\end{align}
Additionally, either the integrand in \refeqn{lcbraggxsdefpositivevarphi} will
be vanishing everywhere, or it will be non-vanishing exactly on a single
sub-interval, $[\phimin,\phimax]$, which will be
identified in \refsec{lcbragg:xseval}.
A final advantage of the chosen coordinate system is that it allows an efficient
pre-check of whether or not a given set of planes will contribute to the
scattering of a particular neutron, by determining whether or not the $z$-range of the Bragg circle overlaps with the $z$-range of
the mosaicity bands. The latter can be trivially pre-calculated at initialisation time and
the former is given by:
\begin{align}
   [\cos(\thetak+\alpha),\cos(\max(0,\thetak-\alpha))]
\end{align}
Which can be calculated cheaply at simulation time by usage of cosine addition
formulas, using $\cos\alpha=\lambda\times(1/2d)$ and
$\sin\alpha=\sqrt{1-\cos^2\alpha}$. In addition to a few basic
arithmetic manipulations, the cost of the pre-check is therefore just a simple square-root
evaluation. To further speed up this critical section of the code, the check is
carried out first using a Taylor expansion approximation of the square-root evaluation.

\subsection{Evaluating the geometrical integral}\labsec{lcbragg:xseval}

In principle \refeqn{lcbraggxsdefpositivevarphi} can be evaluated directly via
numerical integration, using the algorithms described in \refsec{scbragg} to
evaluate $g^\text{SC}_{hkl}$ at each $\varphi$ point. However, for non-vanishing
$\alpha$ and realistic mosaic spreads, this integrand will be
vanishing at most  $\varphi$ values, with non-zero contributions arising
at most in a single narrow region. For numerical integration to proceed reliably and
efficiently, it is therefore necessary to predetermine the $\varphi$-regions in
which the integrand is non-zero analytically, and apply the numerical
integration algorithm only to these regions. In this context it is also worth noting the
importance of being able to evaluate the integrand
of \refeqn{lcbraggxsdefpositivevarphi} with the efficient method
of \refeqn{circleintegralapproxformulawithspline}, thus most of the time
avoiding the computationally demanding scenario of having a numerical integration of an integrand which
is itself evaluated with numerical integration.

Now, a neutron has a non-zero cross section to scatter on a small crystal with rotation
$\varphi$ if and only if:
\begin{align}
           & \left|\alpha-\gamma\right|\le\tau\nonumber\\
\LRA\qquad & \alpha-\tau  \le \gamma \le \alpha+\tau
\end{align}
Since $\gamma\ge0$ by definition, this is equivalent to:
\begin{align}
   \max(0,\alpha-\tau)  \le \gamma \le \alpha+\tau
   \labeqn{lcbragg::gammalimitintermediate}
\end{align}
Now, all three parts of this double inequality resides in $[0,\pi]$, since
by definition $\gamma\le\pi$, $\alpha\le\pi/2$, and $\tau\le\pi/2$
(for the case of $\tau$ this is a deliberate restriction, as discussed in \refsec{scbragg::mosdef}). On
this domain, the cosine function is one-to-one and with non-positive derivative,
so \refeqn{lcbragg::gammalimitintermediate} is equivalent to:
\begin{align}
  \cos(\max(0,\alpha-\tau))  \ge \cos\gamma \ge \cos(\alpha+\tau)
\end{align}
Using \refeqn{lcbragg::cosgamma} this becomes:
\begin{align}
  \cos(\max(0,\alpha-\tau))  \ge \sin\thetak\sin\thetan\cos\varphi+\cos\thetak\cos\thetan \ge \cos(\alpha+\tau)
\end{align}
As the degenerate cases $\thetan\approx0$ and $\thetak\approx0$ are dealt with separately,
it is safe to divide with $\sin\thetak\sin\thetan$:
\begin{align}
  \frac{\cos(\alpha+\tau)-\cos\thetak\cos\thetan}{\sin\thetak\sin\thetan}\le\cos\varphi\le\frac{\cos(\max(0,\alpha-\tau))-\cos\thetak\cos\thetan}{\sin\thetak\sin\thetan}
\end{align}
Since $\varphi\in[0,\pi]$, the range of phi values representing
crystallite orientations with non-zero contributions to the scattering
cross section is thus $\phimin<\varphi<\phimax$, with:
\begin{align}
   \phimin&\equiv\arccos\!\left(\min\!\left(1,\frac{\cos(\max(0,\alpha-\tau))-\cos\thetak\cos\thetan}{\sin\thetak\sin\thetan}\right)\right)\nonumber\\
   \phimax&\equiv\arccos\!\left(\max\!\left(-1,\frac{\cos(\alpha+\tau)-\cos\thetak\cos\thetan}{\sin\thetak\sin\thetan}\right)\right)
   \labeqn{lcbragg::phiminphimax}
\end{align}
Obviously, $g^\text{LC}_{hkl}(\alpha,\hatki)=0$ unless $\phimax>\phimin$,
in which case \refeqn{lcbraggxsdefpositivevarphi} can be replaced by:
\begin{align}
  g^\text{LC}_{hkl}(\alpha,\hatki) &= \frac{1}{\pi}\int_\phimin^\phimax g^\text{SC}_{hkl}(\alpha,\gamma(\thetak,\thetan,\varphi)) d\varphi
  \labeqn{lcbraggxsdefphiminmaxrange}
\end{align}
Now, the code developed in \refsec{scbragg} for evaluation of
$g^\text{SC}_{hkl}(\alpha,\gamma(\thetak,\thetan,\varphi))$ works directly from
$\cos\gamma$ rather than $\gamma$ for reasons of efficiency, which
with \refeqn{lcbragg::cosgamma} translates into a need to provide a value of
$\cos\varphi$ at each evaluated $\varphi$ point. As
in \refsec{scbragg::inteval}, such values are cheaply generated via
trigonometric addition formulas, and the customised implementation of Romberg
integration discussed in \refsec{scbragg::inteval} is reused for efficient
and accurate integration of \refeqn{lcbraggxsdefphiminmaxrange} as well.

\Reffig{lcbraggxspg}
\begin{figure}
  \centering \includegraphics[width=0.99\textwidth]{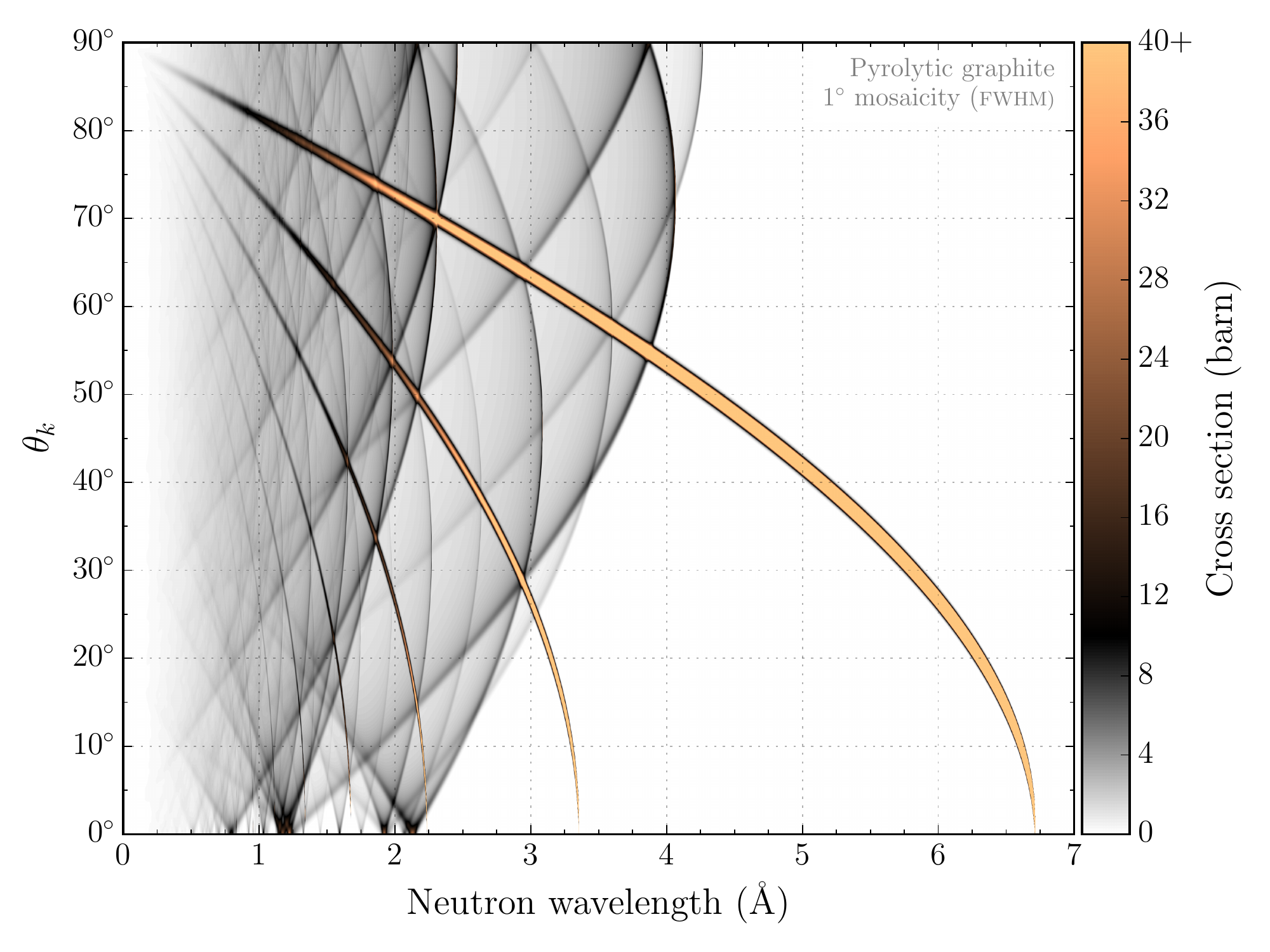}
 \caption{Bragg diffraction cross section in pyrolytic graphite as a function of
  neutron wavelength and incidence angle ($\thetak$) with respect to the lattice
  $c$-axis of the crystal (the axis orthogonal to the average orientation of the
  graphene sheets). The $c$-axis orientation is assumed to be distributed with a
  $1^\circ$ FWHM Gaussian mosaicity, and the cross section values were
  calculated with default settings of \texttt{NCrystal}.}
  \labfig{lcbraggxspg}
\end{figure}
shows the resulting Bragg diffraction cross sections in pyrolytic graphite as a function of
neutron wavelength and $\thetak$, with distinctive structures corresponding to
particular groups of scattering planes. In order to understand how these
structures relate to actual scattering planes, it is instructive to consider a
hypothetical layered single crystal with just a single reflection plane having
$\thetan=60^\circ$. If both ``sides'' of this reflection plane are included,
i.e.\ both $(h,k,l)$ and $(-h,-k,-l)$ are considered (as is always the case
in \texttt{NCrystal}), the resulting cross section structure is shown
in \reffig{lcbragg_singlenormal_thetan60}.
\begin{figure}
  \centering
  \includegraphics[width=0.8\textwidth]{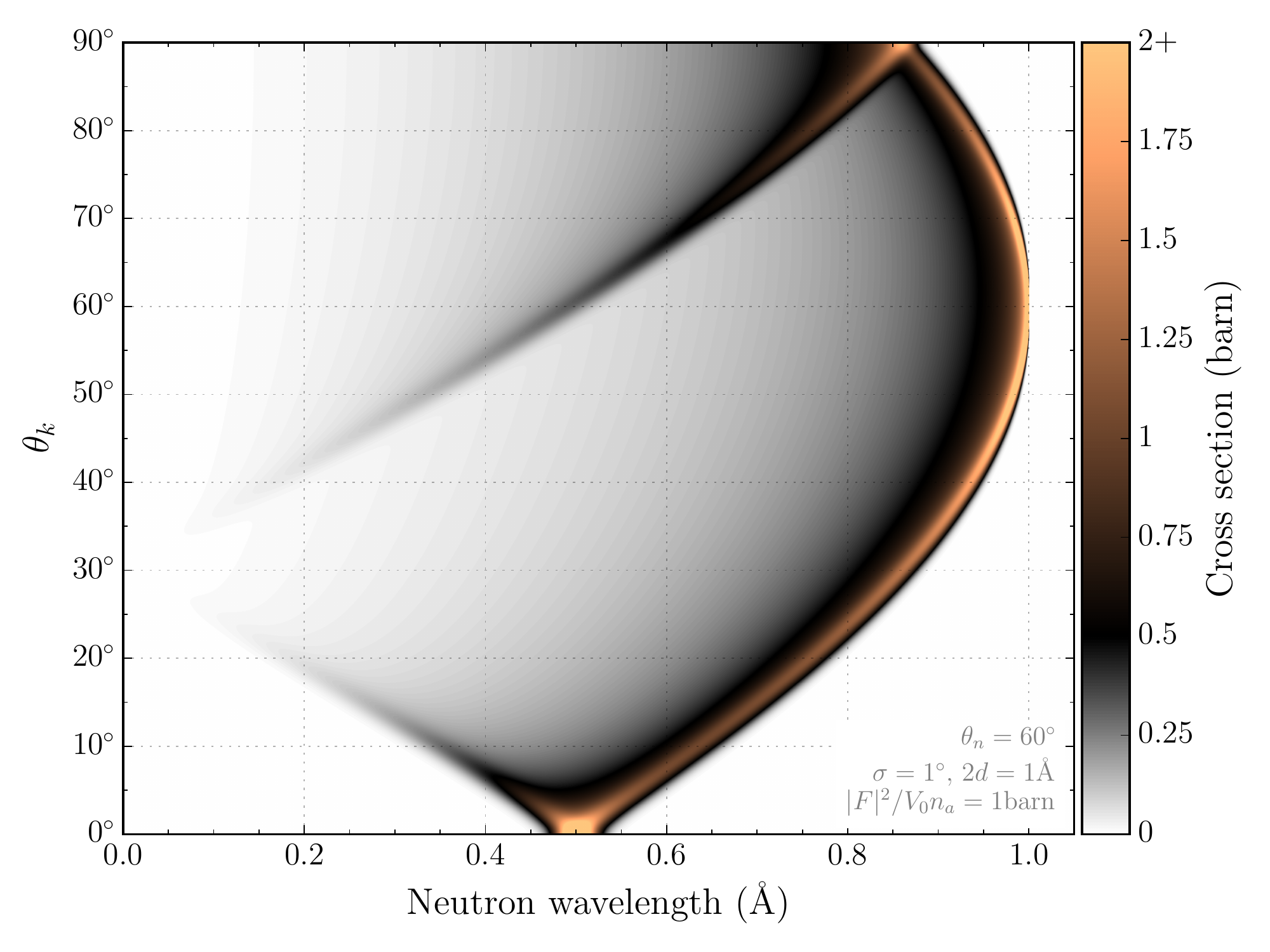}
 \caption{Bragg diffraction cross section in a hypothetical layered single
  crystal containing just a single reflection plane inclined $\thetan=60^\circ$
  with respect to $\hat{L}$, as a function of neutron wavelength and incidence
  angle ($\thetak$). For simplicity, a $1^\circ$ FWHM Gaussian mosaicity is
  assumed, along with $2d_{hkl}=\SI{1}{\angstrom}$ and $|F_{hkl}|^2/V_\text{uc}n_a=\SI{1}{barn}$.}
  \labfig{lcbragg_singlenormal_thetan60}
\end{figure}
The decomposition into contributions
from scattering on $(h,k,l)$ ($\thetan\le\pi/2$) and $(-h,-k,-l)$
($\thetan\ge\pi/2$) is shown
in \reffig{lcbragg_singlenormal_thetan60_decomposed}.
\begin{figure}
  \centering
  \subfloat[\labfig{lcbragg_singlenormal_thetan60_decomposed_noanti}]{\includegraphics[width=0.49\textwidth]%
  {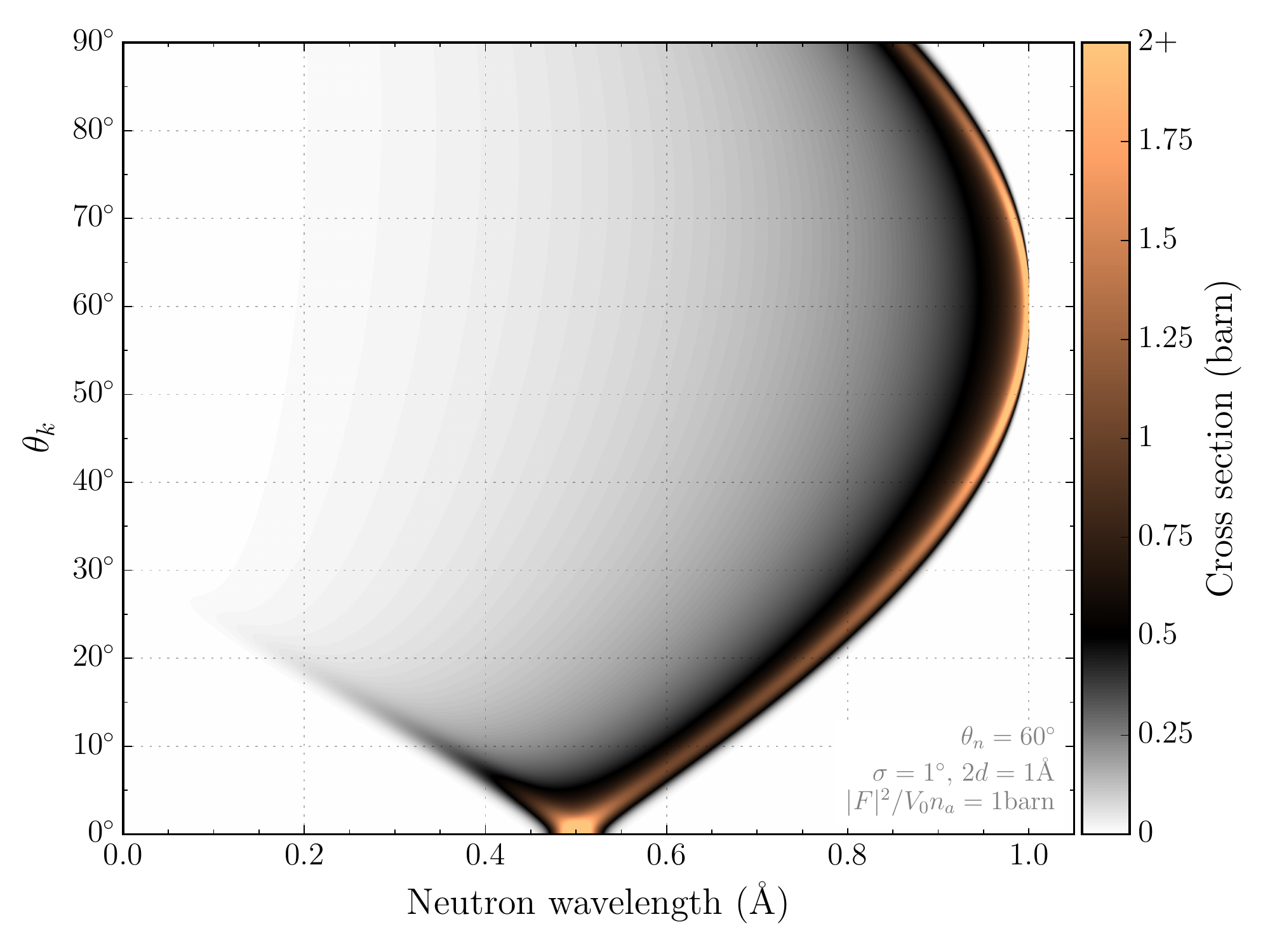}}
  \subfloat[\labfig{lcbragg_singlenormal_thetan60_decomposed_onlyanti}]{\includegraphics[width=0.49\textwidth]%
  {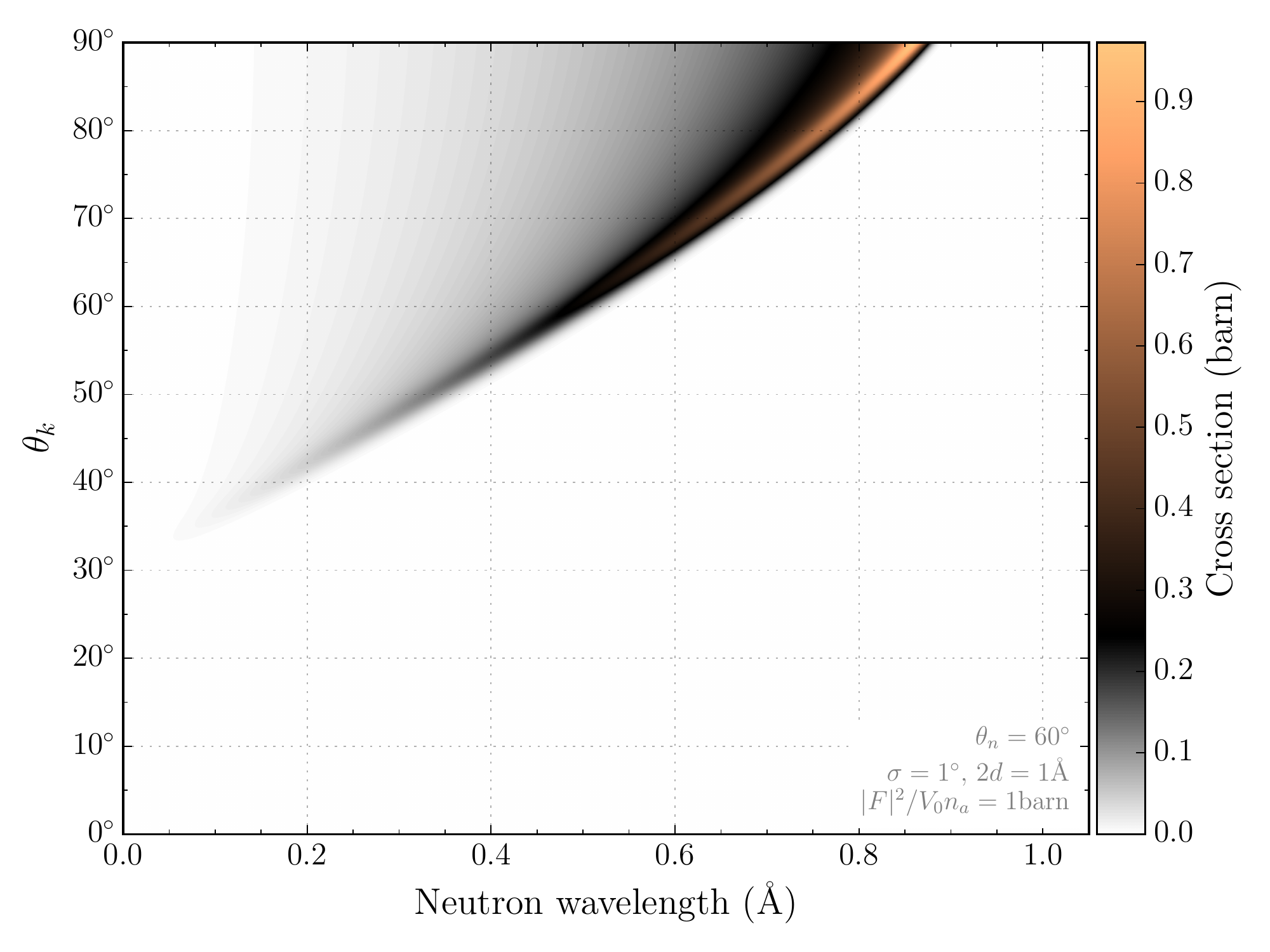}}\\
  \caption{As in \reffig{lcbragg_singlenormal_thetan60}, but showing the
   decomposition into contributions from scattering on the mosaicity band
  centred on $\thetan\in[0,\pi/2]$
   (shown in \refsubfigfrommaincaption{lcbragg_singlenormal_thetan60_decomposed_noanti})
   and  $\pi-\thetan$ (shown in \refsubfigfrommaincaption{lcbragg_singlenormal_thetan60_decomposed_onlyanti})
   respectively.}
  \labfig{lcbragg_singlenormal_thetan60_decomposed}
\end{figure}
Considering first
\reffig{lcbragg_singlenormal_thetan60_decomposed_noanti}, its structure can be understood
(at least in the limit of small mosaicities) by carefully
considering \reffig{lcbraggon3dsphere} while recalling that
$\cos\alpha=\lambda/2d_{hkl}$. The present discussion will limit itself to
consideration of a few important features. Firstly, when $\thetak\approx0$, the
neutron is aligned along $\hat{L}$ and the only contribution comes when
$\alpha\approx\thetan$, which means $\lambda/2d_{hkl}\approx\cos\thetan$,
translating to $\lambda\approx\SI{0.5}{\angstrom}$. Next, in case of
back-scattering, $\lambda\to2d_{hkl}$ and the Bragg circle contracts to a point, the
only contribution comes when $\thetak\approx\thetan$. Keeping
$\thetak\approx\thetan$ but lowering $\lambda$, the Bragg circle will grow and
the integral along it through the mosaic density, $g_{hkl}$, will quickly tend
to a constant, which is reminiscent of --- but not identical to --- the formula
for a crystal powder in \refeqn{geomfactordefpowder}. A constant $g_{hkl}$
inserted into \refeqn{qstrengthandgstrengthfactors} implies
$\sigma^{hkl}\propto\lambda^2/\sqrt{1-(\lambda/2d_{hkl})^2}$, which can be
compared with the equivalent prediction in a powder,
$\sigma^{hkl}\propto\lambda^2$. When $\thetak<\pi/2-\thetan$, the situation is a
bit different: here, $-\hatki$ lies so much ``to the north'' of the mosaicity band in
\reffig{lcbraggon3dsphere}, that when the Bragg circle is at its largest, it
will lie ``to the south'' of the mosaicity band everywhere, which explains the
vanishing cross sections in the bottom left
of \reffig{lcbragg_singlenormal_thetan60_decomposed_noanti}. The ridge along
this triangle shows enhanced cross sections, which can be understood from an
intermittent enhanced overlap between the Bragg circle and the mosaicity band
for rotations around $\varphi=\pi$, just before the Bragg circle extends too
far south. Finally, it might be worth noticing that at $\thetak=90^\circ$, cross
sections will vanish unless the Bragg circle is large enough to reach the
mosaicity band, which happens when $\alpha=\pi/2-\thetan$. I.e, the right-most
ridge in \reffig{lcbragg_singlenormal_thetan60_decomposed_noanti} intersects the
axis at $\thetak=90^\circ$ at the point where $\lambda=2d_{hkl}\sin\thetan$,
which in this case corresponds to $\lambda\approx\SI{0.866}{\angstrom}$.

Moving on to \reffig{lcbragg_singlenormal_thetan60_decomposed_onlyanti} and the
contribution from scattering on the mosaicity band at $\pi-\thetan$, such
scattering is impossible unless $\thetak$ is large enough that the Bragg circle
can reach the mosaicity band on the ``southern hemisphere'' (note that this band is not
explicitly shown in \reffig{lcbraggon3dsphere}). In other words,
contributions only exist in the region defined by $\thetak>\thetan-\alpha$. The
edges of this region intersect the axes at $\lambda=0$ and $\thetak=90^\circ$
in the same locations as the ridges
in \reffig{lcbragg_singlenormal_thetan60_decomposed_onlyanti}, thus completing
the distorted triangular structure visible
in \reffig{lcbragg_singlenormal_thetan60}. The ridge-like structure along
the edge in \reffig{lcbragg_singlenormal_thetan60_decomposed_onlyanti} can again
be understood as coming from an enhanced overlap of the Bragg circle and the
mosaicity band near the edge. Away from the edge in
\reffig{lcbragg_singlenormal_thetan60_decomposed_onlyanti}
and inside the region in the
upper left corner, the cross section will again fall off as
$\sigma^{hkl}\propto\lambda^2/\sqrt{1-(\lambda/2d_{hkl})^2}$.

Qualitatively, the distorted triangular structure seen
in \reffig{lcbragg_singlenormal_thetan60_decomposed} does not depend on the exact
value of $\thetan$, although the particular position of the edges and corners
does of course. In the
limiting cases of $\thetan=0^\circ$ (the $00l$ reflections in pyrolytic
graphite: $002$, $004$, ...), the triangle collapse to a single line as shown in
\reffig{lcbragg_singlenormal_thetan0and90_0}
\begin{figure}
  \centering
  \subfloat[\labfig{lcbragg_singlenormal_thetan0and90_0}]{\includegraphics[width=0.49\textwidth]{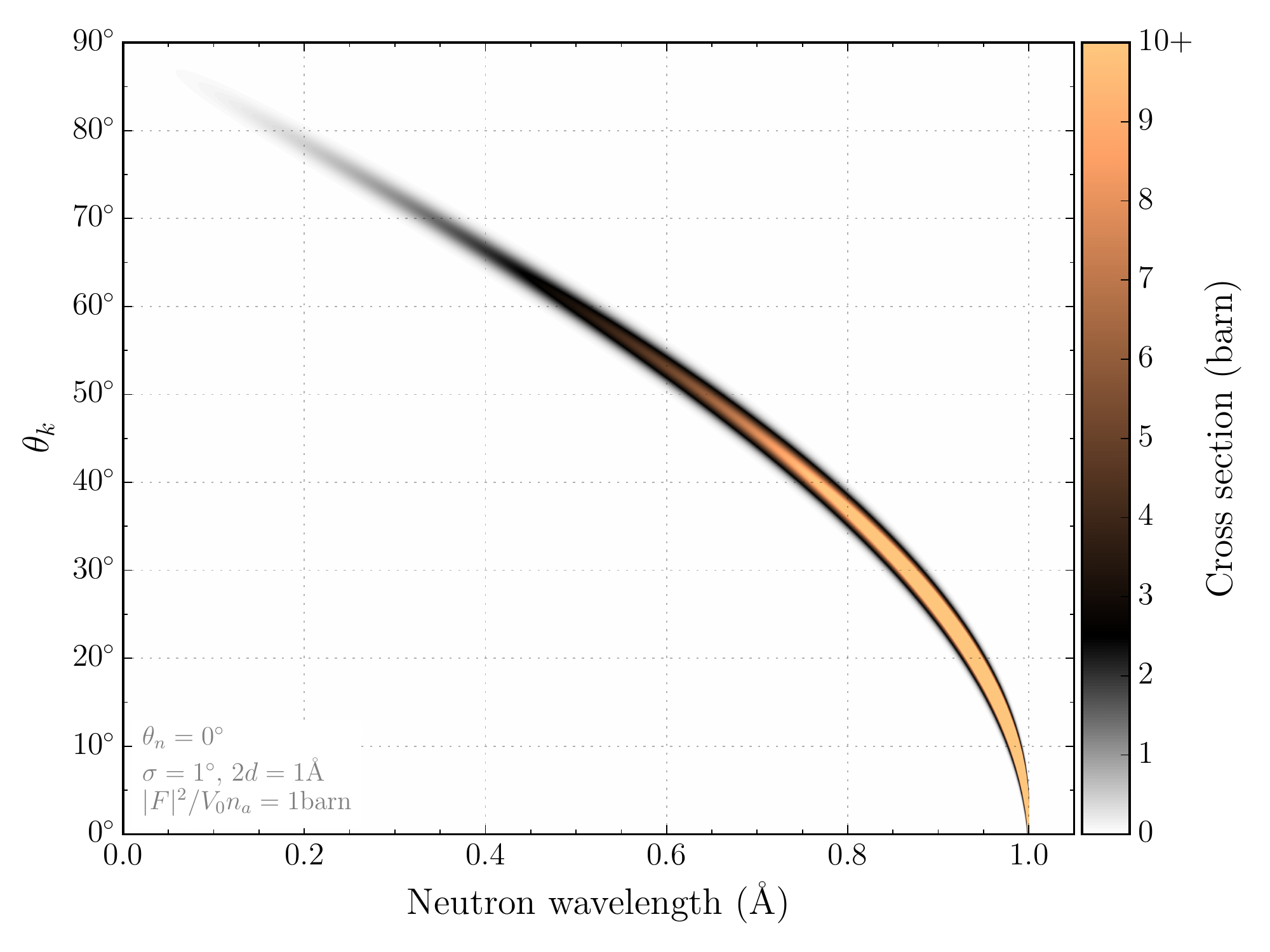}}
  \subfloat[\labfig{lcbragg_singlenormal_thetan0and90_90}]{\includegraphics[width=0.49\textwidth]{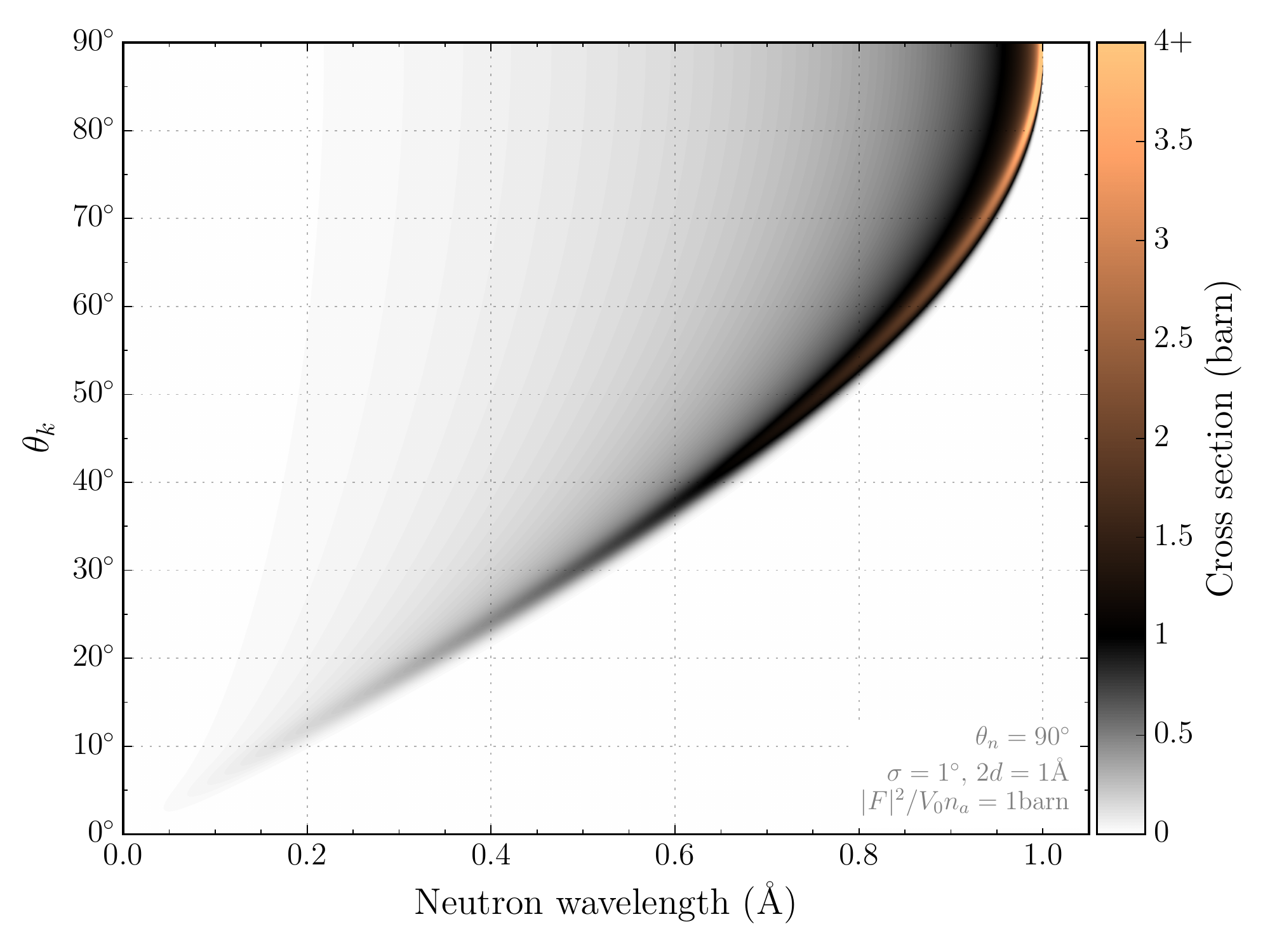}}\\
  \caption{As in \reffig{lcbragg_singlenormal_thetan60}, but for
  $\thetan=0^\circ$ and  $\thetan=90^\circ$, shown
  in \reftwosubfigsfrommaincaption{lcbragg_singlenormal_thetan0and90_0}{lcbragg_singlenormal_thetan0and90_90} respectively.}
  \labfig{lcbragg_singlenormal_thetan0and90}
\end{figure}
--- which essentially is how cross
sections look in an isotropic non-layered single crystal. In the
limiting cases of $\thetan=90^\circ$ (the $hk0$ reflections in pyrolytic
graphite), the triangle again collapses to a single line as shown in
\reffig{lcbragg_singlenormal_thetan0and90_90}, but with a behaviour very
different from that governing isotropic non-layered single crystals.

The structures seen in \reffig{lcbraggxspg} can now be understood as arising
from the various values of $(\thetan,d_{hkl})$ encountered in pyrolytic
graphite. In \refsec{lcbragg::val::frikkee}, the connection to specific $hkl$
values will be explored further.

\subsection{Sampling scattering events}\labsec{lcbragg::sampling}

In case of a scattering event in a layered single crystal, the sampling of an
outgoing neutron direction ($\hatkf$) starts of in the same vein as in the
case of non-layered single crystals discussed
in \refsec{scbragg::genscat}: a set of $hkl$ planes with common $\thetan$ and
$d$-spacing is first sampled randomly among those contributing, with the
sampling being based upon their individual cross section contributions. These
contributions are usually already available from values cached
during a previous cross section calculation. If the selected group of planes has
$\thetan\approx0$, the code
in \refsec{scbragg::genscat} is directly used to sample the scattering, and
otherwise the next step consists of sampling a value of $\varphi$, in essence
determining the exact orientation of the hypothetical small Gaussian mosaic
crystal involved in the scattering. In the degenerate case $\thetak\approx0$,
all crystal orientations contribute equally, and $\varphi$ is picked uniformly
in $[-\pi,\pi]$. Otherwise a value of $\varphi$ must be sampled with weights
proportional to the integrand of \refeqn{lcbraggxsdef}. Technically, this is
done by sampling a value in the interval $[\phimin,\phimax]$
(cf.\ \refeqn{lcbragg::phiminphimax}), and flipping the sign of the sampled
$\varphi$ value 50\% of the time.

The actual sampling on $[\phimin,\phimax]$ will be based on Monte Carlo
acceptance-rejection sampling, as was also the case in \refsec{scbragg::genscat}. This time,
however, the point in $[\phimin,\phimax]$ with the highest contribution is not
known in advance, precluding the simple determination of a constant overlay
function. Instead, an ad hoc overlay function is created on the fly: based on the
evaluation of the curve at 9 evenly spaced points in $[\phimin,\phimax]$, a
piece-wise constant overlay curve is created as illustrated by the examples
in \reftwofigures{lcbraggsampleoverlays1}{lcbraggsampleoverlays2}.
\begin{figure}
  \centering
  \subfloat[\labfig{lcbraggsampleoverlays1a}]{\includegraphics[width=0.79\textwidth]{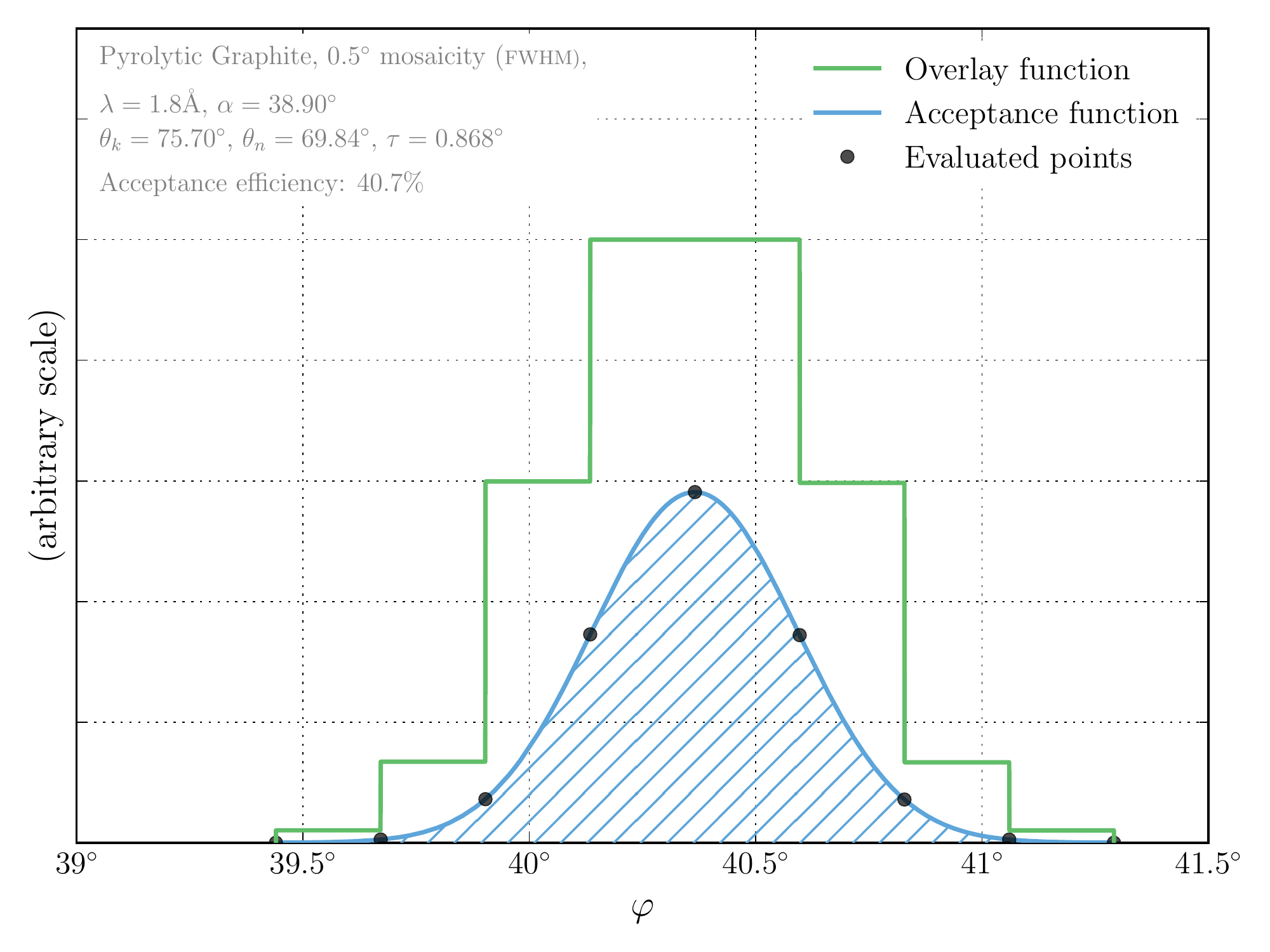}}\\
  \subfloat[\labfig{lcbraggsampleoverlays1b}]{\includegraphics[width=0.79\textwidth]{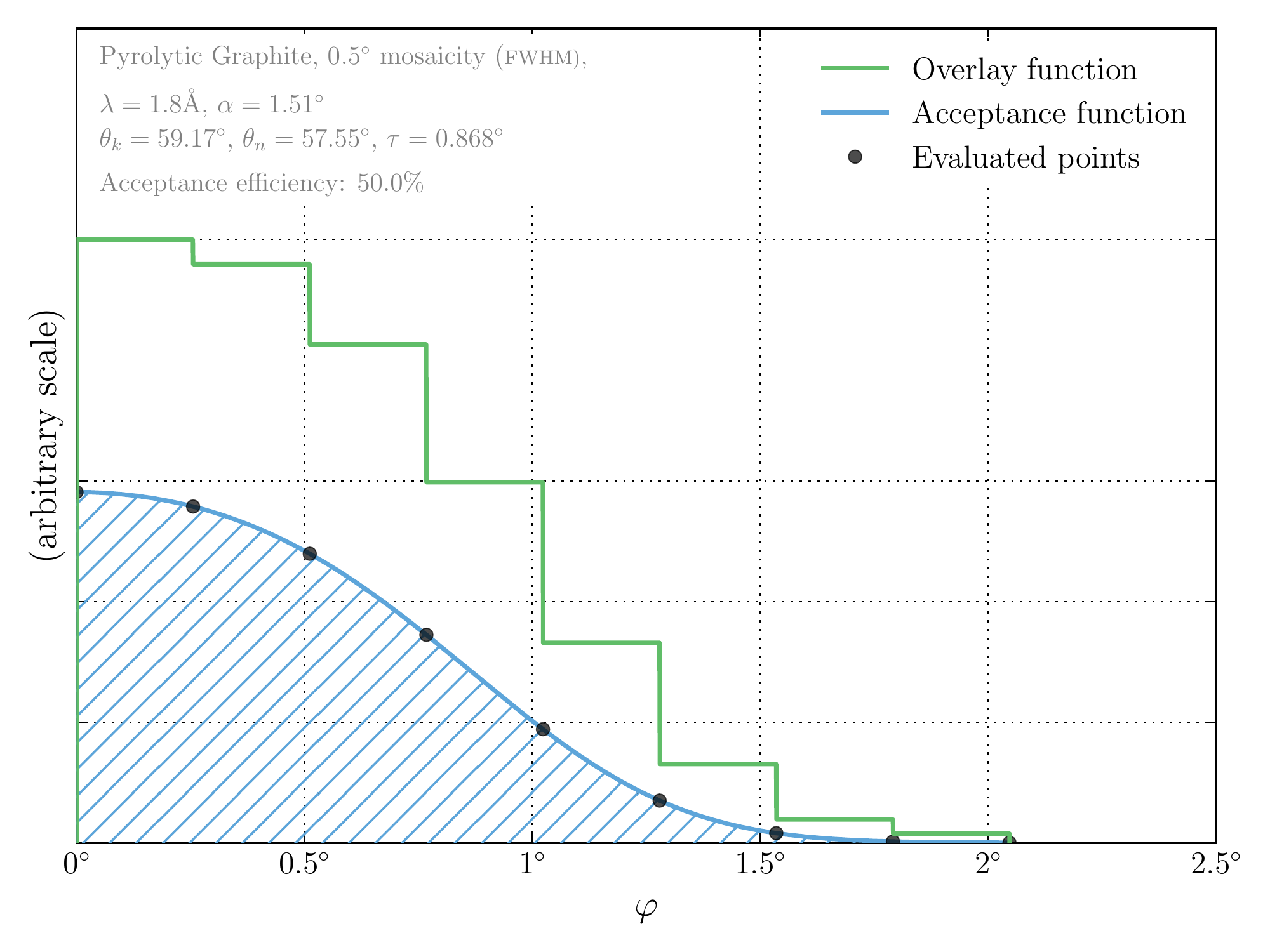}}
  \caption{Examples of overlay and acceptance functions used when sampling
  $\varphi$ values in layered single crystals. The overlay functions are
  constructed on the fly on the basis of 9 evaluations of the acceptance
  functions, shown as black dots. As indicated in the plots, the examples here involves medium mosaic spreads and
  cases without (\refsubfigfrommaincaption{lcbraggsampleoverlays1a}) and
  with (\refsubfigfrommaincaption{lcbraggsampleoverlays1b}) back-scattering.}
  \labfig{lcbraggsampleoverlays1}
\end{figure}
\begin{figure}
  \centering
  \subfloat[\labfig{lcbraggsampleoverlays2a}]{\includegraphics[width=0.79\textwidth]{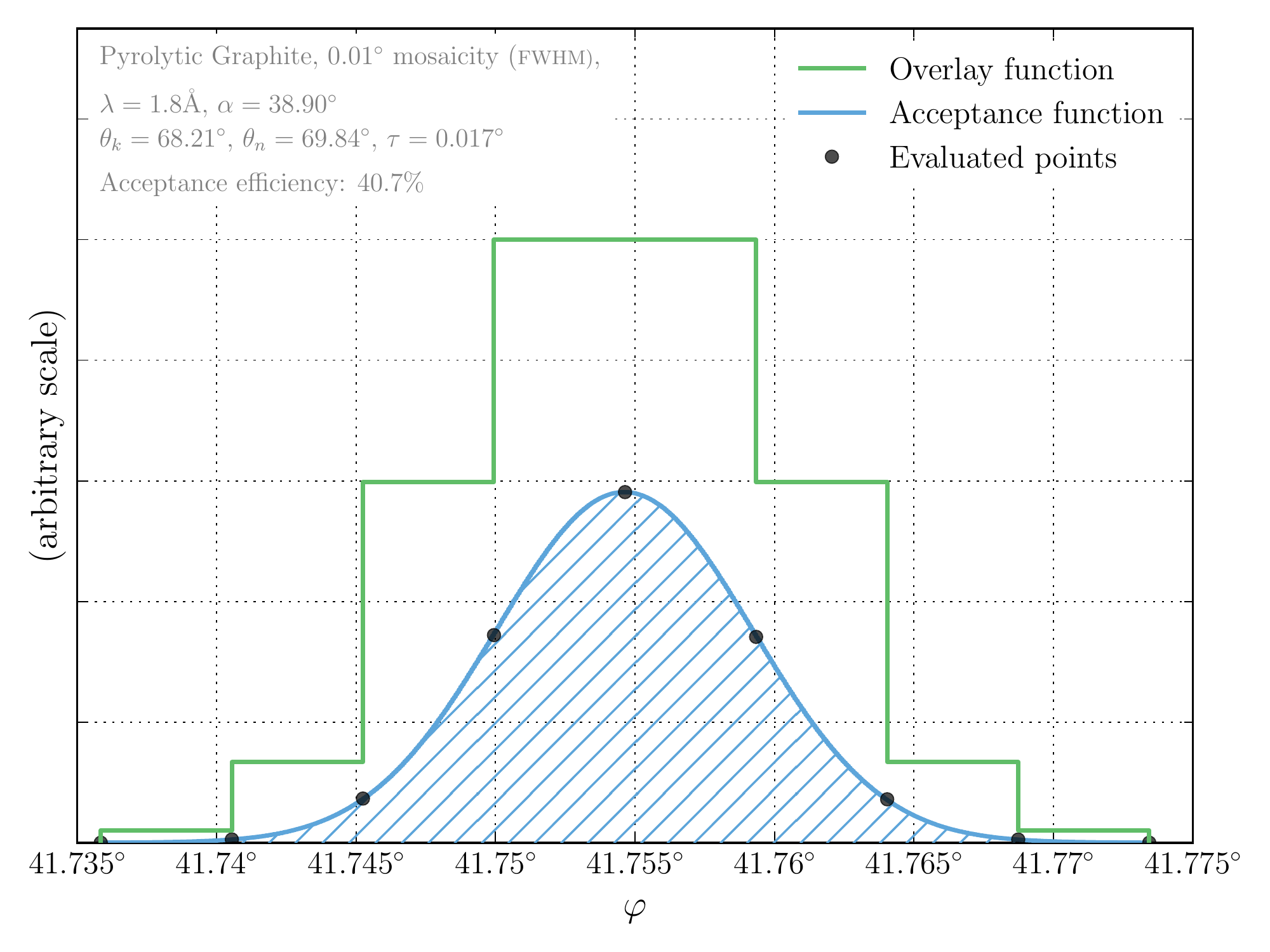}}\\
  \subfloat[\labfig{lcbraggsampleoverlays2b}]{\includegraphics[width=0.79\textwidth]{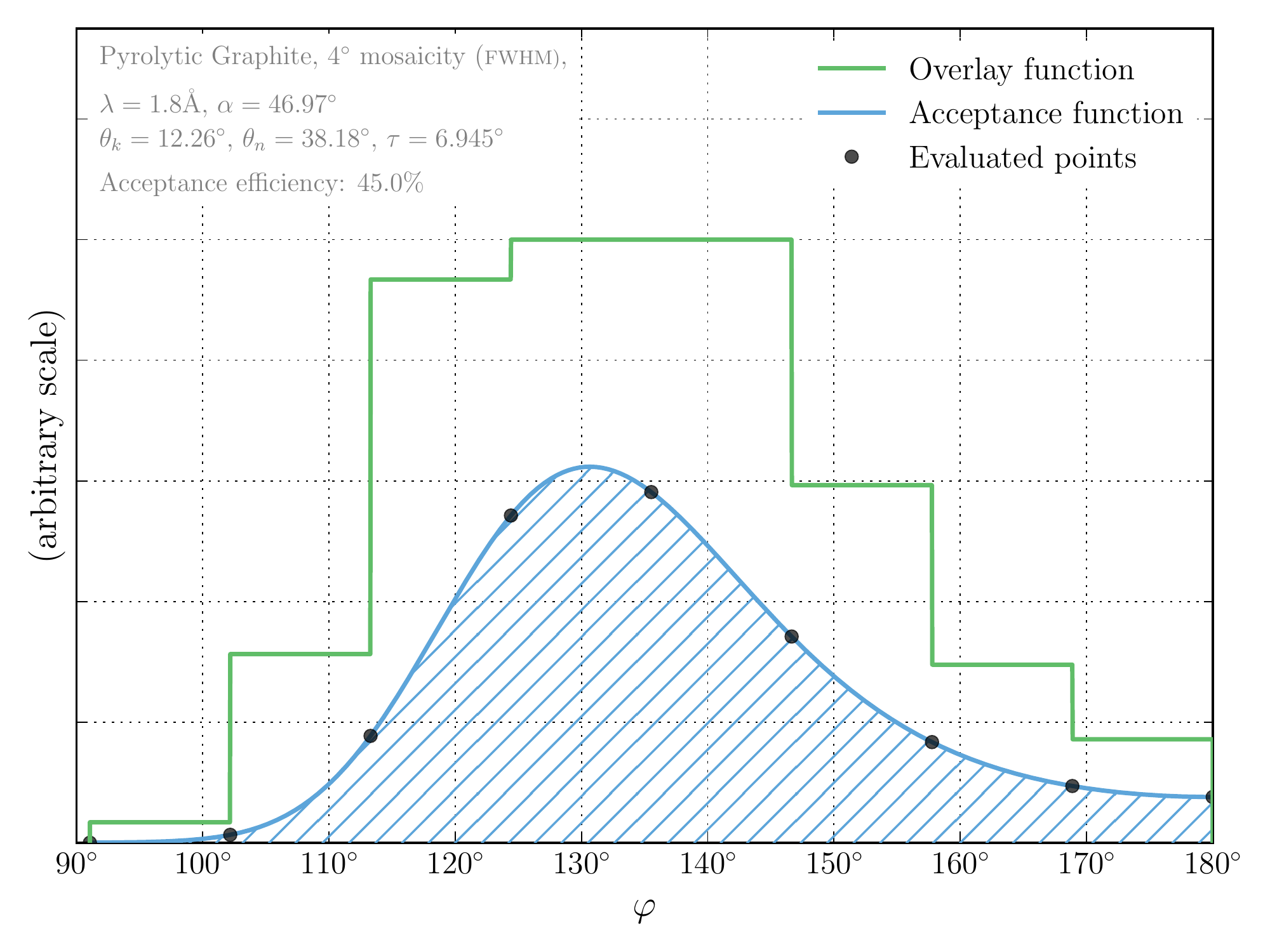}}
  \caption{Examples of overlay and acceptance functions as
  in \reffig{lcbraggsampleoverlays1}, but for examples involving situations with both small
  (\refsubfigfrommaincaption{lcbraggsampleoverlays2a}) and large
  (\refsubfigfrommaincaption{lcbraggsampleoverlays2b}) mosaic spreads.}
  \labfig{lcbraggsampleoverlays2}
\end{figure}
The overlay height in each of the 8 bins is
given as the highest of the two points at the bin's edges for which the curve
was evaluated, with additional safety margins added to account for the cases
where the curve has a local maximum inside the bin. The safety margin is added
as a multiplicative factor of 1.7 and a constant offset of 2\% of the maximal
value seen in all 9 evaluated points. The safety margin is important since the
validity of the resulting distribution of $\varphi$ values is guaranteed if and
only if the overlay curve is everywhere equal to or larger than the
integrand of \refeqn{lcbraggxsdef}. On the other hand the overlay curve should
not be too large, since the acceptance rate and thus computational efficiency,
is given as the ratio of the area under the two curves. The validity of the
constructed overlay curves was verified by numerical investigations of a large
number of scattering scenarios, and additionally the code is written so as to
throw an exception in case an invalid overlay height is ever encountered
(obviously no reports of this has surfaced so far). The acceptance rate seems to
be around 30--40\%, which implies that the sampling requires roughly 11--12
evaluations of the integrand of \refeqn{lcbraggxsdef}: 9 to construct the
overlay curve, and around 2--3 to carry out the sampling.

Having sampled a particular value of $\varphi$ and thereby determined the exact
orientation of the hypothetical small Gaussian mosaic crystal on which to
scatter, the code discussed in \refsec{scbragg::genscat} is used to generate the
actual scattering --- still in the coordinate system
of \reffig{lcbraggon3dsphere}. Finally, the resulting neutron direction,
$\hatkf$ must be rotated into the laboratory frame, which involves the careful
determination of an appropriate rotation matrix based on the values of $\hat{L}$
and $\hatki$ in the laboratory frame.

\subsection{Alternative reference implementation}\labsec{lcbragg:altrefimpl}

For reasons of efficiency, the implementation described in
\refsectionrange{lcbragg::defs}{lcbragg::sampling} groups $hkl$ planes
according to $d$-spacing and $\thetan$, and then treats each such group
separately and in its own coordinate system. This approach takes advantage of
rotational symmetries in these coordinate systems, exploits symmetries between
the $(h,k,l)$ and $(-h,-k,-l)$ reflection planes, and ensures that numerical
integration algorithms will only be deployed on smoothly varying and
non-vanishing functions.

As an alternative to this efficient but admittedly complicated implementation, a
simpler reference model is useful for validation and comparison purposes ---
even if it is prohibitively inefficient for practical usage. Fortunately, such a
model is readily implemented, based again on the notion of modelling a layered
single crystal as consisting of a large number of small non-layered single crystals, rotated uniformly around
$\hat{L}$. In the reference implementation presented here, the small crystals
are instead treated directly with the code for non-layered single crystals
described in \refsec{scbragg}. Thus, with $\{\varphi_j\}$ representing a set of
$N_j$ rotations, ideally very large in number and distributed uniformly over
$[-\pi,\pi]$, the cross section for coherent elastic scattering
of a neutron with wavevector $\vecki$ on a layered single crystal is
modelled as:
\begin{align}
  \sigma^{\text{LC}}(\vecki) = \frac{1}{N_j}\sum_j\left(\sigma^{\text{SC}}(R_{\varphi_j}\vecki))\right)
  \labeqn{lcbraggrefimpldef}
\end{align}
Where $R_{\varphi_j}$ is a rotation operator implementing a rotation of
$\varphi_j$ around $\hat{L}$, with $\varphi=0$ meaning no rotation, and
$\sigma^{\text{SC}}(\vecki)$ is the single crystal cross section. In order to
sample the outcome of a scattering event, a particular $\varphi_j$ is first
sampled according to the contribution of the corresponding term
in \refeqn{lcbraggrefimpldef}, and the non-layered single crystal model is then used to
scatter $R_{\varphi_j}\vecki$ into a final state $\vec{k}_f^\prime$, which
is subsequently transformed back to the original frame to get the actual outcome of the
interaction: $\veckf=R^{-1}_{\varphi_j}\vec{k}_f^\prime$. For simplicity,
and to avoid issues of numerical stability, the rotation operator and its
inverse are implemented using Rodriguez' rotation
formula~\cite{rodriguez1840lois}.

The only remaining issue is how to construct the set of rotational values,
$\{\varphi_j\}$. One approach is the random sampling of $n$ uniformly distributed
values in $[-\pi,\pi]$ --- with a different set generated for each incoming
neutron to reduce statistical bias. Alternatively, the $n$ values can simply be
distributed equidistantly over the interval $[-\pi,\pi]$. In the latter case, it is clear
that for the model to yield reliable results, $n$ must be large enough that
$\pi/n$ will be much smaller than the mosaicity of the crystal. In the former
case of randomised $\varphi$ values, $n$ will likely need to be much higher
since the sampled $\varphi_j$ points tend to cluster in some regions, leaving
other regions with reduced density. On the other hand, one might worry that the
structure inherent to a non-randomised set of $\varphi_j$ values might be
reflected as an unwanted artefact in the final distributions of $\veckf$ values,
but in principle it should not be a significant problem as long as $n$ is high
enough that angular effects at the order of $\pi/n$ are washed out by the
non-vanishing mosaicity effects in the non-layered single crystal models.

Nevertheless, \texttt{NCrystal} supports both models through the \texttt{lcmode}
configuration parameter. If left at the default setting, \texttt{lcmode=0}, the
efficient model described in \refsectionrange{lcbragg::defs}{lcbragg::sampling}
is used. If set to a positive value, \texttt{lcmode=n}, the reference model
described in the present section will be used with $n$ values of $\varphi$, evenly
spaced in $[-\pi,\pi]$. If set to a negative value, \texttt{lcmode=-n}, the
reference model will again be used, but now with the $n$ values of $\varphi$
sampled randomly and independently for each neutron, as described above. It is interesting to note that the \texttt{Single\_crystal}
component~\cite{mcstascompman} of \texttt{McStas} since 2015 supports an
experimental mode for pyrolytic graphite modelling, which essentially
corresponds to \texttt{NCrystal}'s \texttt{lcmode=-1}. As should be clear from
the present discussions, and which will be verified in
section \refsec{lcbragg::val::frikkee}, $|n|=1$ is much too low to provide reliable
results.

In general, the validation work in \refsec{lcbragg::val::refimplcmp} will employ
an evenly spaced distribution of $\varphi_j$ values, with a high value of
$n$. Requiring $2\pi/n$ to be equal to 5\% of the FWHM mosaicity of a crystal,
one finds $n=\num{7.2e3}$ for a mosaicity of $1^\circ$, $n=\num{4.32e5}$ for a
mosaicity of $1'$, and $n=\num{2.592e6}$ for a mosaicity of $10''$. In practice
this limits the usage of the reference models, even for validation work, to
crystals with large mosaicities.

\subsection{Validation}\labsec{lcbragg::validation}

As was the case for the non-layered single crystal model
in \refsec{scbragg::validation}, the work done to validate the layered single
crystal model first consists of a verification
in \reftwosections{lcbragg::val::refimplcmp}{lcbragg::val::consistencychecks}
that the implementation described
in \refsectionrange{lcbragg::defs}{lcbragg::sampling} actually provides results
consistent with the model as originally defined. Next, in \refsec{lcbragg::val::frikkee}, the
model is compared to existing results from the literature. Benchmark numbers for
computational efficiency are included in the general discussion
in \refsec{timing}.

\subsubsection{Comparison with reference implementation}\labsec{lcbragg::val::refimplcmp}

Due to the double-integration involved, it is unfortunately impractical, in terms
of computational time requirements, to proceed in a similar vein as
in \refsec{scbragg::val::mpmathcmp} and provide a high-precision reference
implementation for layered single crystals with \texttt{mpmath}. Instead, the
alternative model described in \refsec{lcbragg:altrefimpl}, available
in \texttt{NCrystal} via the \texttt{lcmode} configuration parameter, will be
used to provide reference results. These are then used to verify the
implementation of the primary model described
in \refsectionrange{lcbragg::defs}{lcbragg::sampling}.

First, \reffig{lcb_xsvsref_thetak40_mos3deg}
\begin{figure}
  \centering
  \includegraphics[width=0.85\textwidth]{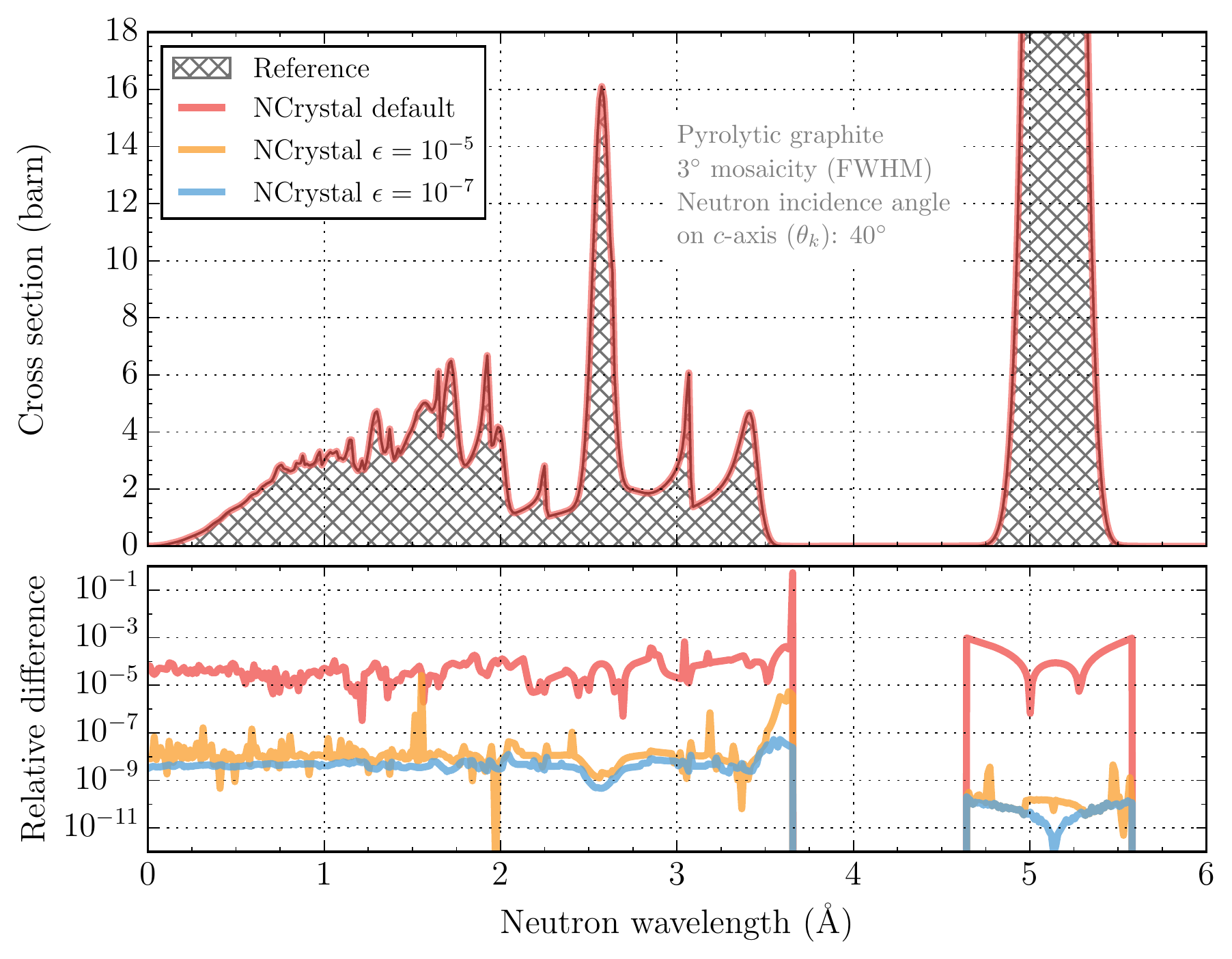}
  \caption{The scattering cross section in pyrolytic graphite as a function of wavelength for
  a FWHM mosaicity of $3^\circ$ and a neutron incidence angle of
  $\thetak=40^\circ$. The reference curve was generated using the model
  described in \refsec{lcbragg:altrefimpl} with luxurious settings: a mosaic
  precision of $\epsilon=\num{e-7}$ (via the \texttt{mosprec} parameter) and
  \num{e7} evenly distributed angular rotations (via the \texttt{lcmode}
  parameter). In order to remove trivial differences by design due to the
  $\epsilon$-dependency of the relative mosaic truncation angle
  (cf.\ \reffig{mosprecntrunc}), a fixed truncation angle of $\tau=5\sigma$ was
  used for all curves. \texttt{NCrystal} results are shown for both default
  ($\epsilon=\num{e-3}$) and increased ($\epsilon=\num{e-5}$ and $\epsilon=\num{e-7}$)
  precision settings. The lower plot shows the relative difference between
  the \texttt{NCrystal} results and the reference curve, which can be taken as a
  measure of their precision.}
  \labfig{lcb_xsvsref_thetak40_mos3deg}
\end{figure}
shows the scattering cross section
as a function of wavelength for a large FWHM mosaicity of $3^\circ$, and a
neutron incidence angle of $\thetak=40^\circ$. As is clear from the
relative difference curves, the results are in good agreement with the
reference, at a level which is everywhere equal to or exceeding the
requested precision. The only exception is a degradation for the
$\epsilon=\num{e-3}$ curve near the edges of the truncated mosaicity --- in
particular around 3.7\AA. As was the case in \refsec{scbragg::val::mpmathcmp},
this degradation is due to artefacts related to the usage of cubic splines.
Also in this case it is deemed to be an acceptable price to pay for the
increased computational efficiency, given that it happens only in regions where
the cross section is almost negligible anyway.

Next, \reffig{lcb_xsvsref_thetak40_mos0d01deg}
\begin{figure}
  \centering
  \includegraphics[width=0.85\textwidth]{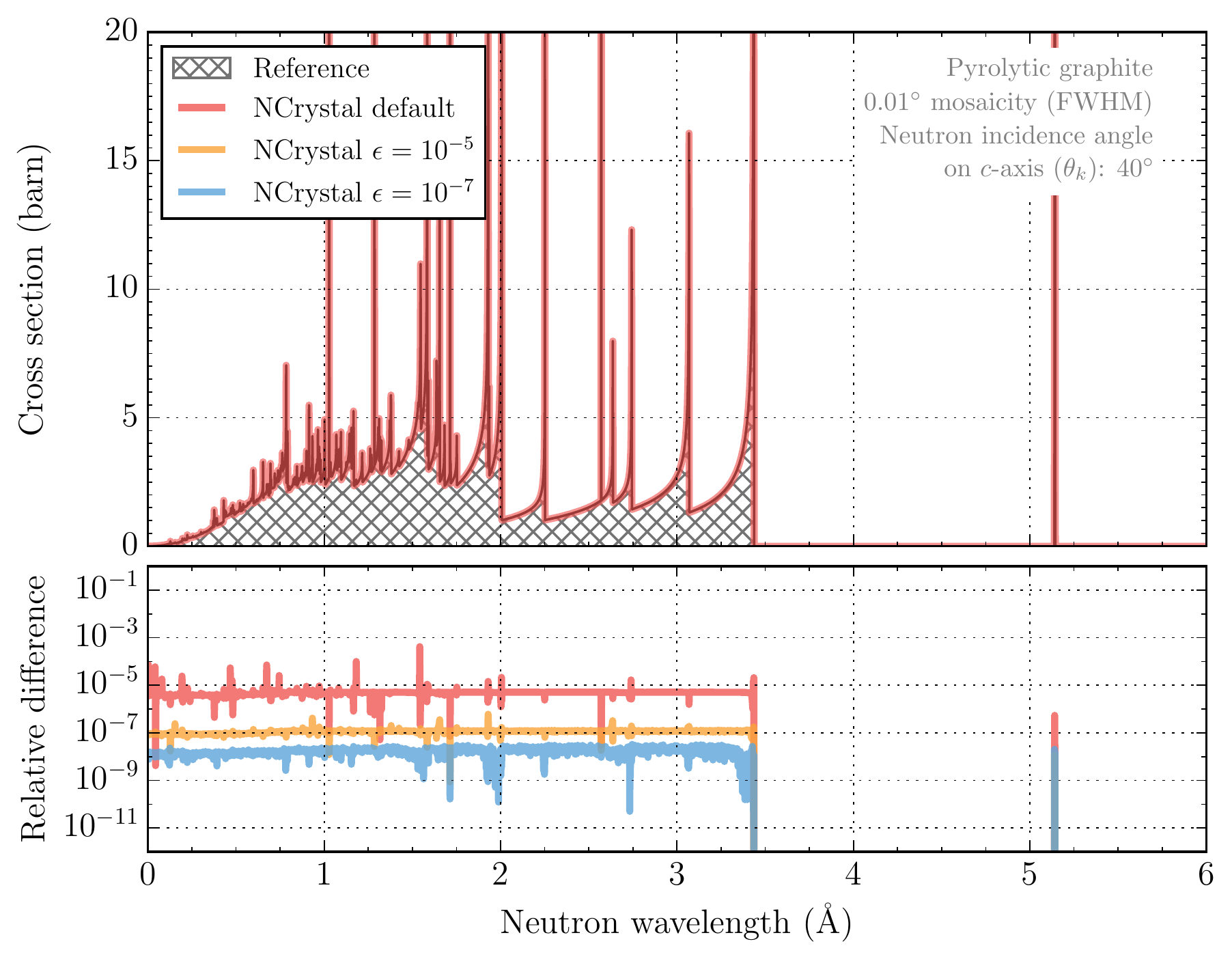}
  \caption{Same curves as in \reffig{lcb_xsvsref_thetak40_mos3deg}, but for a smaller FWHM mosaicity of $0.01^\circ$.}
  \labfig{lcb_xsvsref_thetak40_mos0d01deg}
\end{figure}
shows the same curves, but this
time for a much smaller FWHM mosaic spread of $0.01^\circ$. As expected, the
structure in the curves is more sharply defined at the reduced mosaicity, and
the level of precision in the curves are still at an acceptable level. The
degradation in precision near the edges seems less significant than
in \reffig{lcb_xsvsref_thetak40_mos3deg}, but this is likely just a result of
the finite granularity of the chosen set of wavelength
points. Finally, \reffig{lcb_xsvsref_thetak70_mos0d1deg}
\begin{figure}
  \centering
  \includegraphics[width=0.85\textwidth]{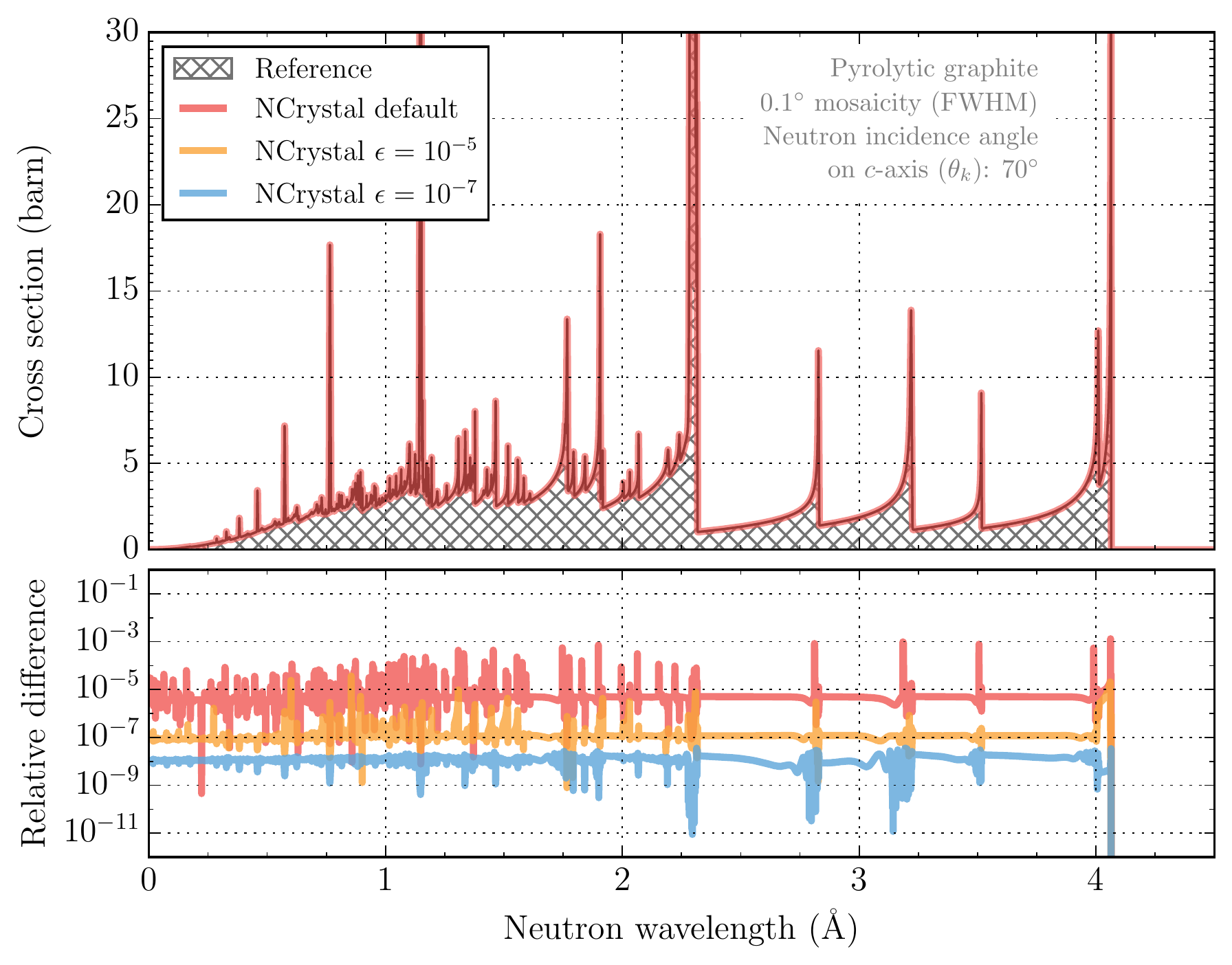}
  \caption{Same curves as in \reffig{lcb_xsvsref_thetak40_mos3deg}, but for a
  neutron incidence of $\thetak=70^\circ$ and a smaller FWHM mosaicity of $0.1^\circ$.}
  \labfig{lcb_xsvsref_thetak70_mos0d1deg}
\end{figure}
shows a similar curve
for an intermediate FWHM mosaic spread of $0.1^\circ$, and a neutron incidence
of $\thetak=70^\circ$. The precision is still observed to be at an acceptable
level, although seemingly with larger fluctuations. Again, this is believed to
be an effect of the cubic spline artefacts, with more clearly defined edges
being crossed when the mosaicity is lower --- yet for a mosaicity of $0.1^\circ$,
the structure is not too fine-grained to be visible on the chosen set of
wavelength points. For reference, comparisons of cross section curves for more
parameters are available in the appendix
in \reffigrange{lcb_xsvsref_appendix_first}{lcb_xsvsref_appendix_last}.

In order to evaluate the scattering event sampling code described
in \refsec{lcbragg::sampling}, both the default \texttt{NCrystal} model and the
reference model (with \num{e4} evenly distributed angular rotations) are used to
sample scattering angles in pyrolytic graphite with a FWHM mosaic spread of
$1^\circ$, for neutron wavelengths uniformly distributed between 2.5{\AA} and
3.5{\AA}, and a fixed neutron incidence of $\thetak=65^\circ$. As can be
inferred from \reffig{lcbraggxspg}, this setup exercises all parts of the
sampling procedure, as it involves scattering on multiple reflection planes, with both
$\thetan=0$, $0<\thetan<\pi/2$, and $\pi/2<\thetan<\pi$ (corresponding
qualitatively
to \refthreefigs{lcbragg_singlenormal_thetan0and90_0}{lcbragg_singlenormal_thetan60_decomposed_noanti}{lcbragg_singlenormal_thetan60_decomposed_onlyanti}
respectively),
and involves reflections both with and without back-scattering. The resulting
distribution of scattered directions by the default \texttt{NCrystal} model is
shown in \reffig{lcbraggscatpattern},
\begin{figure}
  \centering
  \includegraphics[width=0.99\textwidth]{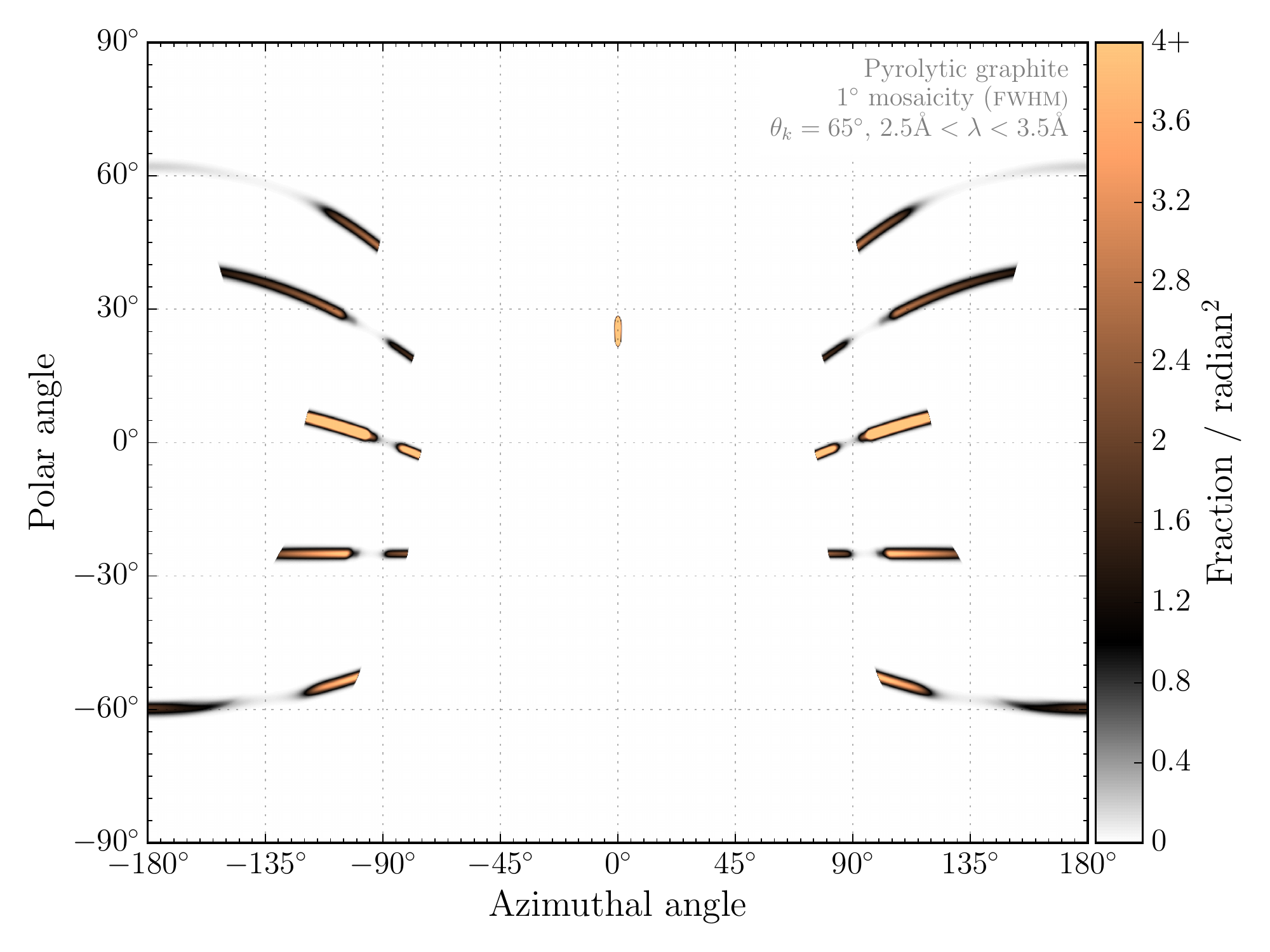}
  \caption{Final direction of \num{e9} neutrons scattered once in pyrolytic
  graphite via Bragg diffraction by \texttt{NCrystal}. Incoming neutron have wavelengths uniformly
  distributed in $[\SI{2.5}{\angstrom},\SI{3.5}{\angstrom}]$ and direction
  $\hatki=(\sin65^\circ,0,\cos65^\circ)$. The resulting distribution of $\hatkf$
  is shown in standard spherical coordinates.}
  \labfig{lcbraggscatpattern}
\end{figure}
which was confirmed visually to be indistinguishable from the same distribution created by the reference model. For
a more quantitative
comparison, \reftwofigures{lcbraggscatpatproj_polar}{lcbraggscatpatproj_azimuthal}
\begin{figure}
  \centering
  \includegraphics[width=0.8\textwidth]{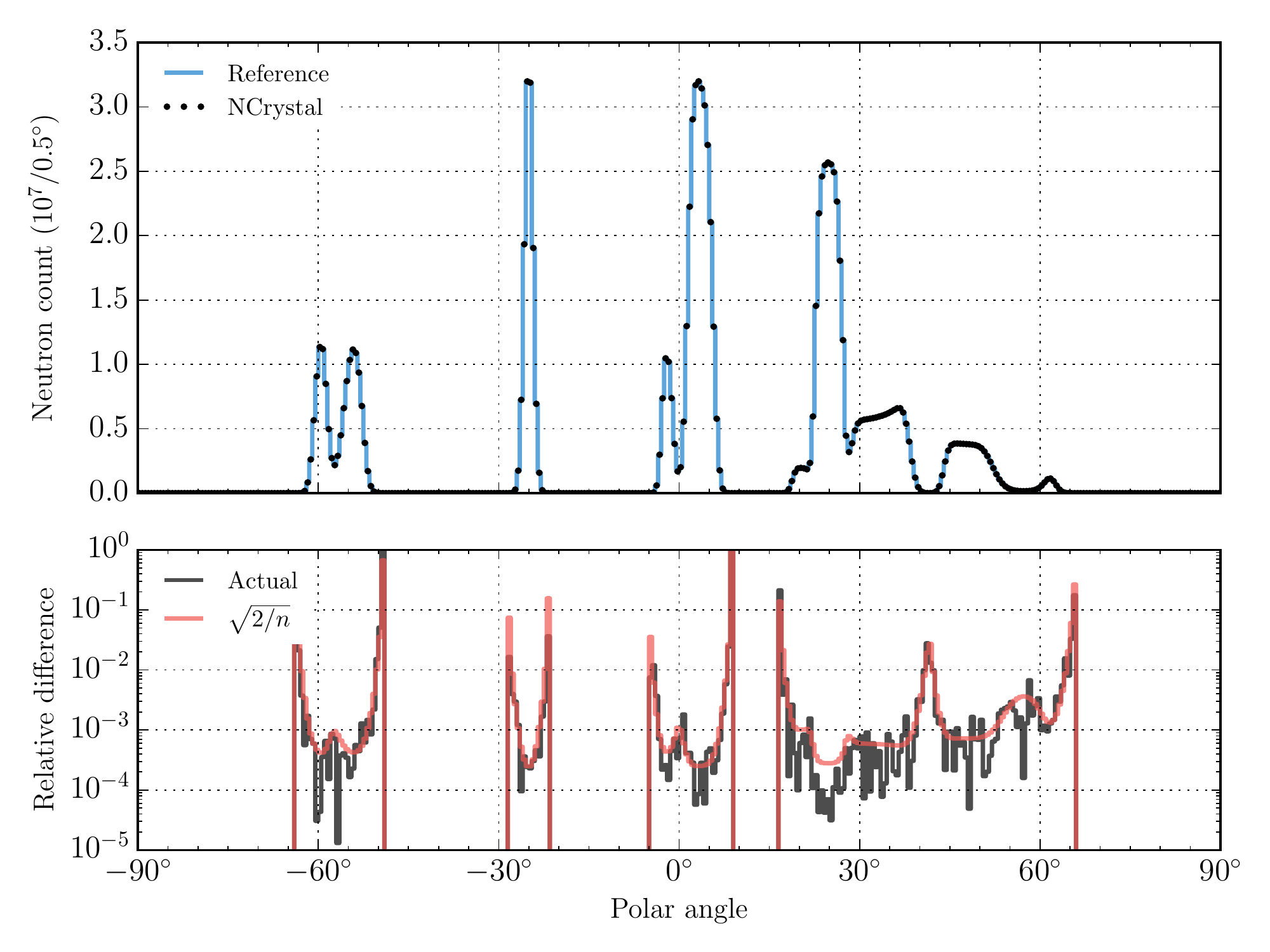}
  \caption{Projection of \reffig{lcbraggscatpattern} showing distribution of
  polar angles, for both the default \texttt{NCrystal} model and the reference
  model. The lower plot indicates the relative level of discrepancy between the
  two models, as well as the expected average level of discrepancy given by
  $\sqrt{2/n}$, where $n$ is the number of neutrons collected in the given bin
  (averaged between the two models).}
  \labfig{lcbraggscatpatproj_polar}
\end{figure}
\begin{figure}
  \centering
  \includegraphics[width=0.8\textwidth]{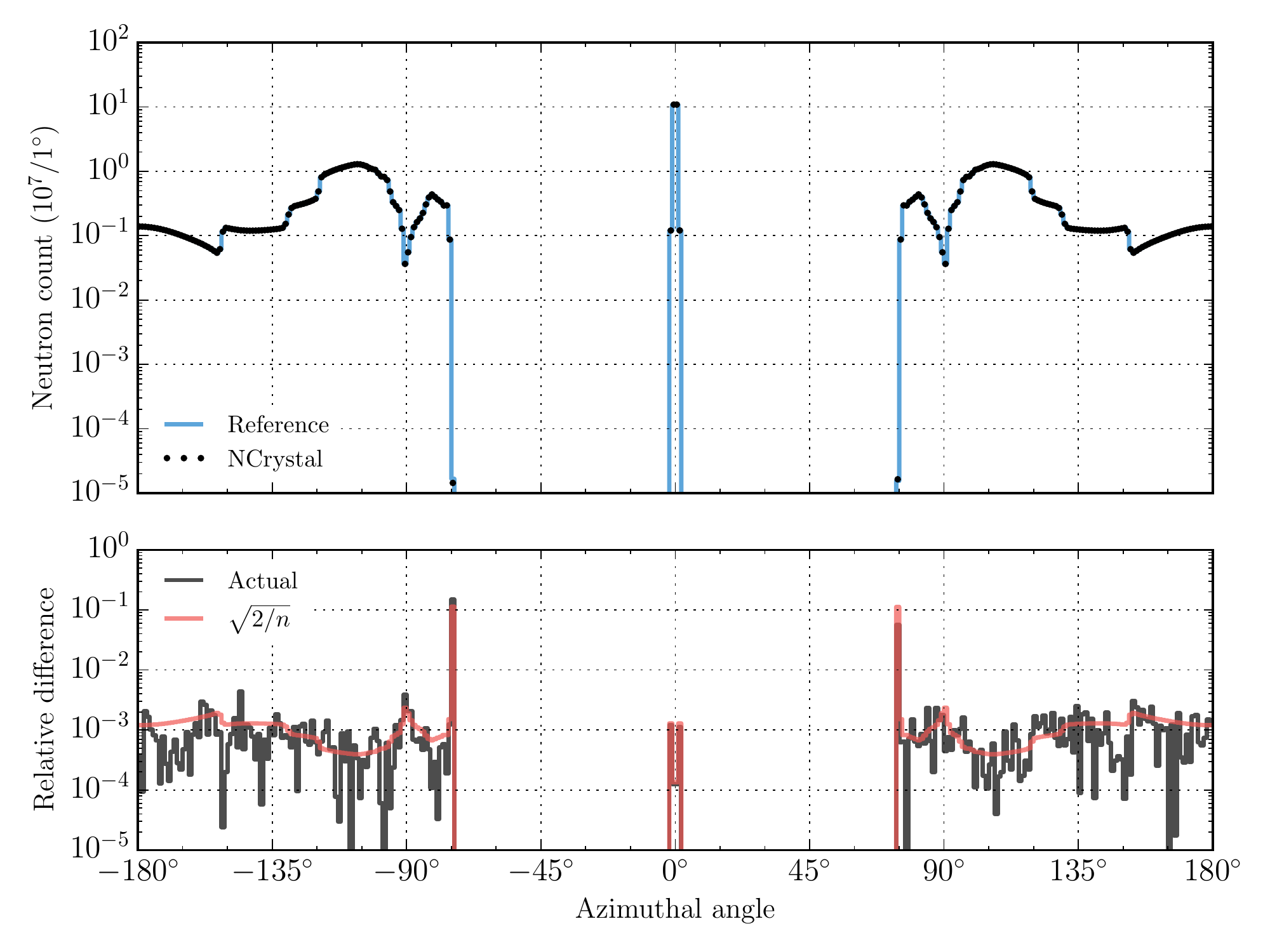}
  \caption{As in \reffig{lcbraggscatpatproj_polar}, but showing the projected
  distribution of azimuthal instead of polar angles.}
  \labfig{lcbraggscatpatproj_azimuthal}
\end{figure}
compares the distribution  of polar and azimuthal scattering angles
respectively. Deviations between the two models are shown to be at a level
consistent with purely statistic fluctuations.

\subsubsection{General consistency checks}\labsec{lcbragg::val::consistencychecks}

If implemented consistently, the layered crystal model should be able to satisfy
the same consistency checks as those carried out for the non-layered single
crystal model in \refsec{scbragg::val::consistencychecks}. Firstly, the ability
to reproduce a ''zig-zag'' walk was verified for a range of scenarios up to
\num{e10} steps in \texttt{NCrystal} simulations. Secondly, \reffig{powderlcbragg}
\begin{figure}
  \centering
  \subfloat[\labfig{powderlcbragg_a}]{\includegraphics[width=0.7\textwidth]{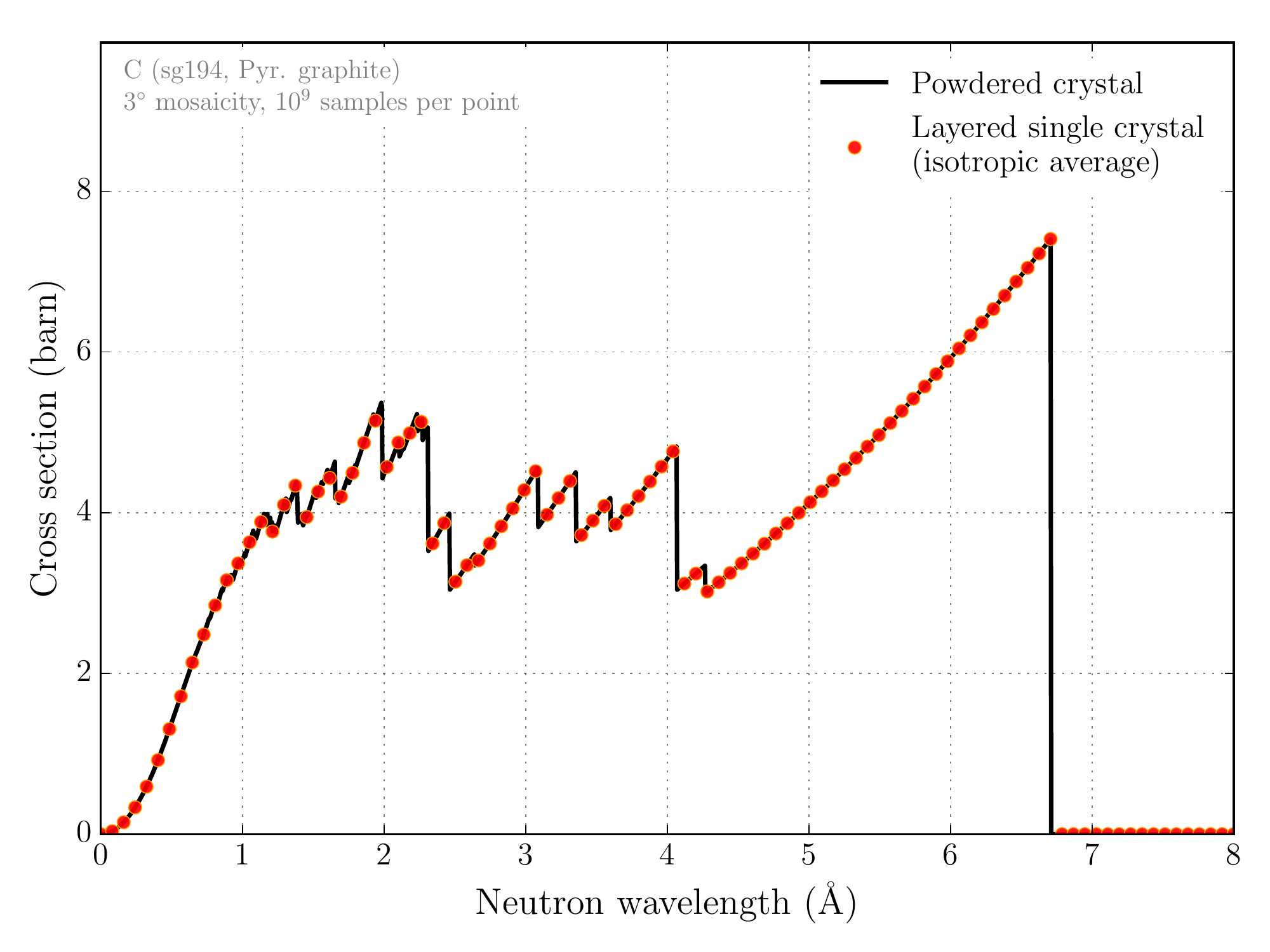}}\\
  \subfloat[\labfig{powderlcbragg_b}]{\hspace*{-0.031\textwidth}\includegraphics[width=0.733\textwidth]{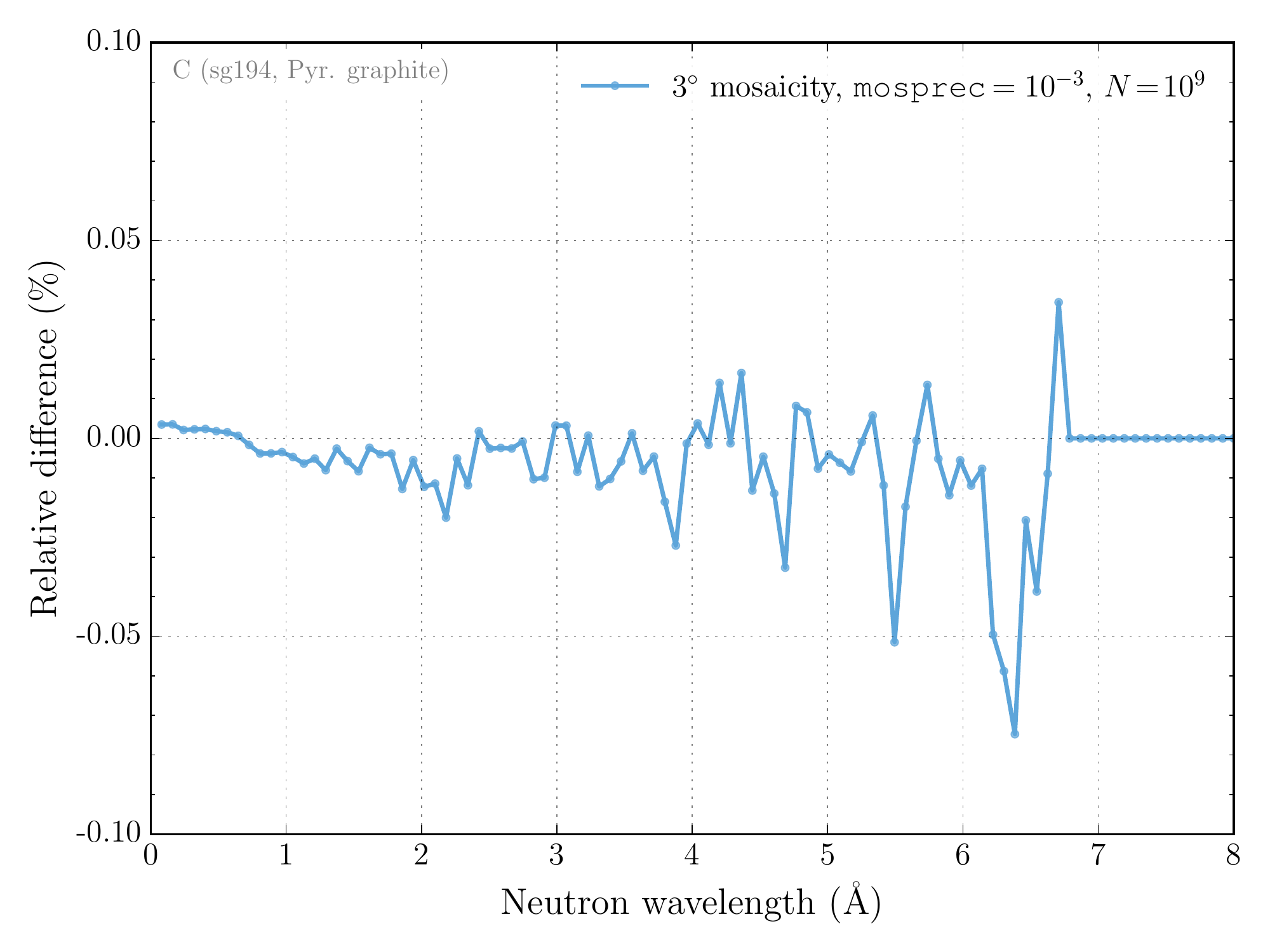}}
  \caption{\Refsubfigfrommaincaption{powderlcbragg_a} shows layered single
  crystal Bragg diffraction cross sections in pyrolytic graphite provided
  by \texttt{NCrystal}, averaged over an isotropic sample of \num{e9} neutrons
  at each wavelength point. For reference, the cross section curve of the same
  crystal as a powder is also shown. \Refsubfigfrommaincaption{powderlcbragg_b} shows
  the relative difference between the isotropically averaged layered single crystal
  numbers, and the equivalent powder cross sections.}
  \labfig{powderlcbragg}
\end{figure}
shows how submitting a layered crystal model of pyrolytic graphite to an
isotropic illumination with neutrons reproduces the same average
wavelength-dependent scattering cross section as found in a crystal powder. For
reasons of computational resources, this was only done for a single large FWHM
mosaicity of $3^\circ$. Despite the large statistical fluctuations at longer
wavelengths (where only few reflection planes contribute), it is still possible
to conclude that the discrepancies in \reffig{powderlcbragg_b} are well below
the requested precision of $\epsilon=\num{e-3}$ at all wavelengths.

\subsubsection{Comparison with existing results}\labsec{lcbragg::val::frikkee}

Unlike the situation concerning non-layered single crystals
(cf.\ \refsec{scbragg::val::cmpothers}), no existing code or analytical models
available to the authors provides reliable and precise results for neutron
scattering on layered crystals like pyrolytic graphite. However,
E.~Frikkee~\cite{frikkee1975} provides analytical predictions of the cross section
structure, parameterising the peak positions visible
in \reffig{lcbraggxspg}. \Reffig{lcbraggxspgwoverlays}
\begin{figure}
  \centering
  \includegraphics[width=0.99\textwidth]{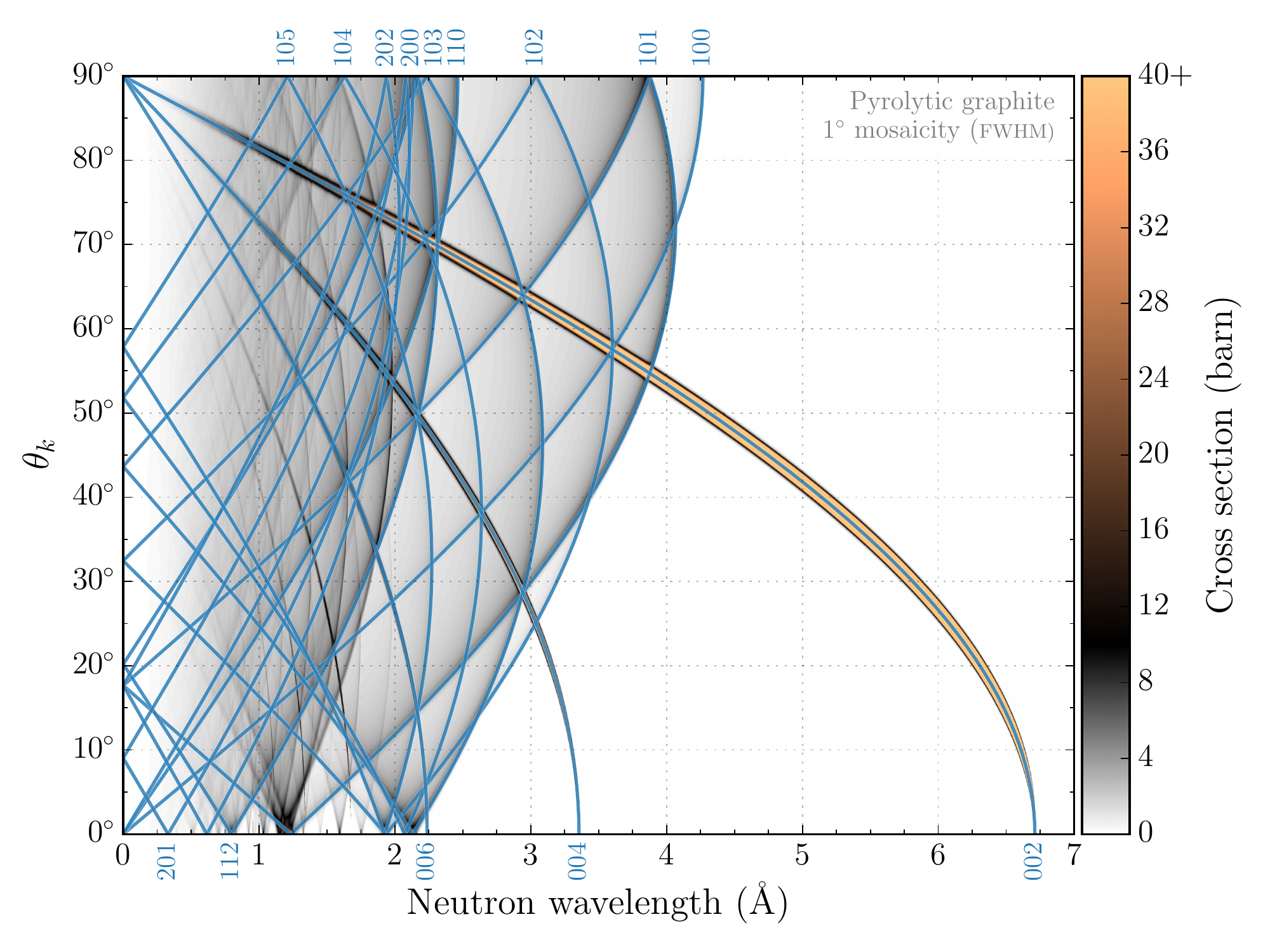}
  \caption{Like \reffig{lcbraggxspg}, but with structures
  predicted by E.~Frikkee~\cite{frikkee1975} indicated in blue for all reflection planes with
  $d_{hkl}>\SI{1}{\angstrom}$.}
  \labfig{lcbraggxspgwoverlays}
\end{figure}
shows the result of
overlaying these curves on the predictions of \texttt{NCrystal}. Excellent agreement is observed.

Additionally, E.~Frikkee also measured a transmission spectrum for a setup in which
monochromatic neutrons impinged upon a \SI{16}{\milli\meter} thick slab of
pyrolytic graphite under various incidence angles. The pyrolytic graphite in
the slab had a $c$-axis orthogonal to the slab surface, and a FWHM mosaic spread
of $1.4^\circ$. Apart from linking qualitative features of the observed
transmission spectrum to scattering on specific $hkl$ reflection planes, E.~Frikkee
was not able to perform a more quantitative analysis of the data. However,
with \texttt{NCrystal} it is now straightforward to reproduce the setup in a
dedicated simulation. The exercise benefits from the fact that E.~Frikkee corrected
the observed transmission spectrum for contaminations due to higher-order
$\lambda/n$ reflections in the upstream monochromator, and had collimators
placed before the slab and between the slab and detectors. The result is shown
in \reffig{lcbragg_pgtransm},
\begin{figure}
  \centering \includegraphics[width=0.99\textwidth]{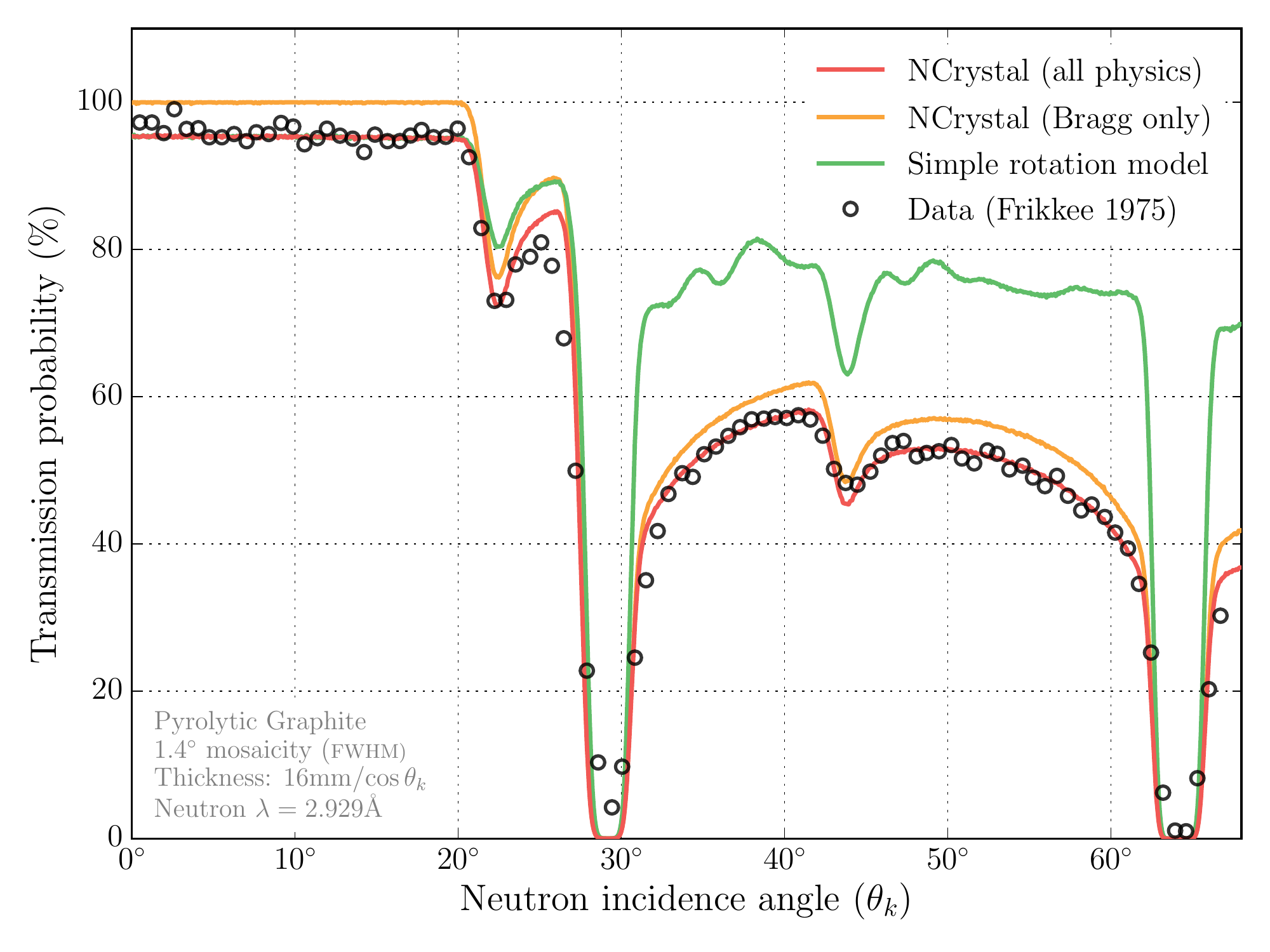}
  \caption{Transmission probabilities for neutrons of
  wavelength \SI{2.929}{\angstrom} impinging on a \SI{16}{\milli\meter} thick
  slab of pyrolytic graphite under various incidence angles. The lattice
  $c$-axis is orthogonal to the slab surface, and the FWHM mosaicity is
  $1.4^\circ$. Measured data points by E.~Frikkee~\cite{frikkee1975}
  are compared to predictions from \texttt{NCrystal} simulations, including for
  reference two non-standard configurations: one with a simpler rotational model
  (\texttt{lcmode=-1}) and one where all physics except Bragg diffraction is disabled.}
  \labfig{lcbragg_pgtransm}
\end{figure}
with a remarkable agreement between the data
points and the predictions of \texttt{NCrystal}. It is worth noting that, as can
be inferred from \reffig{lcbraggxspgwoverlays}, the transmission curve is
expected to be affected significantly by 6 different groups of reflection
planes: $101$, $102$, $103$, $100$, $004$, and $006$. A set which includes
normals whose directions relative to the crystal's $c$-axis are both parallel,
orthogonal, and neither.

For reference, \reffig{lcbragg_pgtransm} also shows the results of disabling
non-Bragg diffraction physics (absorption, and inelastic/incoherent scattering)
in the simulations, proving the necessity of including these components as well
if the most precise predictions are desired. Finally, a ``simple rotation
model'' was also included, based on setting \texttt{lcmode=-1} in
the \texttt{NCrystal} configuration.  As was noted
in \refsec{lcbragg:altrefimpl}, this in principle corresponds to how pyrolytic
graphite is modelled in the \texttt{Single\_crystal} component
of \texttt{McStas}. Unfortunately, but not surprisingly, this model fails to
reproduce the data with any kind of realism --- except of course for the region
below $20^\circ$ where Bragg diffraction is not possible, and the regions around
$29^\circ$ and $63^\circ$ where the physics is respectively dominated by scattering on the
$\thetan=0$ planes $004$ and $006$.

\section{Incoherent elastic scattering}\labsec{incel}

As discussed in more details in~\cite[Sec.~2.3]{ncrystal2019}, the per-atomic
incoherent elastic cross section for scattering a neutron state with wavevector
$\vecki$ into the final state with a wavevector $\veckf$ (where $|\veckf|=|\vecki|\equiv{k}$),
is under the harmonic approximation given by:
\begin{align}
 \frac{d\sigma^{\text{inc,el}}_{\vecki\rightarrow\veckf}}{d\Omega_f}=
\frac{1}{N}\sum_{j=1}^N
\frac{\sigma_j^{\text{inc}}}{4\pi}
{e}^{-2W_j(\vec{Q})}
\labeqn{incelxsdifferential}
\end{align}
Where the summation index $j$ runs over all atomic positions in the unit cell,
the constant $\sigma_j^{\text{inc}}$ is the incoherent scattering cross
section of the atom in question, $\vec{Q}\equiv\veckf-\vecki$ is the momentum
transfer, and $W_j(\vec{Q})$ the Debye-Waller function. True to the nature of
incoherent scattering, \refeqn{incelxsdifferential} shows no interference terms
between atoms at different positions. For isotropic and harmonic
displacements, the Debye-Waller function is given by $\thalf Q^2\delta^2_j$, where $\delta^2_j$ is the
mean-squared displacement of the $j$th atom from its nominal position in
the crystal, as seen over a large ensemble of unit cells. All in all, the contribution of
atoms occupying the $j$th position in the unit cell, is simply given as:
\begin{align}
   \left\{\frac{d\sigma^{\text{inc,el}}_{\vecki\rightarrow\veckf}}{d\Omega_f}\right\}_j=\frac{\sigma_j^{\text{inc}}}{4\pi}{e}^{-Q^2\delta^2_j}
   \labeqn{incelxsdifferential_atomjmsd}
\end{align}
To get the cross section at a given neutron wavelength, $\lambda=2\pi/k$, one
must integrate \refeqn{incelxsdifferential_atomjmsd} over all outgoing directions of
$\veckf$.
Designating the scattering angle as $\theta$, and defining $\mu=\cos\theta$, one finds:
\begin{align}
Q^2 = (\veckf-\vecki)^2 = k_i^2+k_f^2-2\vecki\cdot\veckf = 2k^2(1-\mu).
\end{align}
Now, $d\Omega=\sin{\theta}d{\theta}d{\varphi}=d{\mu}d\varphi$ and integration
over $d\varphi$ merely yields a factor of $2\pi$, so:
\begin{align}
\sigma^{\text{inc,el}}(k) &= \frac{\sigma_\text{inc}}{4\pi}2\pi\int_{-1}^{1} \exp(-2k^2\delta^2(1-\mu)) d\mu\nonumber\\  %
       &= \frac{\sigma_\text{inc}}{4k^2\delta^2}  ( 1 - \exp(-4k^2\delta^2) )\nonumber\\
       &= \sigma_\text{inc} \frac{ 1-\exp(-t)}{t}\labeqn{incelxsintegrated}
\end{align}
Where the parameter $t$, was introduced:
\begin{align}
   t\equiv(2k\delta)^2=\left(\frac{4\pi\delta}{\lambda}\right)^2
\end{align}
For reasons of numerical efficiency and stability, \refeqn{incelxsintegrated} is
in \texttt{NCrystal} evaluated with a third-order Taylor expansion when
$t<0.01$, and with the limiting expression $\sigma_\text{inc}/t$ when $t>24$.
In order to sample a value of $\mu$ for a particular scattering, a probability density proportional to the integrand
in \refeqn{incelxsintegrated} must be used. This implies that a $\mu$ value must be sampled in $[-1,1]$
according to the distribution:
\begin{align}
  P(\mu)= N_t\exp\left(\frac{t\mu}{2}\right)
\labeqn{incelmudist}
\end{align}
With the normalisation factor $N_t=t/(4\sinh(t/2))$. For reasons of numerical
stability and computational efficiency, when $t<0.02$ the
implementation in \texttt{NCrystal} samples $\mu$ via straightforward
acceptance-rejection sampling using a constant overlay function
and a Taylor expansion for evaluating \refeqn{incelmudist}.  For larger values
of $t$ the transformation method is used instead, yielding
$\mu=-1+2t^{-1}\log(1+(e^t-1)R)$ where $R$ is a pseudo-random number distributed
uniformly over the unit interval. As is the case for the implementation of Bragg diffraction in a crystal powder
discussed in \refsec{pcbragg}, the incoherent elastic model is also fast enough
that there is potentially a non-neglible overhead from the construction of complete directional vectors in \texttt{NCrystal}'s vector interface for
scattering event sampling. Consequently, this interface is implemented using the same
efficient methods for vector construction as those discussed in \refsec{pcbragg}.

Coming back to the significance of the parameter $t$: when
$\lambda\gg4\pi\delta$, $t$ approaches 0 and incoherent elastic scattering
becomes isotropic with a wavelength-independent cross section --- which is indeed a widely used
approximation used to model incoherent elastic scattering. The
approximation is, however, worse for materials with large atomic displacements
which could for instance be a result of high material temperature, and it will
always eventually break down when the wavelength of the incoming neutron is
small enough. \Reffig{incelscatangledists}
\begin{figure}
  \centering
  \includegraphics[width=0.75\textwidth]{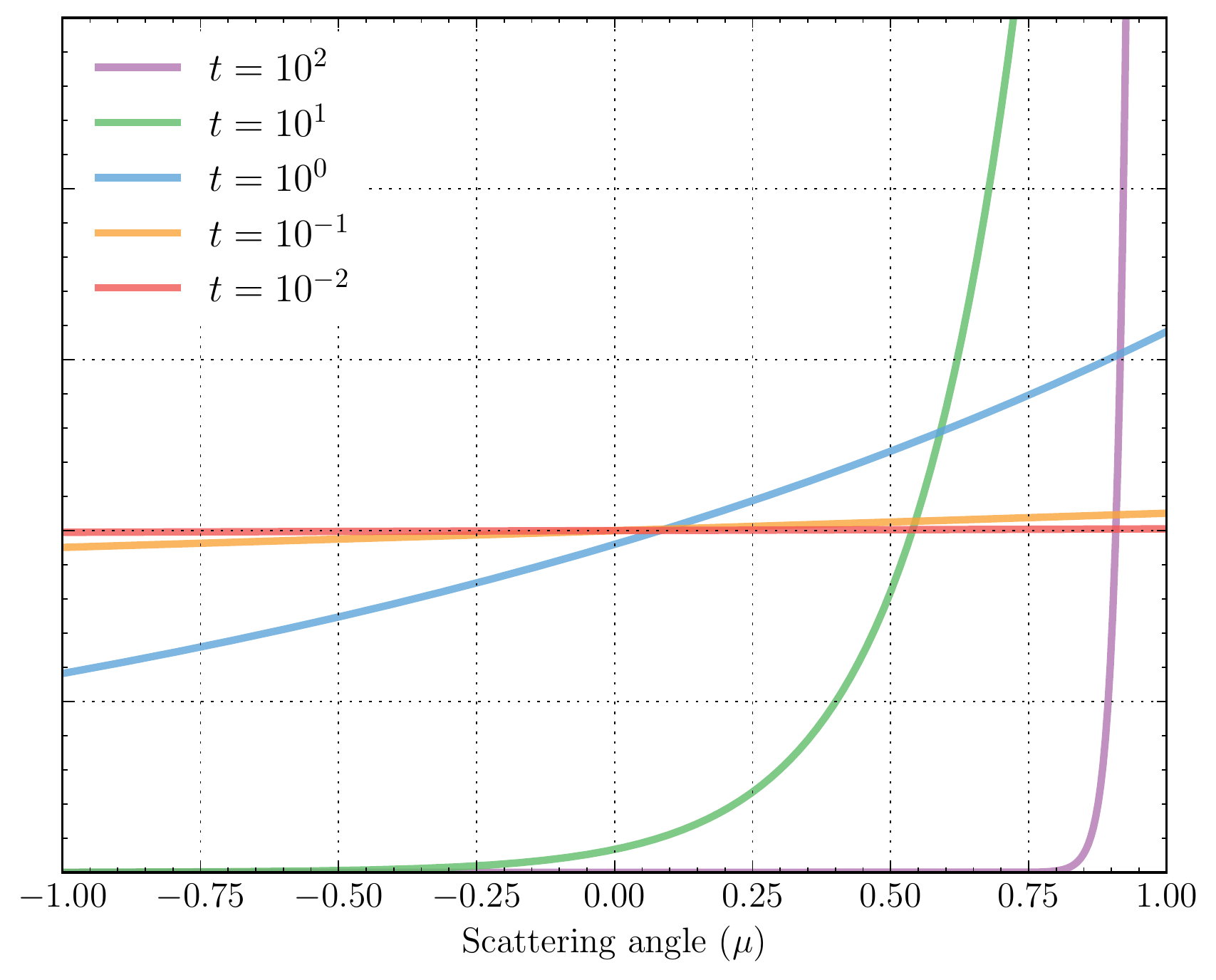}
  \caption{Scattering angle distributions in incoherent elastic interactions for
  various values of $t$. The distributed variable is $\mu=\cos\theta$, so isotropic
  scattering corresponds to a flat distribution in $[-1,1]$.}
  \labfig{incelscatangledists}
\end{figure}
shows scattering angle
distributions in incoherent elastic interactions for various values of $t$: while
the distributions for $t=0.01$ and even $t=0.1$ can be said to be essentially isotropic,
it is clear that forward scattering becomes favoured as $t$ increases beyond
this.  To put this into context, \reffig{inceltfactors}
\begin{figure}
  \centering
  \includegraphics[width=0.9\textwidth]{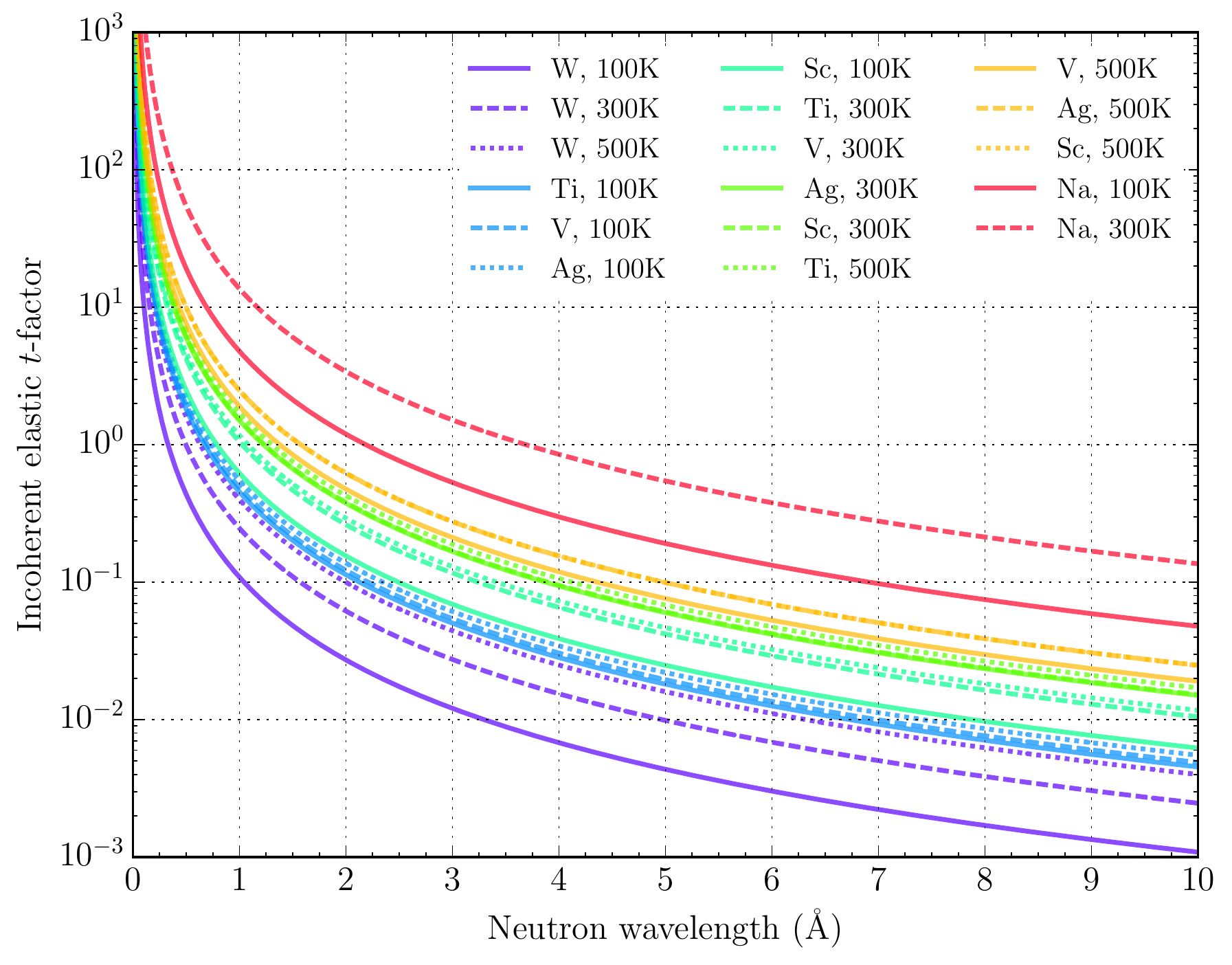}
  \caption{Values of $t=(4\pi\delta/\lambda)^2$ for various materials and temperatures.}
  \labfig{inceltfactors}
\end{figure}
shows $t$-factor values
for different material temperatures in select materials with significant
incoherent elastic cross sections. With the exception of sodium, which has
unusually large atomic displacements, $t<0.01$ is achieved for all investigated
materials when the neutron wavelength is longer than \SI{5}{\angstrom}. However,
for neutrons thermalised at room temperature,
$\lambda=\SI{1.8}{\angstrom}$, this is no longer the case for any of the
investigated materials except tungsten, and it is indeed
prudent to use the more accurate
formulas \reftwoeqns{incelxsintegrated}{incelmudist} to describe the
interactions.

Concerning validation, the modelling based on the
formulas \reftwoeqns{incelxsintegrated}{incelmudist} is relatively trivial. It
was of course verified by comparison with \texttt{mpmath} that the resulting
cross sections and scattering angle distributions are implemented without
technical issues such as numerical
instabilities. Additionally, \reffig{vanadiumnickel}
\begin{figure}
  \centering
  \subfloat[\labfig{vanadiumnickel_v}]{\includegraphics[width=0.8\textwidth]{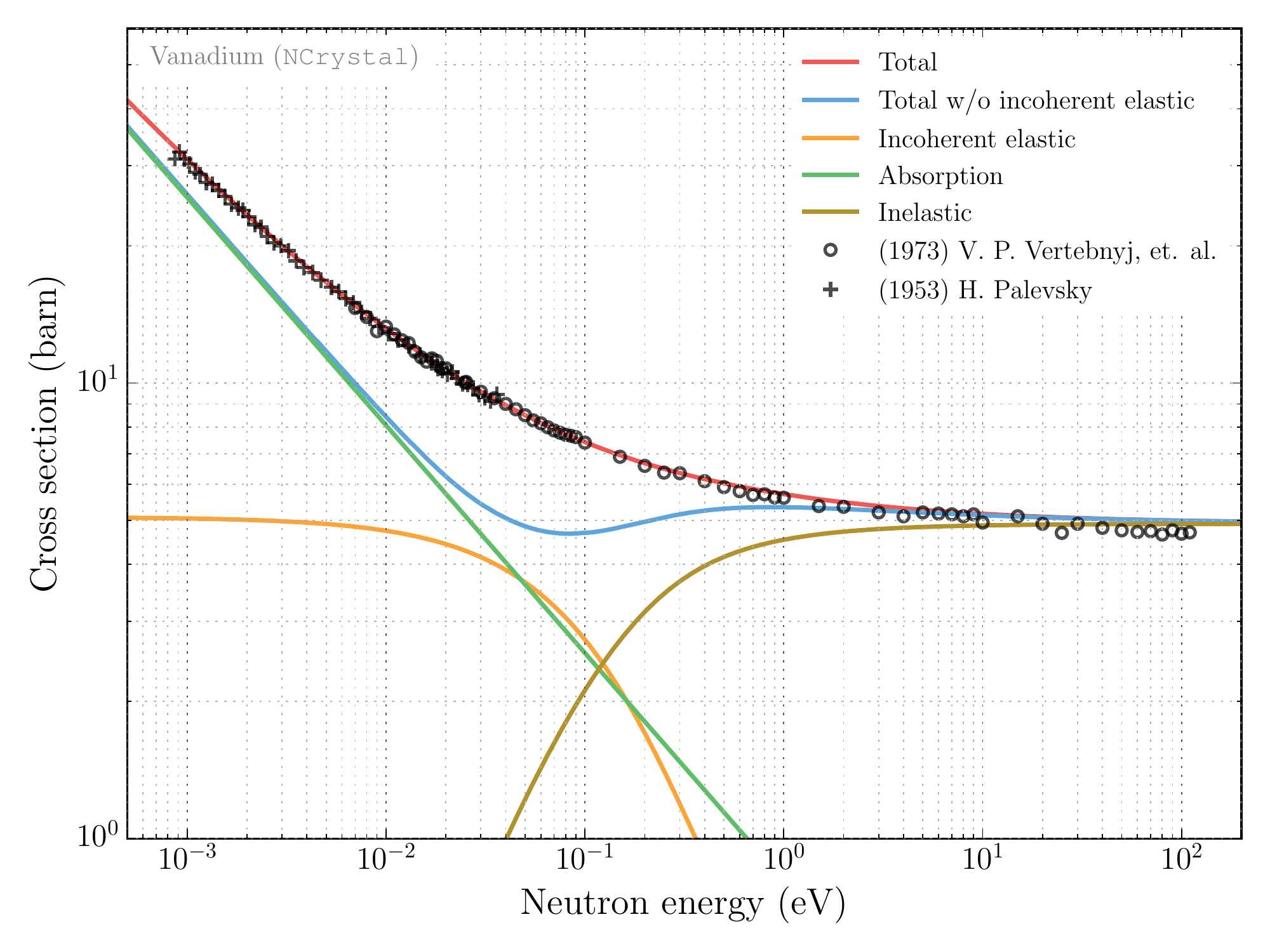}}\\
  \subfloat[\labfig{vanadiumnickel_ni}]{\includegraphics[width=0.8\textwidth]{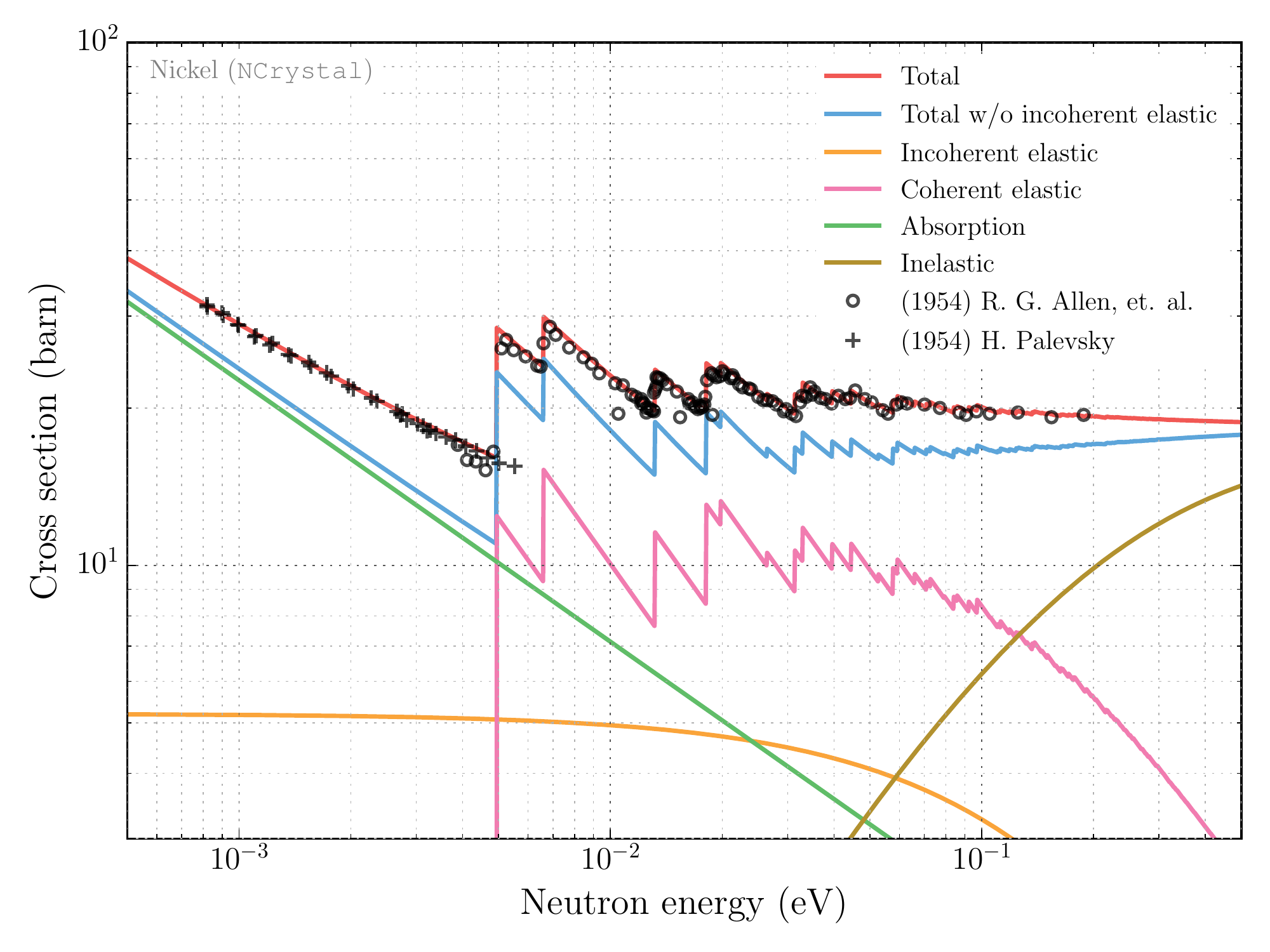}}
  \caption{Comparison of total cross sections predicted
  by \texttt{NCrystal} and experimental data
  for vanadium (\refsubfigfrommaincaption{vanadiumnickel_v})
  and nickel (\refsubfigfrommaincaption{vanadiumnickel_ni}). For the case of
  vanadium the coherent elastic contribution is included in the Total, but not
  shown explicitly as it is
  very low --- with a maximal contribution of \SI{0.03}{barn}
  at \SI{4.5e-3}{\electronvolt}. Measurements were obtained from \texttt{EXFOR}~\cite{exfor2014}.}
  \labfig{vanadiumnickel}
\end{figure}
shows the importance of incoherent elastic scattering for Vanadium and Nickel,
and illustrates how the model as implemented in \texttt{NCrystal} is essential
for reproducing the measured total cross sections. Due to the general low
interest in incoherent elastic scattering in itself, no suitable measurements
were found with which a corresponding comparison of scattering angles could be
performed. In particular, the search for scattering angle data sensitive to
deviations from isotropicity with vanadium samples is complicated by the fact that
many neutron scattering instruments are calibrated with vanadium samples.

\section{Computational efficiency}\labsec{timing}

The computational cost of using \texttt{NCrystal} to provide information about
neutron scatterings is of course highly dependent on the use case: specific
materials, distribution of neutrons, and configuration options can all influence
this greatly. In order to nonetheless gauge this, the present discussion will
use a benchmark in which an isotropic source of neutrons at a given wavelength
interacts in a material which is either a powder, a single crystal, or a layered
single crystal. The direction-dependent interfaces are used in all cases, as
this is what is typically used when \texttt{NCrystal} is embedded in a generic
simulation framework like \texttt{Geant4} or \texttt{McStas}. To enable more
meaningful comparisons, the crystal structure of the material will in all cases
be that of pyrolytic graphite. This is important, since the comparisons would
otherwise be influenced by material-specific details such as the number of
reflection planes and their $d$-spacings (see for
instance~\cite[Fig.~4]{ncrystal2019}). The timings were in all cases carried out
using an otherwise unoccupied node at the ESS-DMSC computing cluster in
Copenhagen on a \SI{2.40}{\giga\hertz} Intel\textsuperscript{\textregistered}
Xeon\textsuperscript{\textregistered} Processor (E5-2680~v4).

\Reffig{timings}
\begin{figure}
  \centering \includegraphics[width=0.99\textwidth]{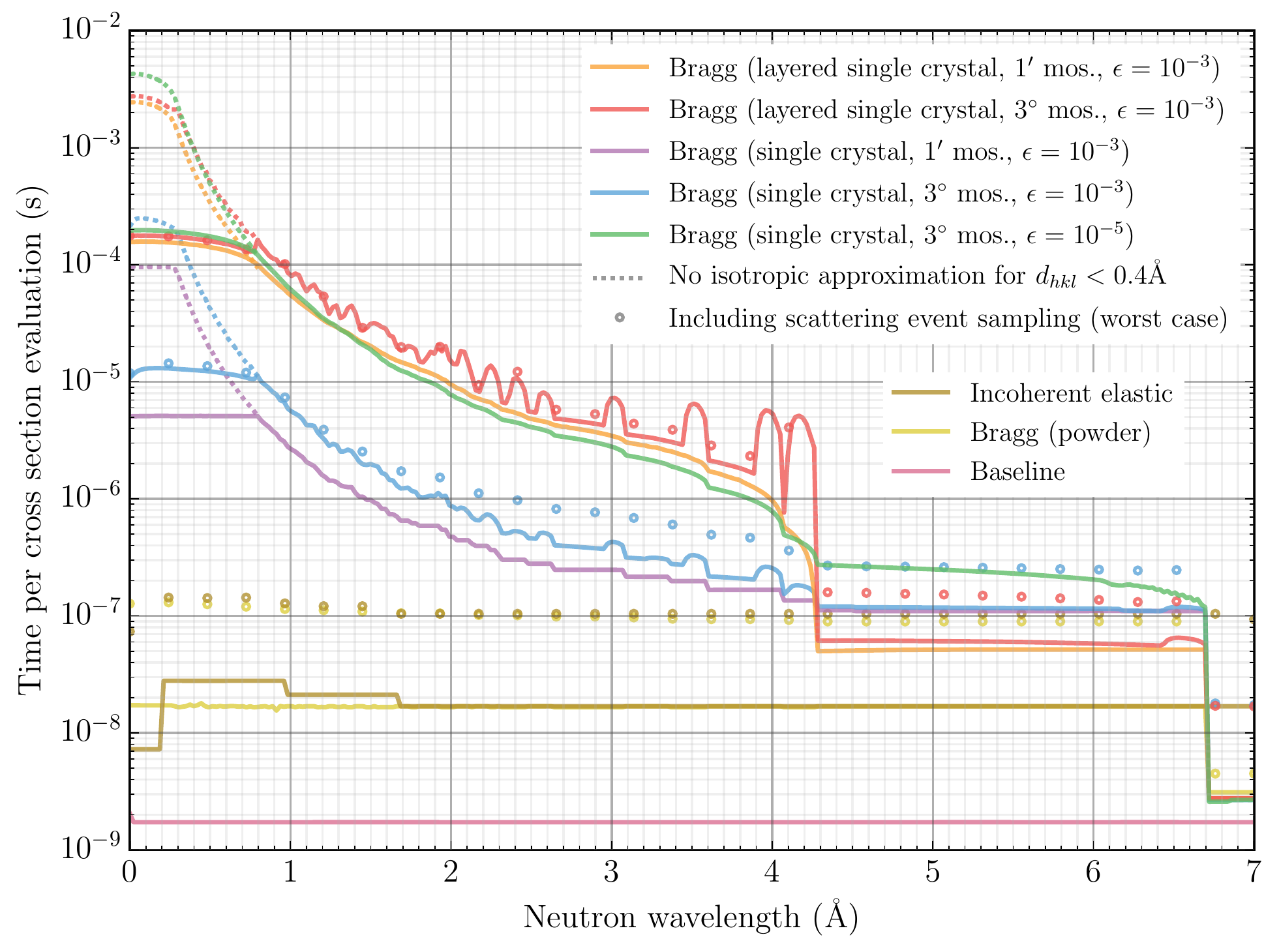}
  \caption{Average computational speed for isotropically distributed neutrons in
  elastic models included with \texttt{NCrystal}, for various parameters as
  indicated. The crystal structure of the material is in all cases that of
  pyrolytic graphite, but with different assumptions of mosaicity distributions
  affecting Bragg diffraction models as indicated. Dashed lines show the effect
  of running with \texttt{sccutoff=0}, and open circles show the effect of
  including exactly one sampling of a scattering event after each cross section
  evaluation.  Where relevant, indicated mosaicities are FWHM values.}
  \labfig{timings}
\end{figure}
shows the result of this timing benchmark, the details of which will be
discussed in the following. The curves show the average time for each cross
section evaluation, while open circles show the average time spent when each
cross section evaluation is followed by exactly one sampling of a scattering
event. This of course represents a worst case estimation of the impact of
scattering event sampling, since in a real simulation the number of cross
section evaluations would often be significantly higher than the number of
scattering events.

Starting from the simplest model and proceeding in the direction of increasing
complexity, the model named \emph{Baseline} is the simplest possible model which just
returns a cross section value of 0 when called. It can provide results at a rate
of \order{\SI{500}{\mega\hertz}}, a rate which primarily represents the overhead
of calling a virtual function in the \texttt{NCrystal} \texttt{C++} interface.

Next, the incoherent elastic model provides cross sections at a fast rate
of \order{\SI{50}{\mega\hertz}}. The small jumps in evaluation times observed at
certain wavelength thresholds are a result of different approximations being used
to evaluate \refeqn{incelxsintegrated} in different energy domains. Incoherent
elastic scattering events can be sampled at a rate
around \order{\SI{10}{\mega\hertz}}, which is seen to be similar to the same rate
for Bragg diffraction in a powder. This is not surprising since, as mentioned
in \refsec{incel}, they employ the same algorithm for finding the directional
vector of the outgoing neutron on the basis of the sampled scattering angle. It
is, however, not inconceivable that a faster algorithm could be adopted for
longer wavelengths where incoherent elastic scattering is almost isotropic.
Given that the incoherent elastic sampling rate is already high, this was not
deemed a high priority so far --- but such an improvement might be
included in a future \texttt{NCrystal} release.

The curve for Bragg diffraction in a powder in \reffig{timings} shows what is
essentially constant performance below the Bragg cutoff of \SI{6.71}{\angstrom},
corresponding to a rate of \order{\SI{50}{\mega\hertz}}. This is not surprising,
given the implementation discussed in \refsec{pcbragg}, whose main overhead is a
binary search in a single contiguous array. As already mentioned, the sampling
speed of \order{\SI{10}{\mega\hertz}} is dominated by the construction of the
directional vector of the outgoing neutron.

Turning next to the single crystal models, layered or not, the situation is more
complicated. Starting at the Bragg cutoff at \SI{6.71}{\angstrom}, more and more
reflection planes are able to satisfy the Bragg condition as the neutron
wavelength decreases. Unlike the case of a powder, each (group of) reflection
planes must be checked separately, leading to a growth all the way down
to \SI{0.8}{\angstrom}, where the isotropic approximation described
in \refsec{scbragg::sccutoff} prevents additional planes from being
considered. The dashed lines in \reffig{timings} show the effect of foregoing
this approximation, leading to a continued growth of evaluation times down
to \SI{0.2}{\angstrom}. The final plateau below \SI{0.2}{\angstrom} is due to
the other threshold in \texttt{NCrystal}, which completely leaves out reflection
planes with $d$-spacing below \SI{0.1}{\angstrom}. It is worth noting that this second threshold
does \emph{not} in general represent a trade-off between realism and
computational efficiency, as it was shown in~\cite[Fig.~6]{ncrystal2019} that
the reflection planes removed by it do not actually contribute significantly to the cross sections.

Next, the differences between the various single crystal curves
in \reffig{timings} can mainly be attributed to two effects. The first of these
is that layered single crystals are about an order of magnitude slower to
process than non-layered single crystals, which is hardly surprising given that
the former must essentially perform a numerical integration of the latter ---
and in most cases the numerical integration needs exactly 17 evaluations of the
integrand. The second effect can be traced down to whether or not the
non-layered single crystal geometrical factors are evaluated with the efficient
approximation \refeqn{circleintegralapproxformulawithspline}, or with a
numerical integration. Numerical integration is required more often when
mosaicities are higher, requested precision ($\epsilon$) is better, and near
back-scattering conditions. Back-scattering conditions occur when the neutron
wavelength is just below the plane thresholds of $2d_{hkl}$, which explains the
series of bumps witnessed in the curves with $\epsilon=\num{e-3}$ and
mosaicities of $3^\circ$. The configuration with $\epsilon=\num{e-5}$ needs
expensive numerical integration in almost all cases, while the configurations
with low mosaicities of $1^\prime$ can do the opposite and use the efficient
approximation in almost all cases.

Although \reffig{timings} show large variations in computational speed for
single crystal models, a few rough conclusions can be drawn. For usual
configurations, cross section evaluations at longer wavelengths,
above \order{\SI{1.5}{\angstrom}}, proceed at a rate
of \order{\SIrange{1}{10}{\mega\hertz}}, and at shorter wavelengths the rate
is \order{\SIrange{0.1}{1}{\mega\hertz}}. If deemed a reasonable trade-off for a
specific use case, it is of course possible to increase the threshold at which
the isotropic approximation is applied, through the \texttt{sccutoff}
configuration parameter, and thus improve the computational speed at lower
wavelengths. Concerning the speed of scattering event sampling, it is in general
seen to be almost negligible compared to the cross section evaluation time. This
is especially true at shorter wavelengths where more planes must be considered,
and is hardly surprising since the code sampling scattering events can benefit
from the work already done when calculating the cross sections, in order to
quickly select just a single plane with which to proceed. As already mentioned,
the treatment of layered single crystals is roughly an order of magnitude slower
than that for single crystals. The exception in \reffig{timings} is for neutron
wavelengths above \SI{4.27}{\angstrom}, where only the $002$ plane
contributes. As this plane is aligned with the lattice $c$-axis, it has
$\thetan=0$, and is in practice treated with code similar to that used in
non-layered single crystals. Finally, it should be noted that the alternative
implementations of layered single crystal code discussed
in \refsec{lcbragg:altrefimpl} would be at the very least 3--4 orders of
magnitude slower than the model used in \reffig{timings}, i.e.\ the one described
in \refsectionrange{lcbragg::defs}{lcbragg::sampling}, thus rendering it
unusable for practical simulation work.

\section{Summary and outlook}\labsec{outlook}

The models presented in \refsectionrange{pcbragg}{incel} cover the capabilities for
elastic thermal neutron scattering in \texttt{NCrystal}
releases \texttt{v1.0.0}--\texttt{v2.2.1}.  Along with the scattering kernel-based
inelastic scattering models introduced in \texttt{NCrystal}
release \texttt{v2.0.0}, the elastic models enable realistic representation of
the most important crystalline materials found at neutron scattering facilities.
The implementations in particular focus on the primary intended usage
of \texttt{NCrystal}: to serve as a physics backend in generic simulation
frameworks like \texttt{Geant4}. Consequently, they focus on ease
of configuration, computational efficiency, and in particular robustness ---
which means that the code provides accurate results for all supported
configurations and neutron state parameters. Thus, the code supports running
with a very large number of reflection planes at shorter wavelengths, both small
and large mosaic spreads in single crystals, and both extreme backwards or
forward scattering. Failure to support some of these cases correctly could lead
to surprises and misleading results for casual users, whose simulations might
just happen to involve one or more of them, which would be unacceptable.

Obviously, the presented models do not account for all effects which one might
find in real crystalline materials, and additional features might in the
future be incorporated in order to reproduce specific characteristics of some
materials --- in particular concerning detailed simulations of objects placed
in the sample position at a neutron scattering instrument, which by design can
be more sensitive to small effects. Examples of such effects might include the
ability to account for effects of the finite crystallite grain sizes,
fluctuations in lattice parameters, site occupancies, chemical disorder (e.g.\
in doped crystals), material strain, longer-scale fluctuations in scattering
length densities (SANS models), and texture effects in polycrystals. Additionally, while support for
multi-phase materials is in principle already included, it currently relies on
user-code for handling the multiple phases. A future release
of \texttt{NCrystal} should make it possible to define multiple phases directly
in the \texttt{NCrystal} configuration strings --- thus ensuring easy and
consistent setup for all users. Finally, it is of course also desirable to
continue to expand
\texttt{NCrystal}'s library of validated material files, and to improve the
performance of existing models.

In addition to developments in capabilities for physics modelling, several
potential improvements to \texttt{NCrystal} of a more technical nature are also
desirable. Firstly, it is planned in the near future to make the framework
multi-thread safe and provide optional interfaces suited for parallel
computing. Secondly, it is obviously very desirable
for \texttt{NCrystal} to be integrated for use as a backend in more simulation
packages than the existing, \texttt{Geant4} and \texttt{McStas}. Obvious
candidates for which some interest has already been expressed in the community
are \texttt{MCNP}~\cite{mcnp5,mcnp6man,mcnp5man,mcnpxman,mcnpx2006},
\texttt{OpenMC}~\cite{openmc}, \texttt{PHITS}~\cite{phits2018}, and
\texttt{VITESS}~\cite{vitess1,vitess2}.

Which of these many additional features will be available in the future, will
depend on the needs of the community as well as the availability of manpower. In
this respect it might be interesting to note that \texttt{NCrystal} release
\texttt{v2.2.0} introduced support for user-contributed plugins, which will
hopefully facilitate contributions to the development of new physics models from
a wider community.

\section*{Acknowledgements}

The work included in \texttt{NCrystal} release \texttt{v1.0.0} was supported in
part by the European Union's Horizon 2020 research and innovation programme
under grant agreement No 676548 (the BrightnESS project). The authors would like
to thank several colleagues for valuable contributions, testing, feedback,
discussions, or other support (in alphabetical order):
M. Bertelsen,
D. Di Julio,
E. Dian,
R. Hall-Wilton,
K. Kanaki,
M. Klausz,
E. Klinkby,
E. B. Knudsen,
J. I. M\'{a}rquez Dami\'{a}n,
V. Maulerova,
A. Morozov,
V. Santoro,
R. Toft-Petersen,
and P. Willendrup.
The authors would also like to acknowledge the ESS-DMSC Computing Centre in
Copenhagen, where the more computationally demanding parts of the validation
work was carried out.

\section*{References}

\bibliography{refs}

\appendix
\setcounter{figure}{0}

\renewcommand*{\thesection}{Appendix \Alph{section}}

\section{Additional reference plots}\labsec{appendix}

For reference, this appendix contains figures for additional parameter values,
complementing the figures presented in
\refthreesections{scbragg::val::mpmathcmp}{scbragg::val::cmpothers}{lcbragg::val::refimplcmp}.

\begin{figure}
  \centering
  \includegraphics[width=0.78\textwidth]{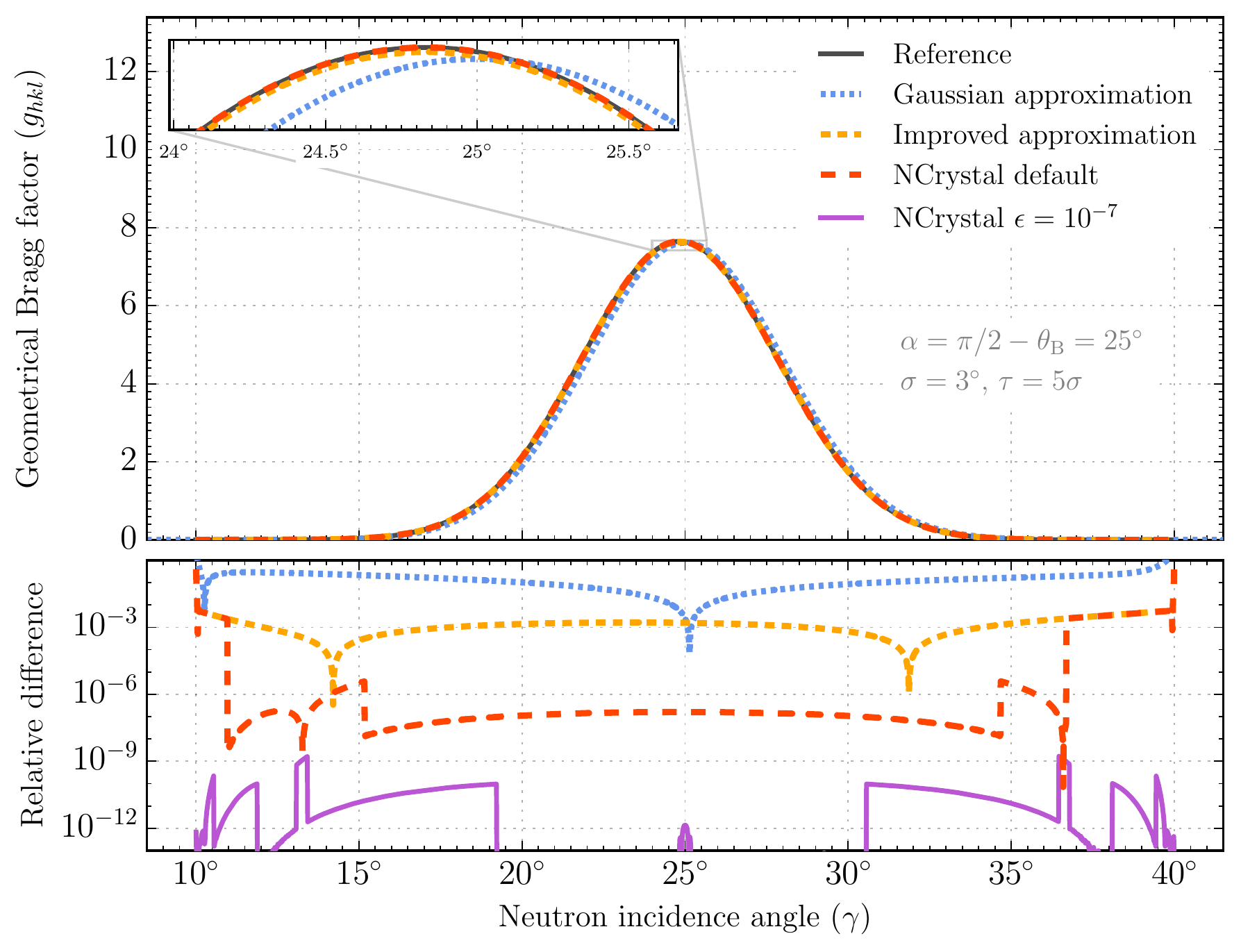}
  \caption{Similar to figure \reffig{gosint_bigmos}, but with an even higher
    mosaicity of $\sigma=3^\circ$ ($7.06^\circ$ FWHM).}
  \labfig{gosint_appendix_first}
\end{figure}

\begin{figure}
  \centering
  \includegraphics[width=0.78\textwidth]{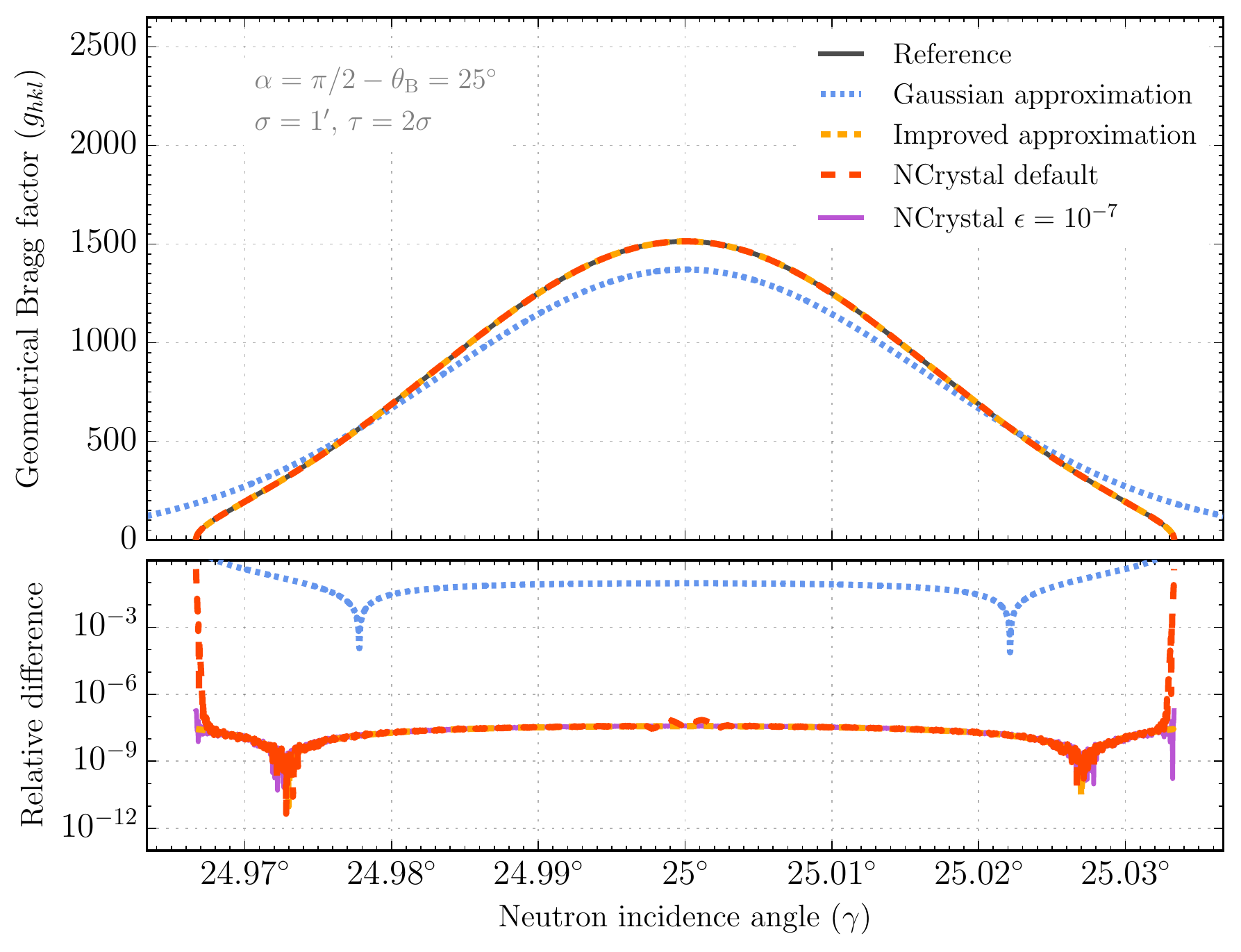}
  \caption{Similar to figure \reffig{gosint_stdcond}, but with a more narrow
    truncation of $\tau=2\sigma$.}
\end{figure}

\begin{figure}
  \centering
  \includegraphics[width=0.78\textwidth]{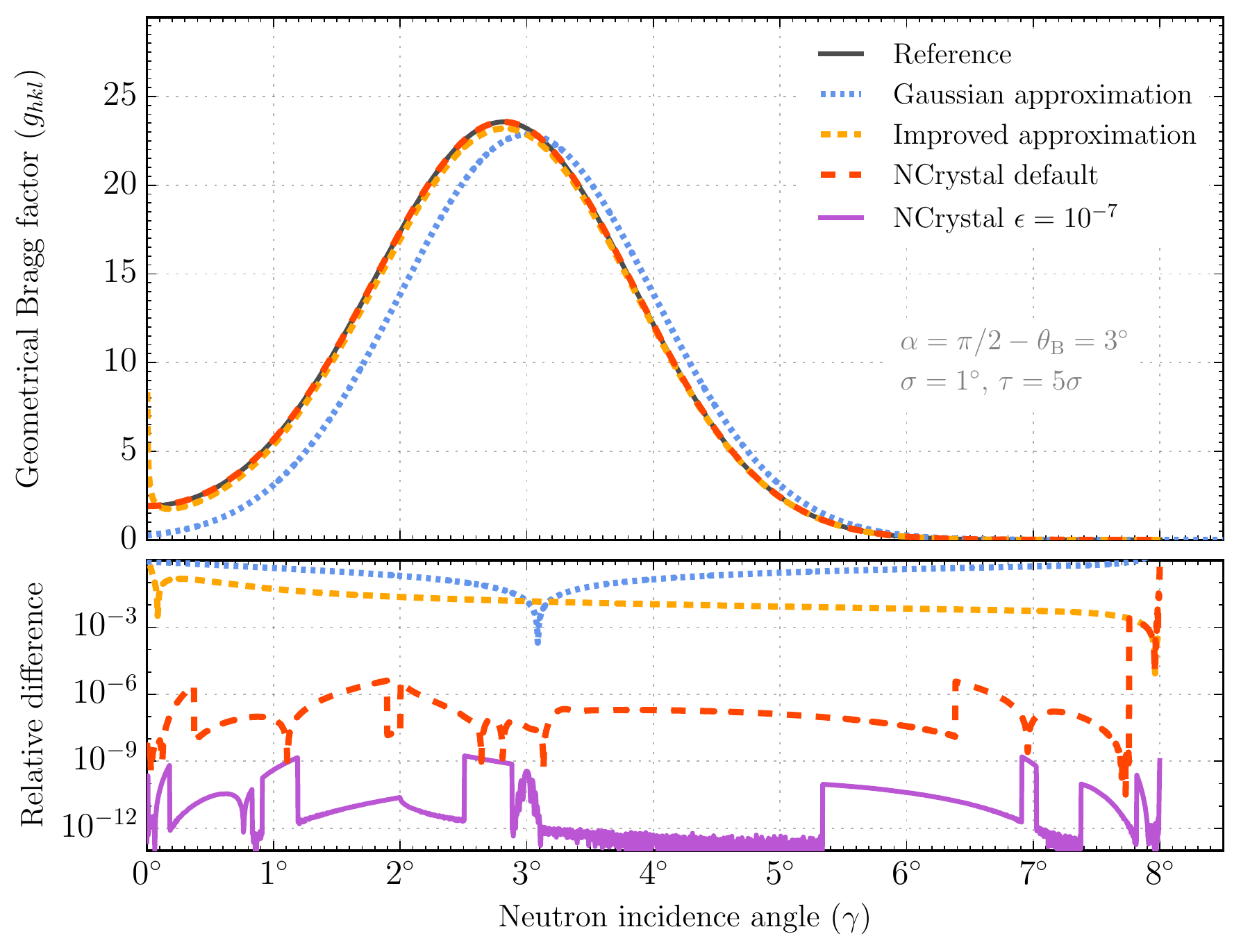}
  \caption{Similar to figure \reffig{gosint_backscat}, but with slightly less
    extreme back-scattering, $\alpha=3\sigma$.}
  \labfig{gosint_appendix_last}
\end{figure}

\begin{figure}
  \centering
  \includegraphics[height=0.6\textwidth]{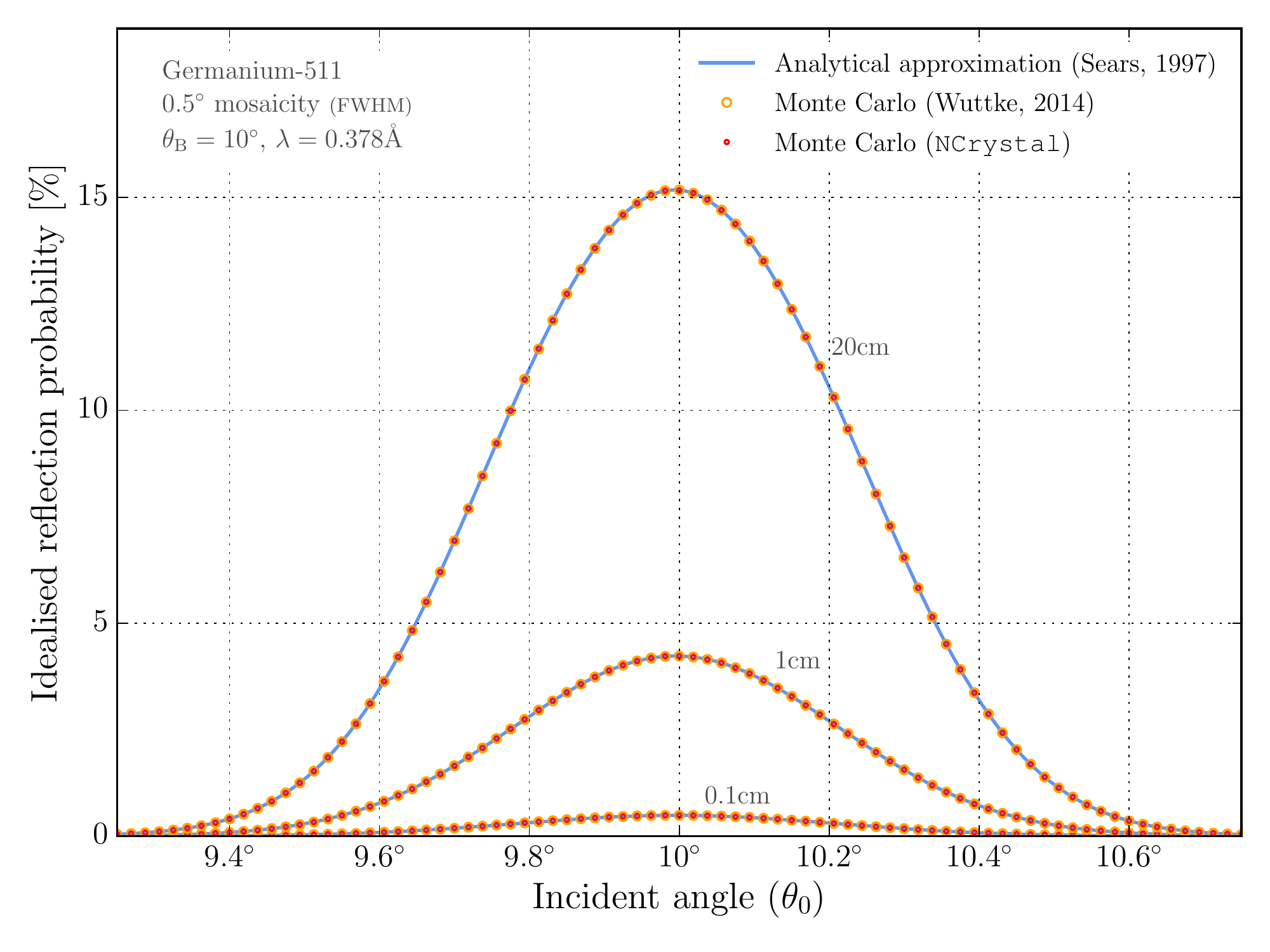}
  \caption{Same curves as in \reffig{wuttkesearsrocking_mos0d5_tb45}, but for a reduced Bragg angle of $10^\circ$.}
  \labfig{appendix_wuttkesearsrocking_the_first}
\end{figure}

\begin{figure}
  \centering
  \includegraphics[height=0.6\textwidth]{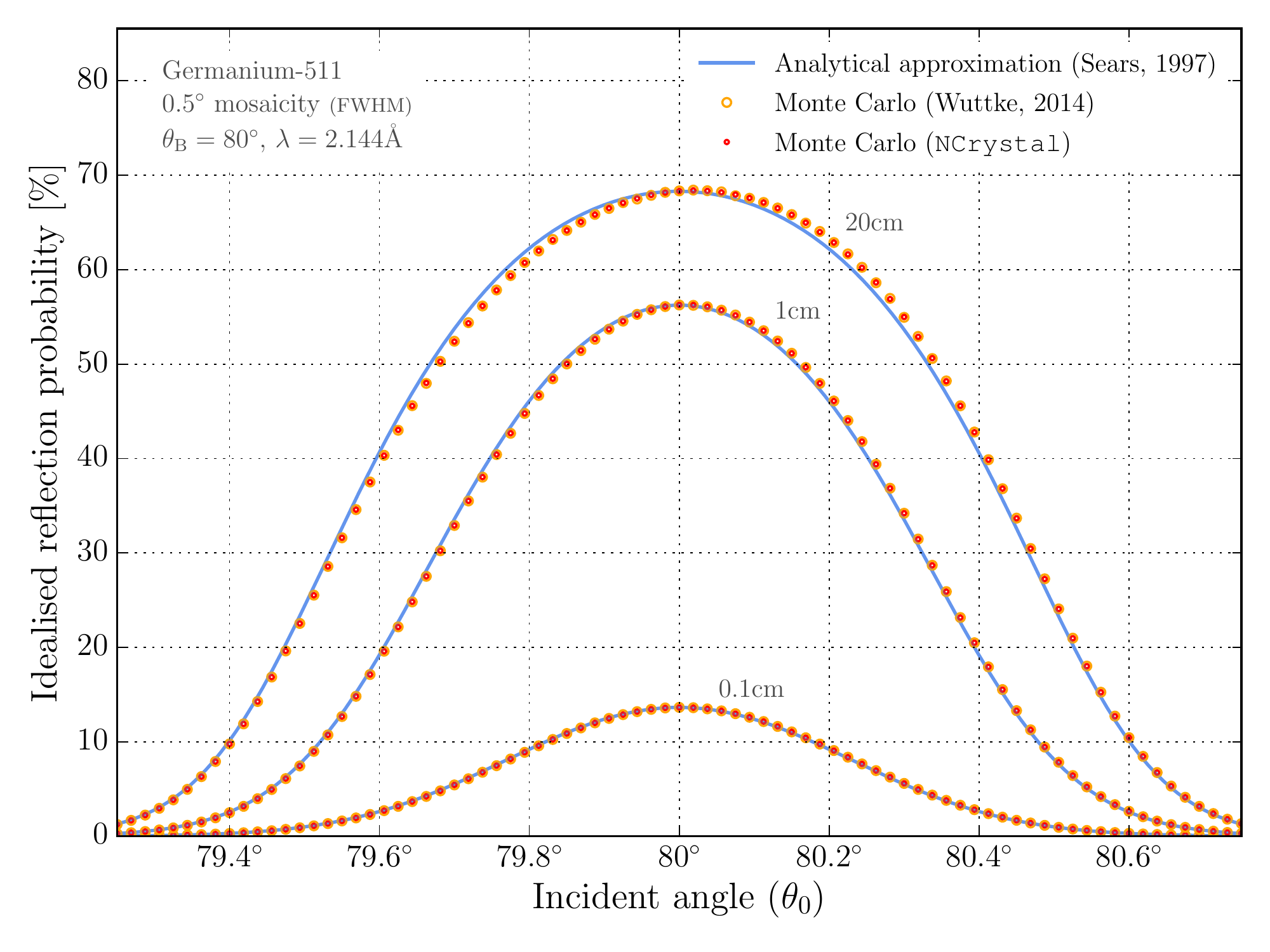}
  \caption{Same curves as in \reffig{wuttkesearsrocking_mos0d5_tb45}, but for
  an increased Bragg angle of $80^\circ$.}
\end{figure}

\begin{figure}
  \centering
  \includegraphics[height=0.6\textwidth]{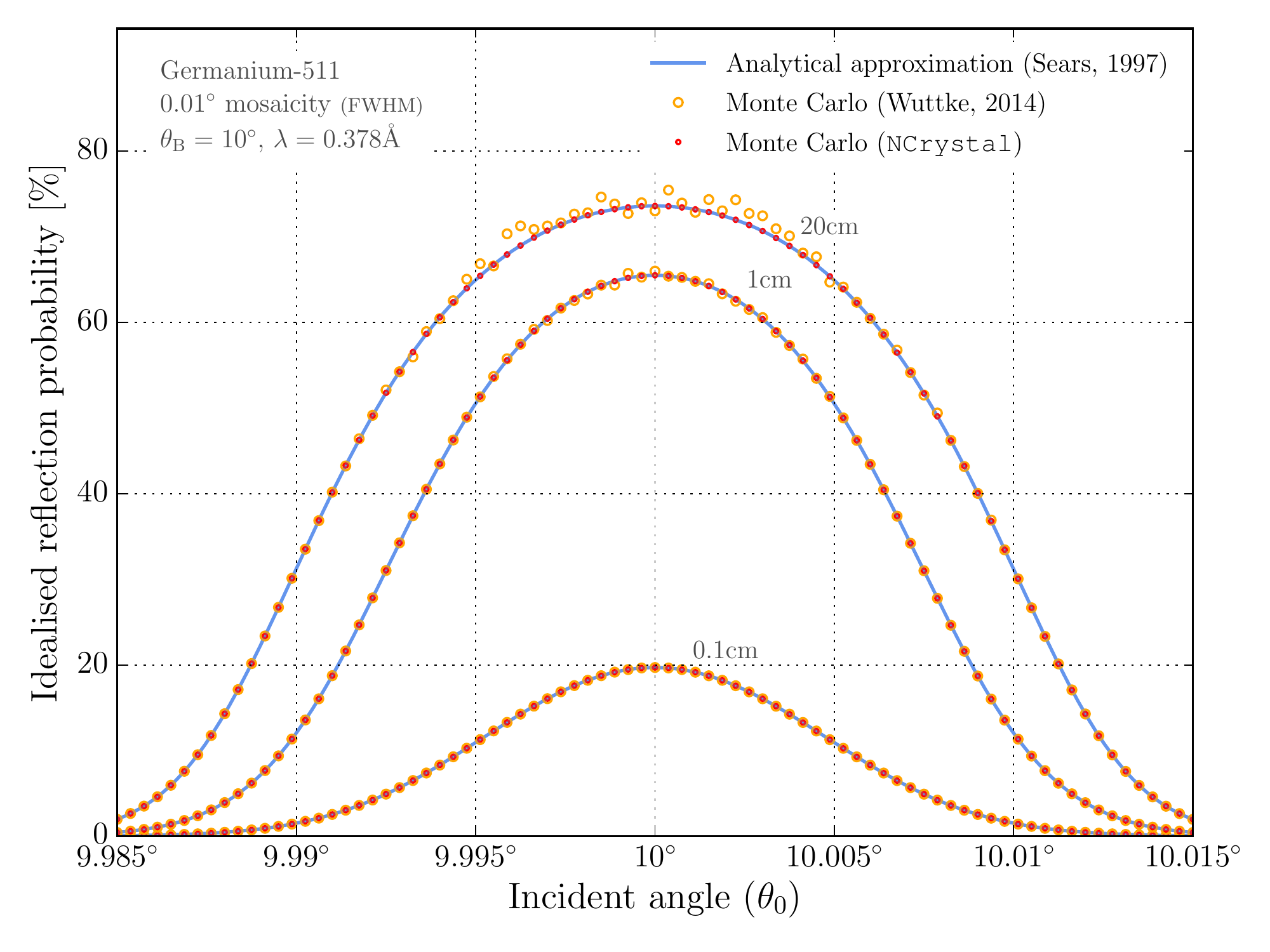}
  \caption{Same curves as in \reffig{wuttkesearsrocking_mos0d5_tb45}, but for a
    smaller FWHM mosaicity of $0.01^\circ$ and a reduced Bragg angle of $10^\circ$.}
\end{figure}

\begin{figure}
  \centering
  \includegraphics[height=0.6\textwidth]{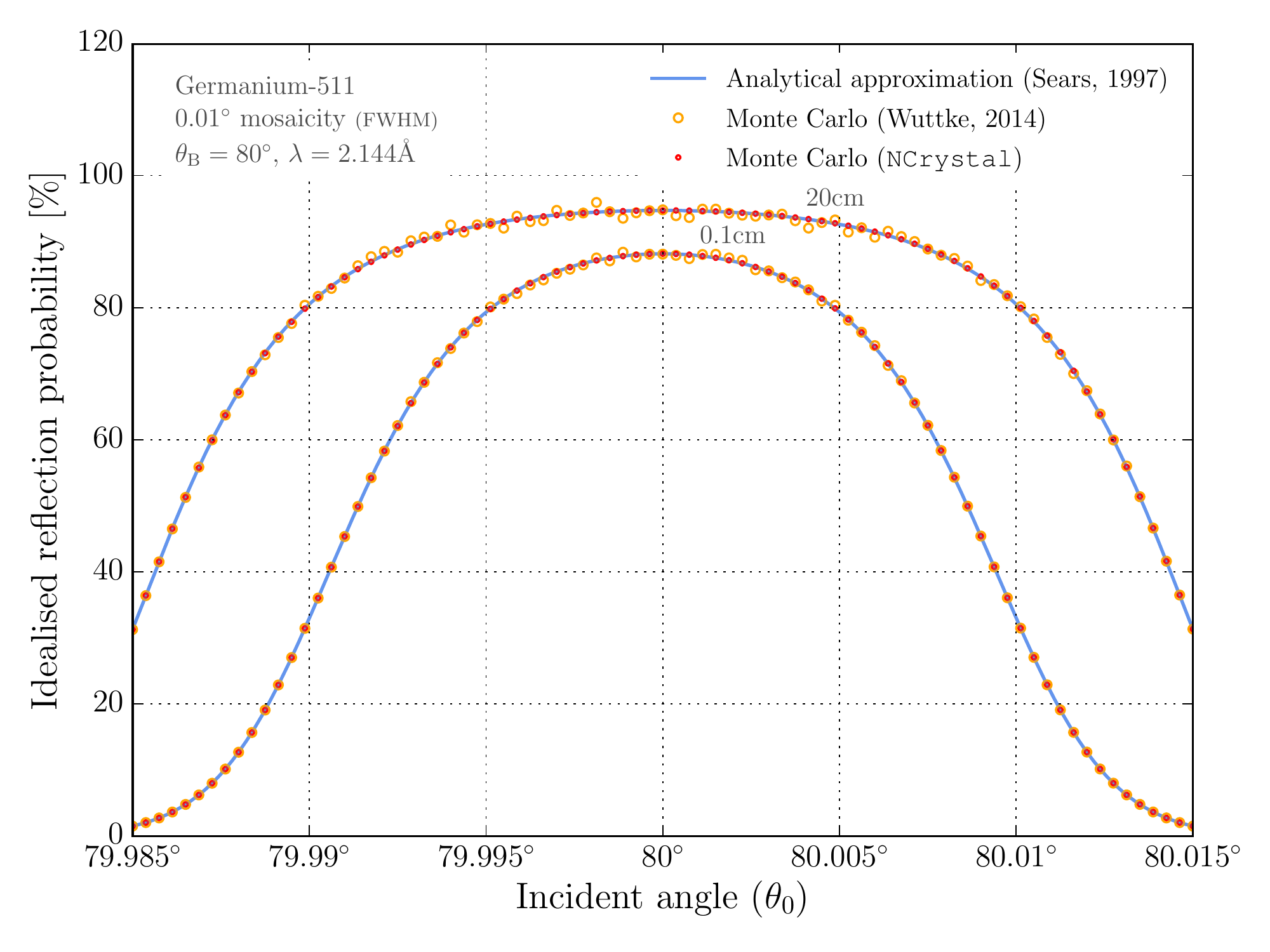}
  \caption{Same curves as in \reffig{wuttkesearsrocking_mos0d5_tb45}, but for a
  smaller FWHM mosaicity of $0.01^\circ$ and an increased Bragg angle of
  $80^\circ$. For clarity the curve for a slab
  thickness of $\SI{1}{\centi\meter}$ is not shown, due to overlaps with the
  $\SI{20}{\centi\meter}$ curve.}
\end{figure}

\begin{figure}
  \centering
  \includegraphics[height=0.6\textwidth]{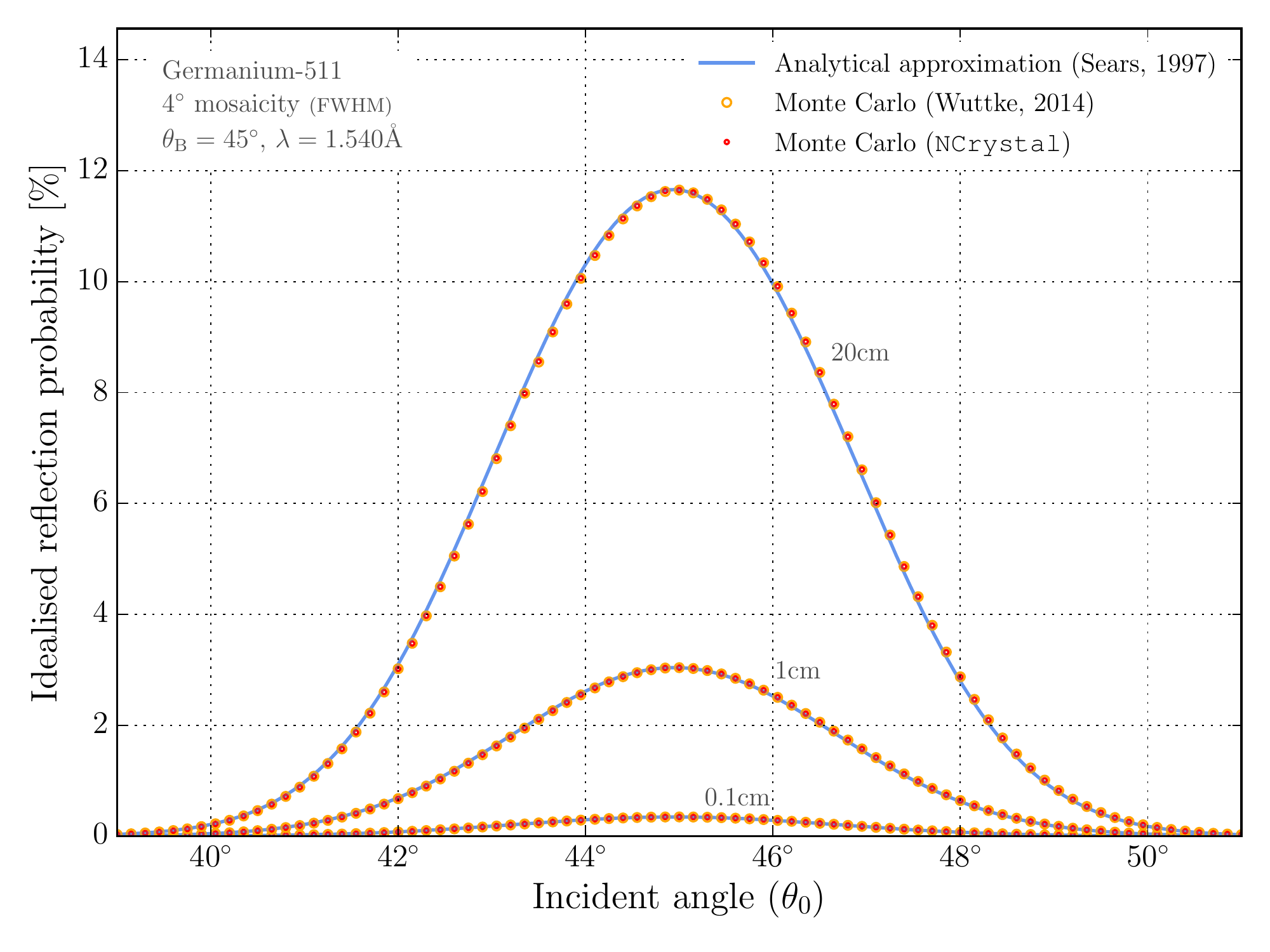}
  \caption{Same curves as in \reffig{wuttkesearsrocking_mos0d5_tb45}, but for a
  larger FWHM mosaicity of $4^\circ$.}
\end{figure}

\begin{figure}
  \centering
  \includegraphics[height=0.6\textwidth]{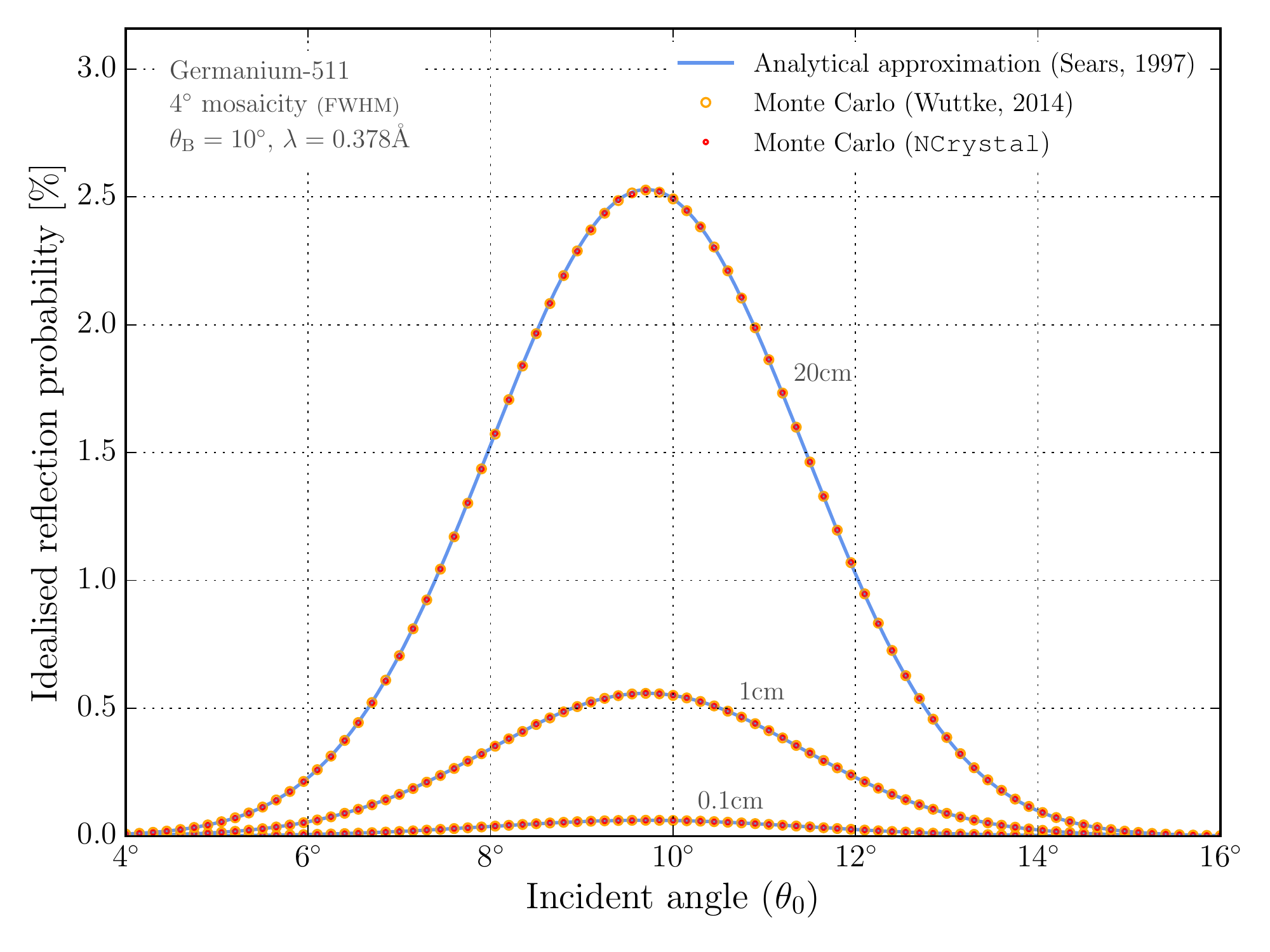}
  \caption{Same curves as in \reffig{wuttkesearsrocking_mos0d5_tb45}, but for a
  larger FWHM mosaicity of $4^\circ$ and a reduced Bragg angle of $10^\circ$.}
  \labfig{appendix_wuttkesearsrocking_the_last}
\end{figure}

\begin{figure}
  \centering
  \includegraphics[height=0.6\textwidth]{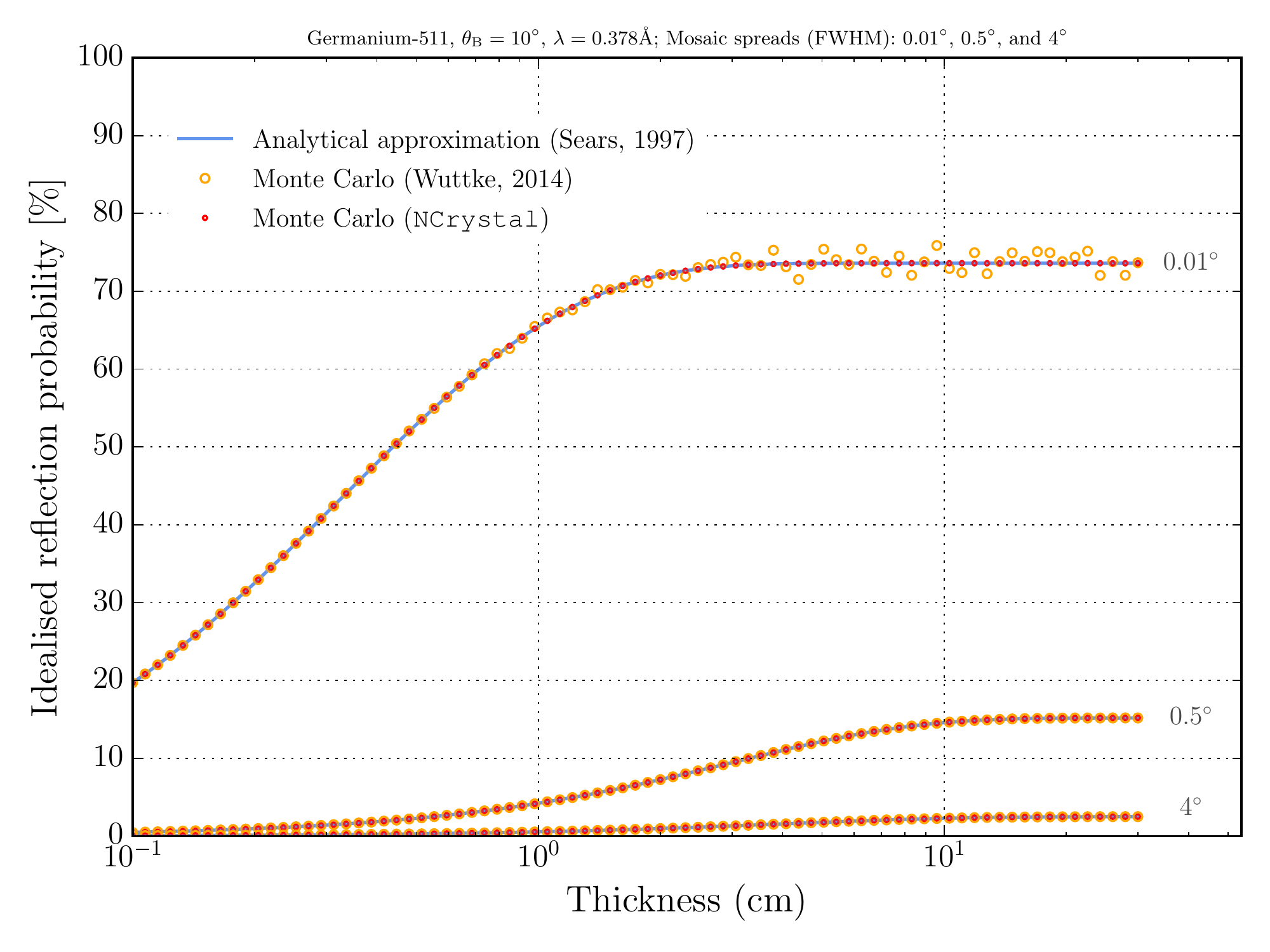}
  \caption{Idealised reflection probabilities as in
    \reffig{wuttkesears_reflvthickness_thetabragg80}, but for neutron incidence
    angle and Bragg angle both $10^\circ$.}
  \labfig{appendix_wuttkesears_reflvthickness_tb10}
\end{figure}

\begin{figure}
  \centering
  \includegraphics[height=0.6\textwidth]{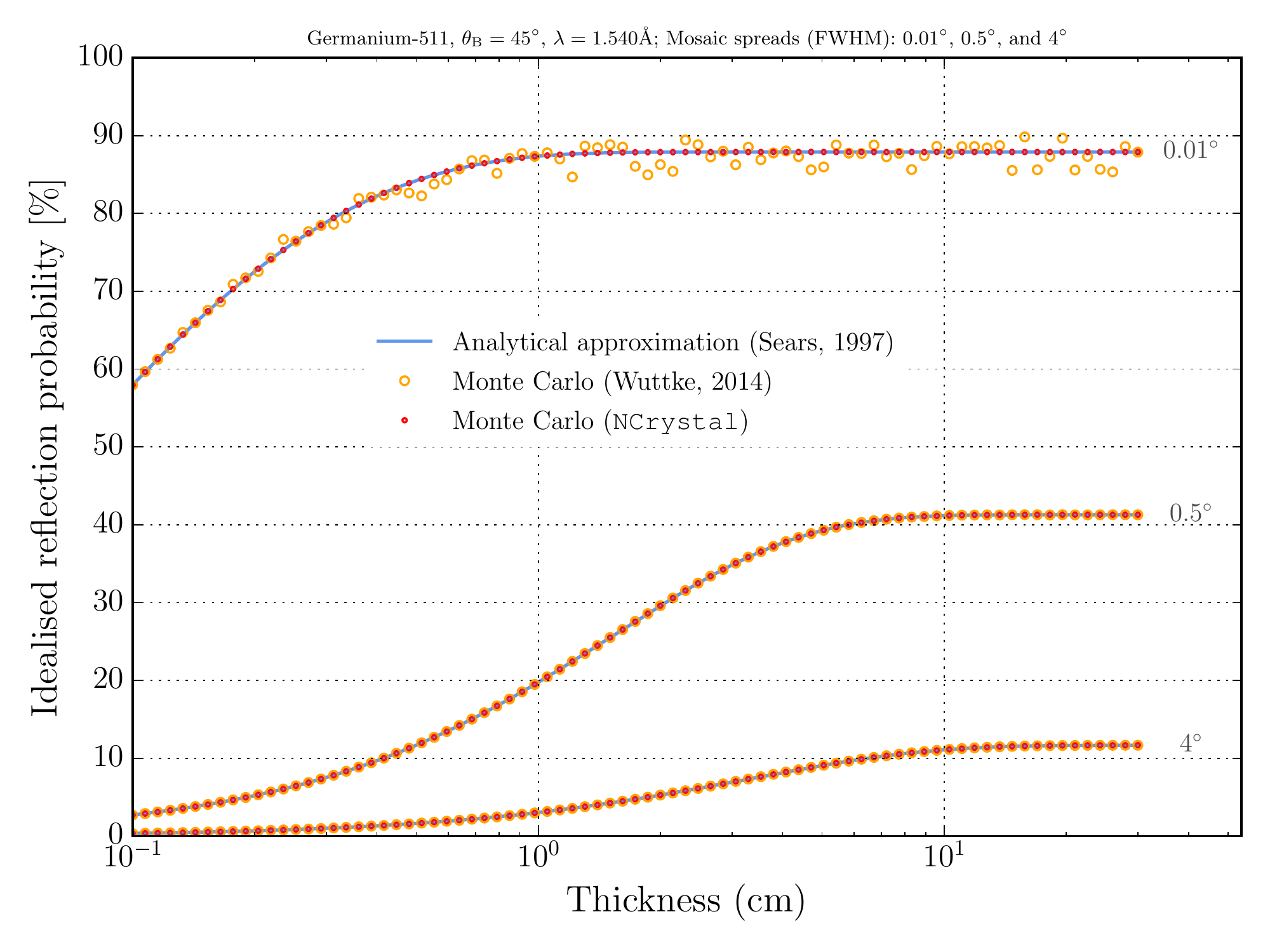}
  \caption{Idealised reflection probabilities as in
    \reffig{wuttkesears_reflvthickness_thetabragg80}, but for neutron incidence
    angle and Bragg angle both $45^\circ$.}
  \labfig{appendix_wuttkesears_reflvthickness_tb45}
\end{figure}

\begin{figure}
  \centering
  \includegraphics[width=0.85\textwidth]{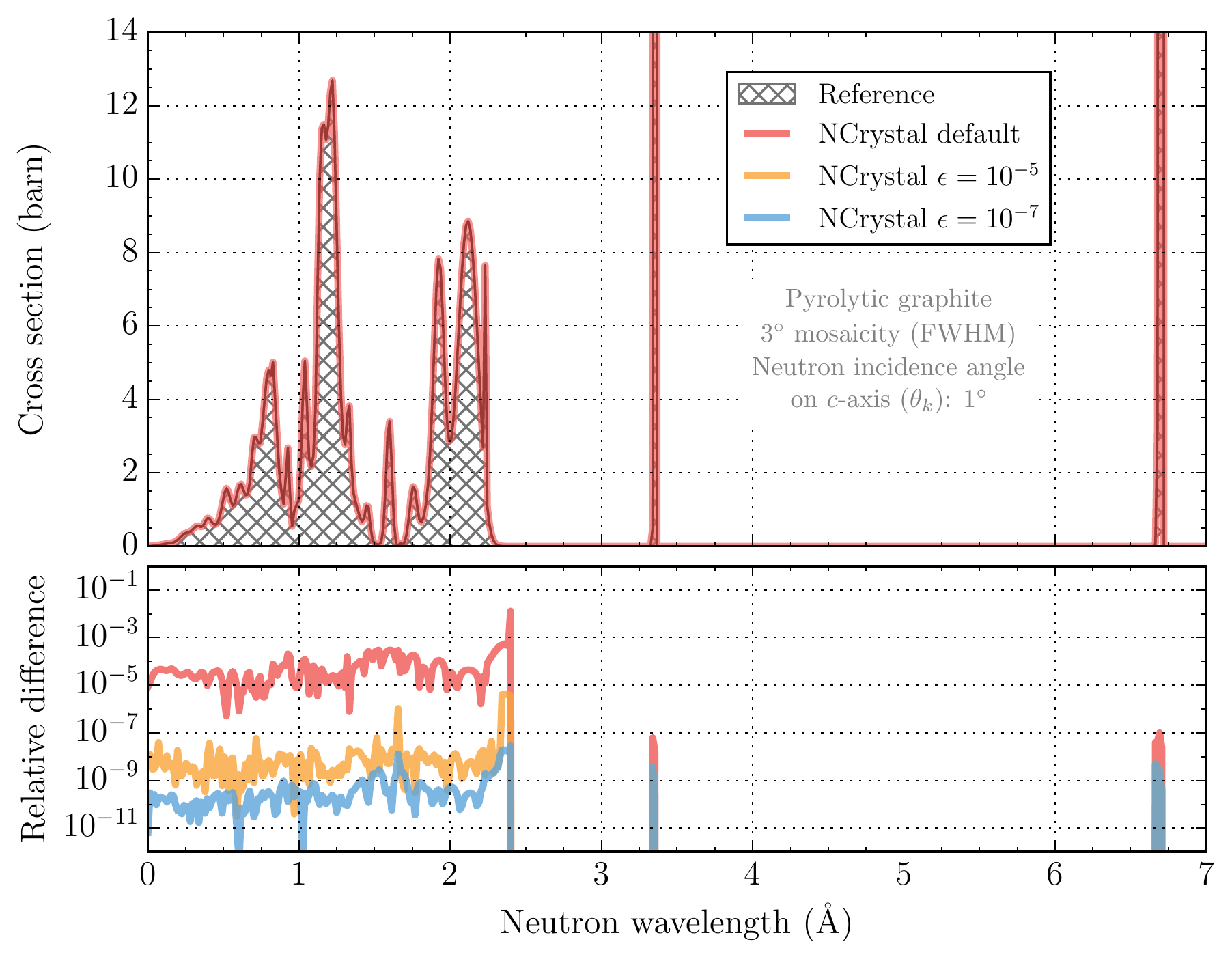}
  \caption{Same curves as in \reffig{lcb_xsvsref_thetak40_mos3deg} but for a neutron incidence of
  $\thetak=1^\circ$.}
  \labfig{lcb_xsvsref_appendix_first}
\end{figure}

\begin{figure}
  \centering
  \includegraphics[width=0.85\textwidth]{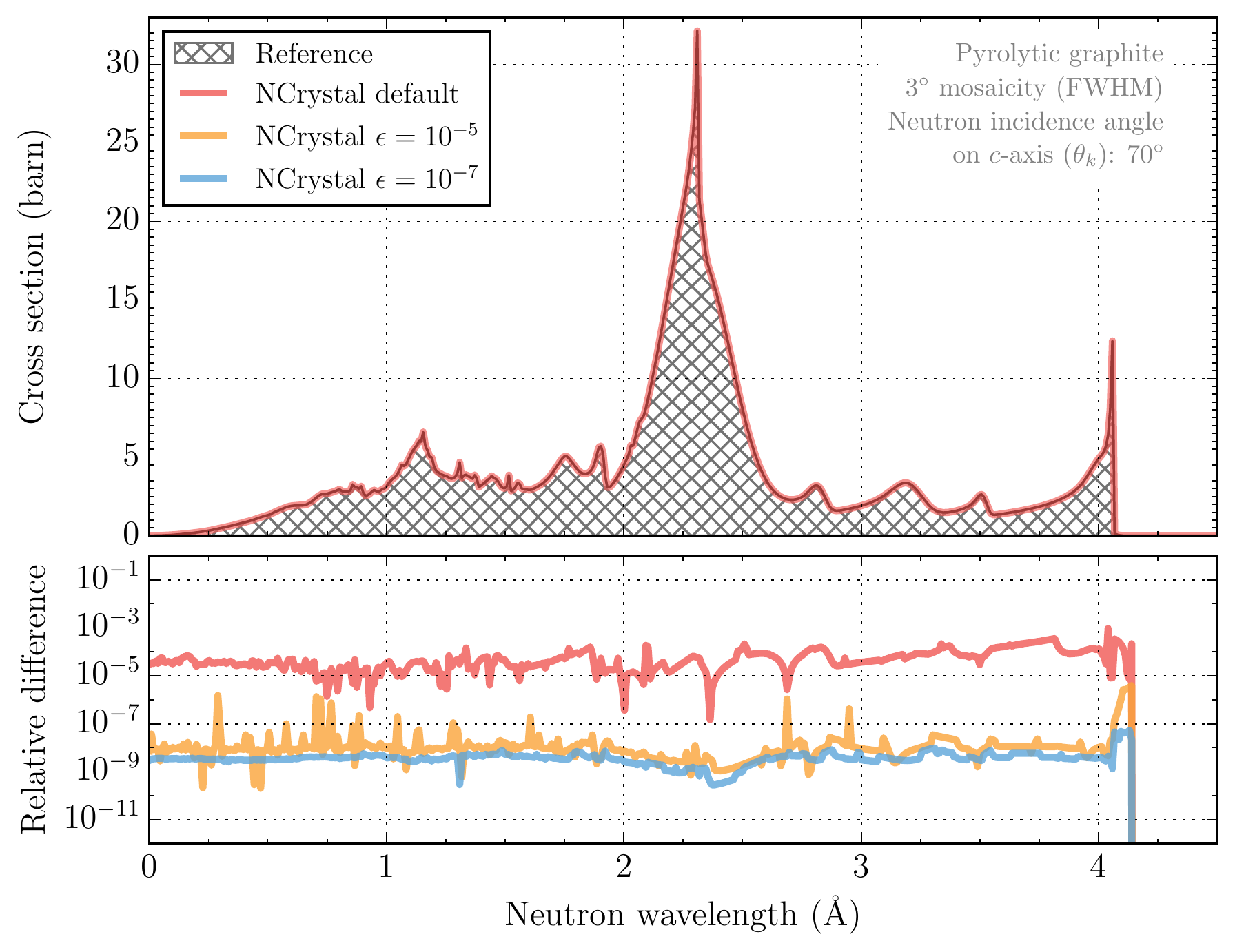}
  \caption{Same curves as in \reffig{lcb_xsvsref_thetak40_mos3deg} but for a neutron incidence of
  $\thetak=70^\circ$.}
\end{figure}

\begin{figure}
  \centering
  \includegraphics[width=0.85\textwidth]{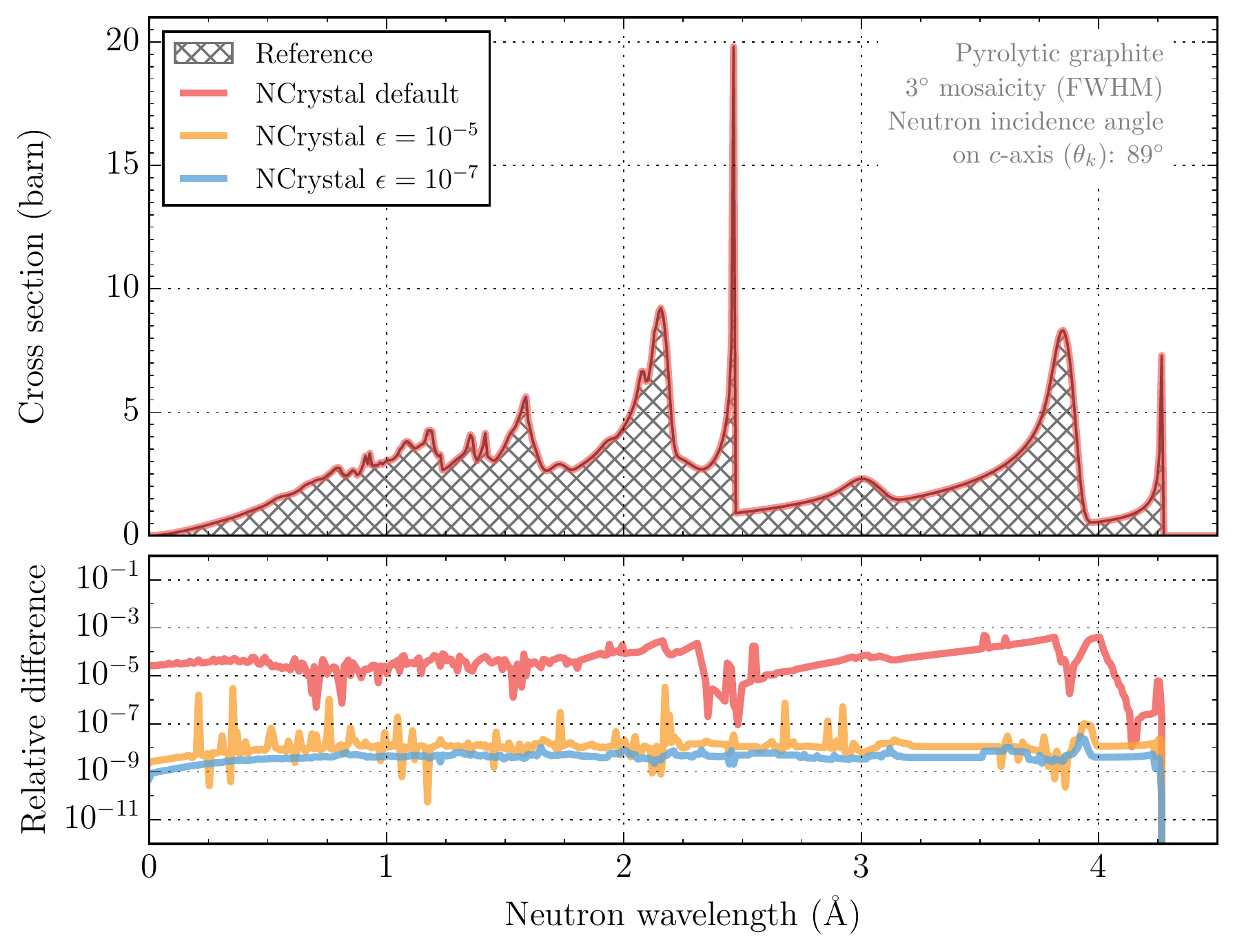}
  \caption{Same curves as in \reffig{lcb_xsvsref_thetak40_mos3deg} but for a neutron incidence of
  $\thetak=89^\circ$.}
\end{figure}

\begin{figure}
  \centering
  \includegraphics[width=0.85\textwidth]{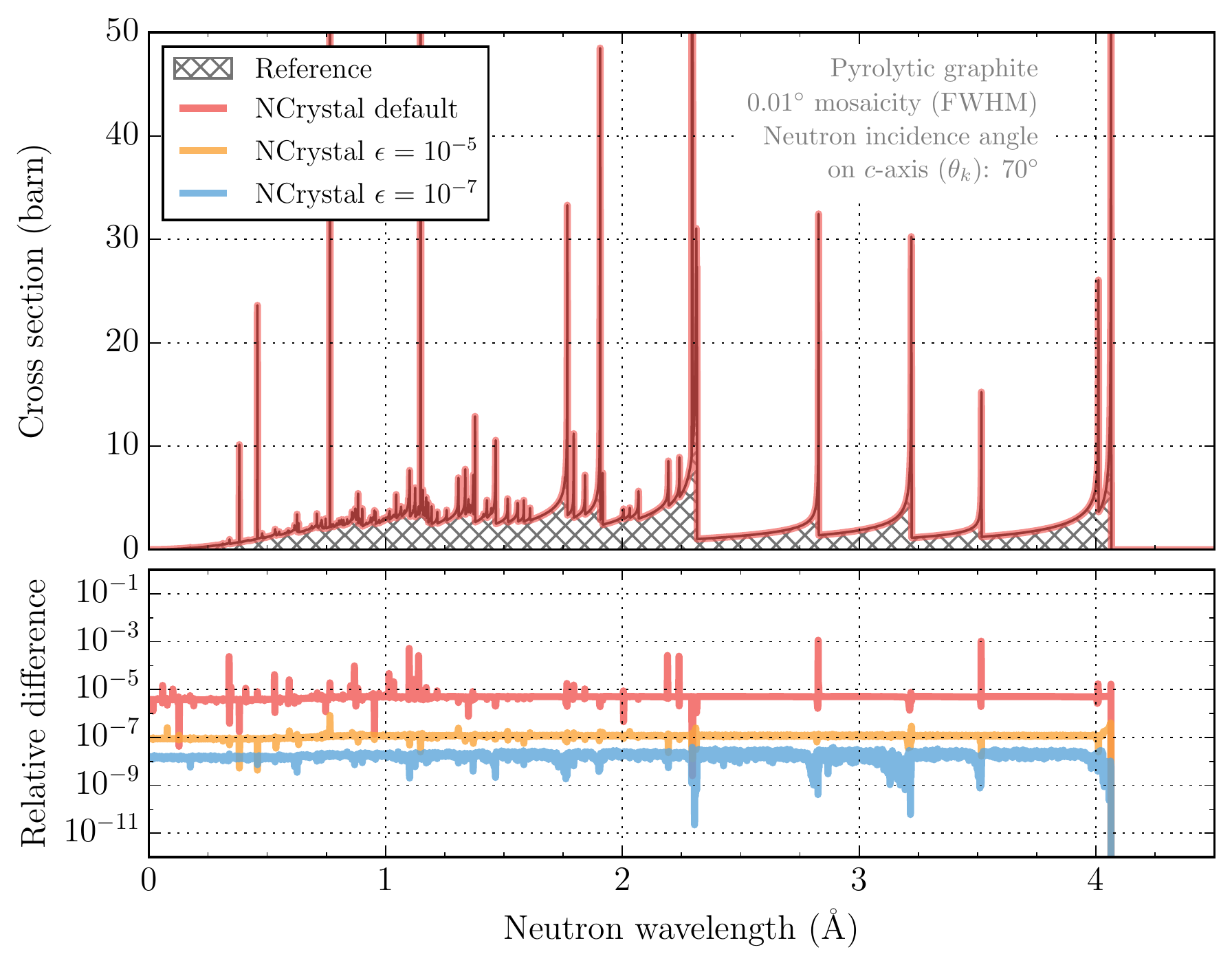}
  \caption{Same curves as in \reffig{lcb_xsvsref_thetak40_mos3deg} but for a neutron incidence of
  $\thetak=70^\circ$ and a FWHM mosaicity of $0.01^\circ$.}
\end{figure}

\begin{figure}
  \centering
  \includegraphics[width=0.85\textwidth]{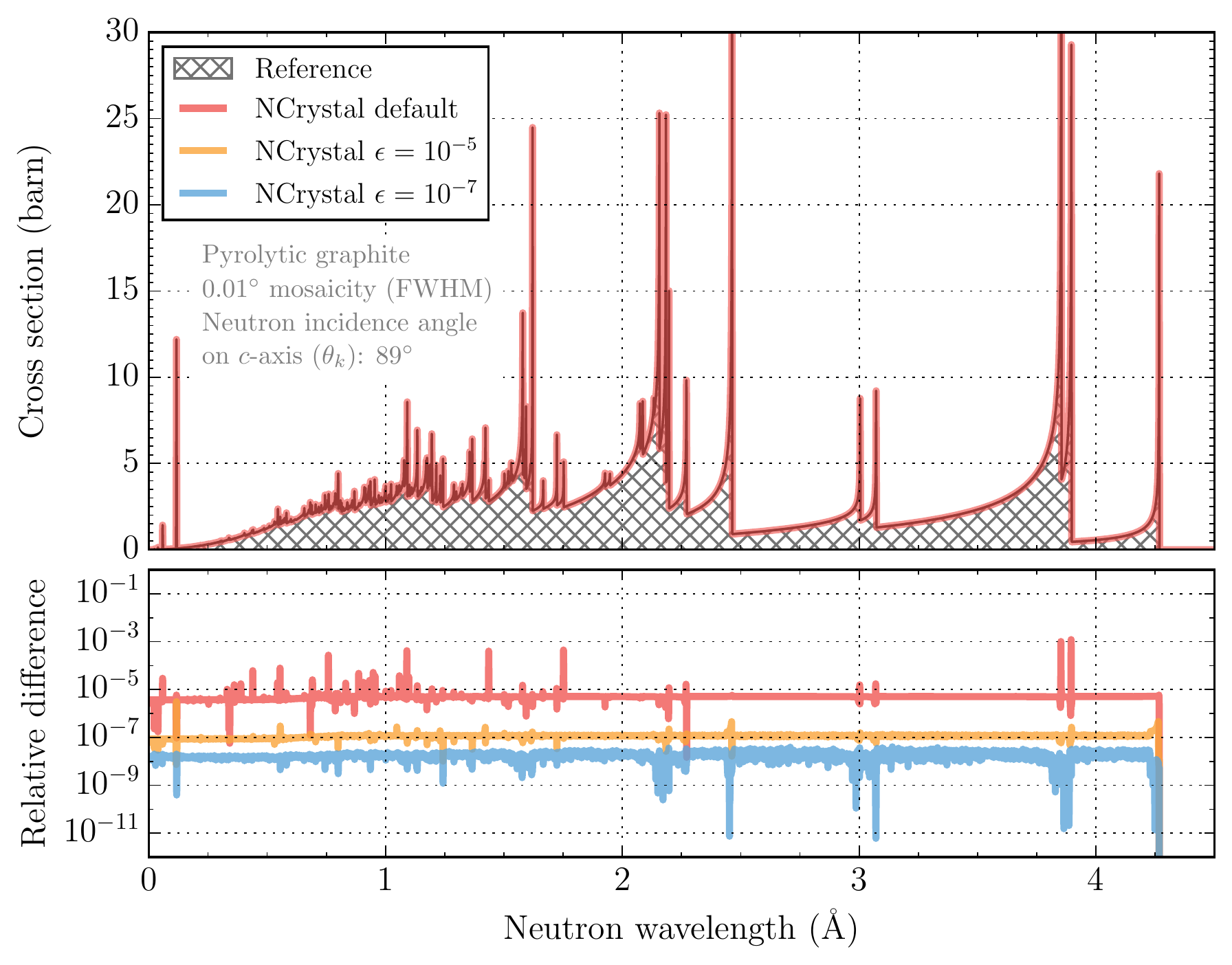}
  \caption{Same curves as in \reffig{lcb_xsvsref_thetak40_mos3deg} but for a neutron incidence of
  $\thetak=89^\circ$ and a FWHM mosaicity of $0.01^\circ$.}
  \labfig{lcb_xsvsref_appendix_last}
\end{figure}

\end{document}